\documentclass[
 aip,
 jmp,
 amsmath,
 amssymb,
]{revtex4-1}

\usepackage{graphicx}
\usepackage{dcolumn}
\usepackage{bm}
\usepackage{ulem} 

\begin{document}


\title[Path probability ratios for Langevin dynamics]{Path probability ratios for Langevin dynamics - exact and approximate}


\author{S. Kieninger}
\affiliation{Department of Biology, Chemistry, Pharmacy, Freie Universit\"{a}t Berlin, Arnimallee 22, D-14195 Berlin, Germany}

\author{B. G. Keller}
\email[]{bettina.keller@fu-berlin.de}
\affiliation{Department of Biology, Chemistry, Pharmacy, Freie Universit\"{a}t Berlin, Arnimallee 22, D-14195 Berlin, Germany}

\date{\today}


\begin{abstract}
Path reweighting is a principally exact method to estimate dynamic properties from biased simulations - provided that the path probability ratio matches the stochastic integrator used in the simulation. 
Previously reported path probability ratios match the Euler-Maruyama scheme for overdamped Langevin dynamics.
Since MD simulations use Langevin dynamics rather than overdamped Langevin dynamics, this severely impedes the application of path reweighting methods. 
Here, we derive the path probability ratio $M_L$ for Langevin dynamics propagated by a variant of the Langevin Leapfrog integrator.
This new path probability ratio allows for exact reweighting of Langevin dynamics propagated by this integrator. 
We also show that a previously derived approximate path probability ratio $M_{\mathrm{approx}}$ differs from the exact $M_L$ only by $\mathcal{O}(\xi^4\Delta t^4)$, and thus yields highly accurate dynamic reweighting results.
($\Delta t$ is the integration time step, $\xi$ is the collision rate.)
The results are tested and the efficiency of path-reweighting is explored using butane as an example.
\end{abstract}


\maketitle 



\section{Introduction}
Molecular dynamics are astonishingly complex, and occur on a wide range of length and timescales \cite{Gopich:2020, Lane:2013, Dror:2012}. 
To elucidate the mechanisms by which different parts of a molecular system interact and how macroscopic properties arise from these interactions, 
molecular dynamics (MD) simulations have become an indispensable tool \cite{Barros2020, Harpole2018, Cournia2017, Badaoui2018, Joswig2020, Mey2020}.
Because the timescales covered by MD simulations are often orders of magnitude lower than the slowest timescale of the system, a wide variety of enhanced sampling techniques have been developed which distort the dynamics of the simulation such that rare molecular transitions occur more frequently. 
This can be achieved by raising the temperature, or by adding a bias to the potential energy function \cite{Tuckerman2010, Frenkel2002}. 
How to extract the correct values of dynamical properties (mean-first passage times, residence times, binding rates or transition probabilities) from these accelerated dynamics is an open question, and a very active field of research.
The goal of dynamical reweighting methods is to estimate dynamical properties of the system at a target state $\widetilde S$ from a trajectory generated at simulation state $S$. 
$S$ could correspond to a higher temperature, or to a biased potential. 
Starting points for the derivation of dynamical reweighting methods are Kramers rate theory \cite{deOliveira:2007, Tiwary2013, Valsson2016, Casasnovas:2017}, the likelihood function for estimating the transition probabilities from MD trajectories \cite{Wu2014, Mey:2014, Wu2016, Stelzl2017}, or a discretization of the Fokker-Planck equation \cite{Bicout:1998, Rosta2014, Badaoui2018, Donati:2018b} .
The methods differ in the ease of use and the severity of the assumptions they make \cite{Kieninger2020}.
A principally exact formalism to reweight dynamic properties are path reweighting methods, which have been reported already early on \cite{Zuckerman:1999, Woolf:1998, Zuckerman:2000, Xing:2006, Adib2008}.
In path reweighting methods the trajectory generated at state $S$ is split into short paths $\omega$.
Then the path probability $\widetilde P_L(\omega; \Delta t| (x_0,v_0))$ of a given $\omega$ at the target state $\widetilde S$ is calculated by reweighting the path probability $P_L(\omega; \Delta t| (x_0,v_0))$ of $\omega$ at the simulation state $S$
\begin{eqnarray}
    \widetilde{P}_L(\omega; \Delta t| (x_0,v_0)) \approx M \cdot P_L(\omega; \Delta t| (x_0,v_0)) \, . \label{equ:path_reweighing}
\end{eqnarray}
$(x_0,v_0)$ is the initial state of the path $\omega$, and $\Delta t$ is the integration time step. 
$M(\omega)$ is the path probability ratio or reweighting factor. 
Eq.~\ref{equ:path_reweighing} is exact if the path probability ratio $M = \widetilde P_L(\omega; \Delta t| (x_0,v_0)) /P_L(\omega; \Delta t| (x_0,v_0))$ is derived from the numerical integration scheme used to generate $\omega$.
The mathematical basis for path reweighting methods is the Girsanov theorem \cite{Girsanov1960, Oeksendal2003}, or else they can be derived from the Onsager-Machlup action \cite{Onsager1953, Woolf:1998, Zuckerman:1999, Zuckerman:2000, Xing:2006}.
A prerequisite for path reweighting is that a stochastic integrator is used in the MD simulation, e.g. a Langevin thermostat. 
However, it has been challenging to apply path reweighting to simulations of large molecular systems. 
For example, the variance of the reweighting estimators increase rapidly with increasing path length, 
such that for long paths reweighting becomes inefficient compared to direct simulation of the target state. 
Combining path reweighting techniques with Markov state models (MSMs) alleviates this problem. 
\cite{Prinz:2011c, Schuette:2015, Donati2017, Donati2018}.
In MSMs \cite{Huisinga2003, Swope2004, Buchete:2008, Keller:2010, Prinz2011, Prinz:2011b, Husic:2018} the dynamics of the system is represented by transitions between discrete states in the conformational space of the molecular system, where the lag time $\tau$ of the transition is much shorter than the slow timescales of the system. 
Thus, only short paths of length $\tau$ are needed to estimate and reweight the transition probabilities. 
Second, a number of technical difficulties arise. 
The path probability ratio $M$ decreases exponentially with the path length $\tau$, such that the standard numerical accuracy is quickly exceeded. 
This problem can be solved by using high precision arithmetic libraries \cite{Donati2018}.
To calculate the path probability ratio $M$, one needs to know the trajectory and the random numbers of the stochastic integrator at every integration time step. 
Writing this information to disc at every integration time step is not a workable option.
We therefore proposed to calculate the path reweighting factor ``on-the-fly'' during the simulation, and to write out intermediate results at regular intervals, e.g. whenever the positions are written to disc. 
The additional storage requirements and computational costs for the ``on-the-fly''-calculations are negligible compared to the overall cost of the simulation \cite{Donati2017, Donati2018}. 
Having solved the technical challenges, we tested the path reweighting method on several peptides using path lengths of up to $\tau = 600\, \mathrm{ps}$ \cite{Donati2017, Donati2018}. 
Applications to larger systems and longer path lengths are likely within reach.

Yet, the equation for the path probability ratio $M$ poses a barrier to a more widespread use of path reweighting methods.
Because $M$ is derived from the stochastic integration scheme used to simulate the system, one cannot readily apply a path probability ratio derived for one integration scheme to a simulation generated by another integration scheme. 
In temperature reweighting, i.e.~when simulation and target state differ in the temperature,
only the random term of the stochastic integrator is effected by the change in temperature. 
Path probability ratios for temperature reweighting have been constructed by rescaling the normal distributions of the random or noise terms of the stochastic integration scheme \cite{Chodera:2011, Prinz:2011c}.
In potential reweighting, i.e.~when simulation and target state differ in the potential energy function, one needs to account for changes in the drift terms of the stochastic integration scheme. 
The path probability ratio $M_o$ for the Euler-Maruyama scheme for overdamped Langevin dynamics has been reported multiple times \cite{Woolf:1998, Zuckerman:1999, Zuckerman:2000, Schuette:2015}. 
However, the dynamics of large molecular systems is better reproduced by Langevin dynamics, and MD programs implement a wide variety of Langevin integration schemes \cite{vanGunsteren:1981, Brunger1984, Stoltz:2007, Bussi2007, Ceriotti2009, Izaguirre2010, Goga:2012, Leimkuhler:2013, Sivak:2014, Fass2018}.
The time-continuous Onsager-Machlup action for Langevin dynamics has been reported \cite{Xing:2006}, but to the best of our knowledge path probability ratios for Langevin integration schemes $M_L$ have not yet been reported.
Thus, exact path reweighting for Langevin dynamics has not been possible, so far.

In refs.~\onlinecite{Donati2017, Donati2018}, we demonstrated that path reweighting can be applied to biased simulations of large molecular systems nonetheless. 
We used an approximate path probability ratio $M_{\mathrm{approx}}$ which is based on the path probability ratio for the Euler-Maruyama scheme, but uses the random numbers that are generated during the Langevin MD simulation. 
We tested $M_{\mathrm{approx}}$ extensively, and for low-dimensional model systems and for molecular systems this approximate path probability ratio yielded very accurate results.
In these two publications we used a variant of the Langevin Leapfrog integration scheme developed by J. A.~Izaguirre, C. R.~Sweet, and V.S.~Pande \cite{Izaguirre2010} to propagate the system. 
Both the Langevin Leapfrog integration scheme and its variant are implemented in OpenMM \cite{Eastman2017} (see appendix \ref{appendix:derivation_ISP_scheme}).
We will abbreviate the variant by ``ISP scheme''.
In this contribution, we derive the path probability ratio $M_L$ for Langevin dynamics propagated by a variant of the Langevin Leapfrog integrator\cite{Izaguirre2010}. 
$M_L$ allows for exact reweighting of Langevin dynamics (section \ref{sec:Langevin}). 
We analyze why $M_{\mathrm{approx}}$ is an excellent approximation to $M_L$ (section \ref{sec:Approximate_path_probability_ratio}), and we discuss whether there are scenarios in which $M_o$ is a viable approximation to $M_L$ (section \ref{sec:Comparing_Langevin_and_overdamped_Langevin_dynamics}).
The general framework of the path reweighting equations, and the corresponding equations for the Euler-Maruyama scheme are summarized in sections \ref{sec:preliminaries} and \ref{sec:overdamped_Langevin}.
Section \ref{sec:methods} reports the computational details.

\section{Path reweighting}
\label{sec:preliminaries}
The path probability $P(\omega; \Delta t  | (x_0, v_0))$ is the probability to generate a time-discretized path $\omega=(x_0, x_1, \dots, x_n)$ starting in a pre-defined initial state $(x_0,v_0)$ at the simulation potential $V(x)$.
The notation emphasizes that the probability is conditioned on an initial state $(x_0, v_0)$ and that the path has been generated with a fixed time step $\Delta t$, whereas $\omega$ is the argument of the function.
In short, $P(\omega; \Delta t  | (x_0, v_0))$ maps a path in position space to  a probability.
Its functional form depends on the integration scheme used to generate $\omega$ and the potential energy function. 
The path probability ratio is the ratio between the probability $\widetilde P(\omega; \Delta t  | (x_0, v_0))$ to generate a path $\omega$ at a target potential
\begin{eqnarray}
    \widetilde{V}(x) &=& V(x) + U(x) \,  
\label{equ:def_perturbed_pot}
\end{eqnarray}
and the probability $P(\omega; \Delta t  | (x_0, v_0))$ to generate the same path $\omega$ at the simulation potential $V(x)$
\begin{eqnarray}
    M(\omega; \Delta t  | (x_0, v_0)) &=& \frac{\widetilde P(\omega; \Delta t  | (x_0, v_0))}{P(\omega; \Delta t  | (x_0, v_0))} \,.
\label{eq:M_omega}    
\end{eqnarray}
The potential energy function $U(x)$ is usually called perturbation or bias.
In integration schemes for stochastic dynamics, random numbers are used to propagate the system.
If a single random number is drawn per integration step, then the probability to generate $\omega$ is equal to the probability $P(\eta)$ to generate the corresponding random number sequence $\eta =(\eta_0, \eta_1, \dots, \eta_{n-1})$
\begin{eqnarray}
    P(\omega; \Delta t  | (x_0, v_0)) = P(\eta) \, ,
\label{eq:P_omega_equals_P_eta}
\end{eqnarray}
where $\omega$ and $\eta$ are linked by the equations for the integration scheme.
Since the random numbers $\eta_{k}$ are drawn from a Gaussian distribution with zero mean and unit variance, the functional form of $P(\eta)$ is
\begin{equation}
    P(\eta) = N \exp \left( - \frac{1}{2}\sum_{k=0}^{n-1}\eta_{k}^2 \right) \, , \quad N= \left(\frac{1}{2 \pi} \right)^{\frac{n}{2}}\, . 
\label{equ:Gaussian}
\end{equation}

$P(\eta)$ is a function that maps a random number sequence to a probability.
One can interpret eq.~\ref{eq:P_omega_equals_P_eta} as a change of variables from $\omega$ to $\eta$, where the change is defined by the equations for the integration scheme.

Suppose $\eta$ is the random number sequence needed to generate $\omega$ at a simulation potential $V(x)$. 
To generate the same path at a target potential $\widetilde V(x)$, one would need a different random number sequence 
$\widetilde \eta = (\widetilde\eta_0, \widetilde\eta_1, \dots, \widetilde\eta_{n-1})$ with 
\begin{eqnarray}
    \widetilde \eta_k &= & \eta_k +\Delta \eta_k \, .
\label{eq:eta_plus_Delta_eta}
\end{eqnarray}
$\Delta \eta_k$ is the random number difference, and it depends on the integration scheme and the difference between the two potentials.
The random number probability ratio is the ratio between the probability of drawing $\eta$ and the probability of drawing $\widetilde \eta_k$
\begin{eqnarray}
    \frac{P(\widetilde \eta)}{P(\eta)}
    &=& \frac{N \exp \left( - \frac{1}{2}\sum\limits_{k=0}^{n-1}(\eta_{k} +\Delta \eta_k)^2 \right)}{N \exp \left( - \frac{1}{2}\sum\limits_{k=0}^{n-1}\eta_{k}^2 \right)} \cr
    &=& \exp \left( - \sum\limits_{k=0}^{n-1} \eta_{k} \cdot \Delta \eta_{k} \right) 
    \cdot \exp \left( - \frac{1}{2} \sum\limits_{k=0}^{n-1} (\Delta \eta_{k})^2 \right) \, . 
\label{eq:M_eta}
\end{eqnarray}

Mathematically the following has happened in the previous paragraph. 
The path $\omega$ remained unchanged.
The functional form of the path probability has changed: $\widetilde P(\omega; \Delta t |(x_0, v_0))$, because the potential energy enters the equations for the integration scheme.
Likewise, the change of variables from $\omega$ to $\widetilde \eta$ has changed. 
The functional form of the random number probability remains the same (eq.~\ref{equ:Gaussian}).
The analogon to eq.~\ref{eq:P_omega_equals_P_eta} for the target potential is
\begin{eqnarray}
    \widetilde{P}(\omega; \Delta t  | (x_0, v_0)) &=& P(\widetilde \eta) \, ,
\end{eqnarray}
where $\omega$ and $\widetilde \eta$ are linked by the equations for the integration scheme using $\widetilde V(x)$. 
Given the two changes of variables for the simulation and the target potential, the path probability ratio (eq.~\ref{eq:M_omega}) and the random number probability ratio (eq.~\ref{eq:M_eta}) are equal.
Note that eq.~\ref{eq:M_omega} is a ratio of two different functions that have the same argument $\omega$, 
whereas eq.~\ref{eq:M_eta} is the ratio of the same function with different arguments $\eta$ and $\widetilde \eta$.
Eq.~\ref{eq:M_eta} is of little practical use, because $\widetilde \eta$ is not available from a simulation at the simulation state. 
However, the random number difference $\Delta \eta_k$ can be expressed as a function of $\omega$, and the random number probability ratio can thus be expressed as a function of $\omega$ and $\eta$
\begin{eqnarray}
    M(\omega, \eta; \Delta t  | (x_0, v_0)) &=& \frac{P(\widetilde \eta)}{P(\eta)}     \, .
\end{eqnarray}
For a path $\omega$ and the corresponding random number sequence $\eta$ that was used to generate this path, we will use the following equality
\begin{eqnarray}
    M(\omega, \eta; \Delta t|(x_0, v_0))  &=& M(\omega; \Delta t  | (x_0, v_0))
\label{eq:M_omega_equals_M_eta} \, .
\end{eqnarray}

The functional form and the value of the properties introduced in this section depend strongly on the integration scheme. 
In section \ref{sec:overdamped_Langevin}, we summarize the equations for the Euler-Maruyama scheme for overdamped Langevin dynamics. 
In section \ref{sec:Langevin}, we derive the corresponding equations for the ISP integration scheme for Langevin dynamics. 
(See Table \ref{tab:equation_index}).
Throughout the manuscript, properties associated to Langevin dynamics are subscripted with $L$, and properties associated to overdamped Langevin dynamics are subscripted with $o$.
\begin{table}[h]
    \centering
    \begin{tabular}{l l |c c }
    \hline
                                &&Overdamped Langevin    &Langevin\\
    \hline                        
    equation of motion              &&eq.~\ref{equ:OLD_1D}            &eq.~\ref{equ:LD_1D} \\
    integration scheme              &&eq.~\ref{equ:Euler-Maruyama_1D} &eqs.~\ref{equ:iteration_scheme_1D_1}, \ref{equ:iteration_scheme_1D_2}\\
    path probability                &$P(\omega; \Delta t  | (x_0, v_0))$    
                                    &eq.~\ref{equ:multiple_step_path_prob_odamped_langevin} &eq.~\ref{equ:multiple_step_path_prob} \\
    path probability ratio          &$M(\omega; \Delta t  | (x_0, v_0))$    
                                    &eq.~\ref{equ:M_odamped_Langevin_from_path}    &eq.~\ref{equ:M_Langevin_from_path}\\
    random number                   &$\eta_k$       
                                    &eq.~\ref{equ:odamped_rand_numb_simulated_sys}               &eq.~\ref{equ:rand_numb_Langevin_simulated_sys}\\    
    random number difference        &$\Delta \eta_k$       
                                    &eq.~\ref{equ:delta_eta_overdamped_Langevin}    &eq.~\ref{equ:delta_eta_Langevin} \\ 
    random number probability ratio &$M(\omega, \eta ; \Delta t  | (x_0, v_0))$       &eq.~\ref{M_odamped_Langevin_from_rand_numb} &eq.~\ref{M_Langevin_from_rand_numb} \\ 
    \hline
    \end{tabular}
    \caption{References to the equations for the properties introduced in section \ref{sec:preliminaries}.}
    \label{tab:equation_index}
\end{table}

\section{Overdamped Langevin dynamics \label{sec:overdamped_Langevin}}

 
\subsection{Equation of motion and integration scheme}
Consider a one particle system that moves in a one-dimensional position space with temperature $T$ and potential energy function $V$.
The overdamped Langevin equation of motion is 
\begin{align}
\dot{x}(t) = - \frac{\nabla V(x(t))}{\xi m} ~+~ \sqrt{\frac{2k_B T}{\xi m}} \, \eta(t) \, , \label{equ:OLD_1D}
\end{align}
with particle mass $m$, position $x$,  velocity $v=\dot{x}$ and Boltzmann constant $k_B$. 
$x(t) \in \Omega_o$ is the state of the system at time $t$,
where $\Omega_o \subset \mathbb{R}$ is the state space of the system.
The collision rate $\xi$ (in units of s$^{-1}$) models the interaction with the thermal bath.
$\eta(t) \in \mathbb{R}$ describes an uncorrelated Gaussian white noise with unit variance centered at zero, which is scaled by the 
volatility $\sqrt{\frac{2k_B T}{\xi m}}$.
A numerical algorithm to calculate an approximate solution to eq.~\ref{equ:OLD_1D} is the Euler-Maruyama integration scheme\cite{Oeksendal2003, Rabee2014}
\begin{eqnarray}
x_{k+1} = x_k - \frac{\nabla V(x_k)}{\xi m} \, \Delta t + \sqrt{\frac{2k_B T}{\xi m}} \, \sqrt{\Delta t} \, \eta_{o,k} \, , \label{equ:Euler-Maruyama_1D}
\end{eqnarray}
where $\Delta t$ is the time step, $x_k$ is the position, and $\eta_{o,k}$ is the random number at iteration $k$. The random numbers are drawn from a Gaussian distribution with zero mean and unit variance. 
For $k=0,\dots,n-1$, eq.~\ref{equ:Euler-Maruyama_1D} yields a time-discretized overdamped Langevin path \mbox{$\omega_o= (x_0, x_{1}, \dots, x_{n})$} which starts at the pre-defined initial position $x_0$. 
Note that, while the state of the system at iteration $k$ is defined by the position $x_k$ 
the progress to $x_{k+1}$ depends on $x_k$, and on the value of the random number $\eta_{o,k}$.
The random number sequence that was used to generate a specific $\omega_o$ is denoted by $\eta_o = (\eta_{o,0},\dots, \eta_{o,n-1})$.


\subsection{Path probability and path probability ratio}
The probability to observe a path $\omega_o$ generated by the Euler-Maruyama scheme (eq.~\ref{equ:Euler-Maruyama_1D}) is \cite{Donati2017, Chow2015, Bressloff2014, Adib2008}
\begin{eqnarray}
&P_o(\omega_o;\Delta t | x_0) 
= \Bigg[  \sqrt{\frac{\xi m}{4 \pi k_B T \Delta t}} \Bigg]^n \cdot \exp \left( - \frac{\xi m}{4 k_B T \Delta t} \sum\limits_{k=0}^{n-1} \left( x_{k+1} - x_k + \frac{\Delta t}{\xi m} \nabla V(x_k) \right)^2 \right) \, . \label{equ:multiple_step_path_prob_odamped_langevin}
\end{eqnarray}
For the Euler-Maruyama scheme, the path probability $P_o(\omega_o;\Delta t | x_0)$ does not depend on the initial velocity, hence we dropped $v_0$ in the notation.
But it does depend on the potential energy function $V(x)$
that has been used in eq.~\ref{equ:Euler-Maruyama_1D} to generate the path $\omega_o$. 
The path probability that the same path $\omega_o$ has been generated at a target potential $\widetilde V(x)$ (eq.~\ref{equ:def_perturbed_pot}) 
is $\widetilde{P}_o(\omega_o;\Delta t | x_0)$, which is obtained by replacing the potential $V(x)$ with $\widetilde V(x)$ in eq.~\ref{equ:multiple_step_path_prob_odamped_langevin}.
The ratio between the two path probabilities is
\begin{eqnarray}
    M_{o}(\omega_o; \Delta t|x_0) 
    &=& \frac{\widetilde{P}_o(\omega_o;\Delta t | x_0)}{P_o(\omega_o;\Delta t | x_0)} \cr
    &=& \exp \left( -\frac{\sum\limits_{k=0}^{n-1} (x_{k+1} - x_k) \left( \nabla \widetilde{V}(x_k) - \nabla V(x_k) \right)}{2 k_B T} \right) \cr
    &&\times \exp \left( - \frac{\sum\limits_{k=0}^{n-1} \left( \nabla \widetilde{V}^2(x_k) - \nabla V^2(x_k) \right) \Delta t}{4 k_B T \xi m} \right) \, . \label{equ:M_odamped_Langevin_from_path}
\end{eqnarray}
Eq.~\ref{equ:M_odamped_Langevin_from_path} is a function of the path $\omega_o$ and and does not depend on the random number sequence $\eta_o$ explicitly. 
It is equivalent to eq.~B4 in ref.~\onlinecite{Donati2017}.

\subsection{Random numbers and random number probability ratio}
\label{sec:Odamped_Langevin_rand_numb_sequence}
Given $\omega_o$, the sequence of random numbers $\eta_{o}$ that was used to generate $\omega_o$ at the simulation potential $V(x)$ can be back-calculated by rearranging eq.~\ref{equ:Euler-Maruyama_1D} for $\eta_{o,k}$
\begin{eqnarray}
    \eta_{o,k} &=& \sqrt{\frac{\xi m}{2 k_B T \Delta t}} \left( x_{k+1} - x_k + \frac{\nabla V(x_k)}{\xi m} \Delta t \right) \, . 
\label{equ:odamped_rand_numb_simulated_sys}
\end{eqnarray}
We remark that the path-probability (eq.~\ref{equ:multiple_step_path_prob_odamped_langevin}) 
can formally be derived by inserting  eq.~\ref{equ:odamped_rand_numb_simulated_sys} 
into eq.~\ref{equ:Gaussian}.
Since eq.~\ref{equ:odamped_rand_numb_simulated_sys} defines a coordinate transformation from $x_k$ to $\eta_{o,k}$, one needs to normalize with respect to the new coordinates in order to obtain the correct normalization constant.
The random number sequence $\widetilde \eta_o$ needed to generate $\omega_o$ at a target potential $\widetilde V(x)$ is calculated by inserting eq.~\ref{equ:def_perturbed_pot} into eq.~\ref{equ:odamped_rand_numb_simulated_sys}
\begin{eqnarray}
    \widetilde{\eta}_{o,k} 
    &=& \sqrt{\frac{\xi m}{2 k_B T \Delta t}} \left( x_{k+1} - x_k + \frac{\nabla V(x_k)}{\xi m} \Delta t \right) 
    + \,  \sqrt{\frac{\Delta t}{2 k_B T \xi m}} \nabla U(x_k) \nonumber \\
    &=& \eta_{o,k} + \Delta \eta_{o,k} \, . \label{equ:P2}
\end{eqnarray}
Eq.~\ref{equ:odamped_rand_numb_simulated_sys} defines the change of variables from $\omega$ to $\eta_o$ for the Euler-Maruyama scheme at the simulation potential. 
Likewise eq.~\ref{equ:P2} defines the change of variables from $\omega$ to $\widetilde \eta_o$ at the target potential.
The random number difference is
\begin{equation}
    \Delta \eta_{o,k} = \sqrt{\frac{\Delta t}{2 k_B T \xi m}} \nabla U(x_k) \, . \label{equ:delta_eta_overdamped_Langevin}
\end{equation}
It depends on the perturbation $U(x)$, but not on the simulation potential $V(x)$.
Inserting $\Delta \eta_{o,k}$ (eq.~\ref{equ:delta_eta_overdamped_Langevin}) into eq.~\ref{eq:M_eta} yields the random number probability ratio
\begin{eqnarray}
    &&M_{o}(\omega_o, \eta_o; \Delta t|x_0)  \cr
    &=& \exp \left(-\sum\limits_{k=0}^{n-1} \sqrt{\frac{ \, \Delta t}{2k_B T \xi m}}  \nabla U(x_k)\cdot \eta_{o,k} \right) \cdot
        \exp \left(-\frac{1}{2}\sum\limits_{k=0}^{n-1} \frac{\Delta t}{2k_B T \xi m} (\nabla U(x_k))^2 \right) \, . \label{M_odamped_Langevin_from_rand_numb}
\end{eqnarray}
Because of eq.~\ref{eq:M_omega_equals_M_eta}, eq.~\ref{equ:M_odamped_Langevin_from_path} and eq.~\ref{M_odamped_Langevin_from_rand_numb} are equal. 
However, the two probability ratios use different time-series and different information on the system to evaluate the path probability ratio. 
To evaluate eq.~\ref{equ:M_odamped_Langevin_from_path}, one needs 
the path $\omega_o$, the simulation potential $V(x)$, and the target potential $\widetilde V(x)$.
To evaluate eq.~\ref{M_odamped_Langevin_from_rand_numb}, one needs 
the path $\omega_o$, the random number sequence for the simulation potential $\eta_o$, and the perturbation $U(x)$.
Because $U(x)$ often only affects a few coordinates of the systems, 
i.e.~it is low-dimensional, 
eq.~\ref{M_odamped_Langevin_from_rand_numb} is computationally more efficient. 
Besides the force calculation $-\nabla V(x)$ needed to generate the path $\omega_o$, 
it requires an additional force calculation $-\nabla U(x)$ only along the coordinates that are affected by the perturbation. 
By contrast, eq.~\ref{equ:M_odamped_Langevin_from_path} requires an additional force calculation 
on the entire system $-\nabla \widetilde V(x)$.
%

\section{Langevin dynamics}
\label{sec:Langevin}
%
%
\subsection{Equation of motion and integration scheme}
Consider a one particle system that moves in a one-dimensional position space with temperature $T$ and potential energy function $V$.
The Langevin equation of motion is
\begin{equation}
m \ddot x(t) = - \nabla V(x(t)) - \xi m \dot x(t) + \sqrt{2 k_B T \xi m} \, \eta(t) \, , \label{equ:LD_1D}
\end{equation}
with particle mass $m$, position $x$, velocity $v=\dot{x}$, acceleration $a=\ddot{x}$, and Boltzmann constant $k_B$. 
The state of the system at time $t$ is determined by the position and the velocity 
$(x(t), \dot{x}(t)) \in \Omega_L$, 
where $\Omega_L \subset \mathbb{R}^2$ is the state space of the system.
The collision rate $\xi$ (in units of s$^{-1}$) models the interaction with the thermal bath. $\eta \in \mathbb{R}$ describes an uncorrelated Gaussian white noise with unit variance centered at zero, which is scaled by the volatility $\sqrt{2 k_B T \xi m}$.
A numerical algorithm to calculate an approximate solution to eq.~\ref{equ:LD_1D} is the ISP scheme \cite{Izaguirre2010}
\begin{eqnarray}
x_{k+1} &=& x_k + \exp \left( - \xi \, \Delta t \right) \, v_k \Delta t - \bigg[ 1 - \exp \left( - \xi \, \Delta t\right) \bigg] \, \frac{\nabla V(x_k)}{\xi m} \Delta t \cr
&& + \sqrt{\frac{k_B T}{m} \, \bigg[ 1 - \exp \left( - 2 \xi \, \Delta t\right) \bigg]} \, \eta_{L,k} \, \Delta t \label{equ:iteration_scheme_1D_1} \\
v_{k+1} &=& \frac{x_{k+1} - x_k}{\Delta t} \, ,
\label{equ:iteration_scheme_1D_2}
\end{eqnarray}
where $\Delta t$  is the time step, $x_k$ is the position, $v_k$ is the velocity, and $\eta_{L,k}$ is the random number at iteration $k$ (see appendix \ref{appendix:derivation_ISP_scheme}). 
The random numbers are drawn from a Gaussian distribution with zero mean and
unit variance.
For $k=0,\dots,n-1$, eqs.~\ref{equ:iteration_scheme_1D_1} and \ref{equ:iteration_scheme_1D_2} yield a time-discretized Langevin path $\omega_L = ( (x_0,v_0), (x_{1},v_{1}),\dots, (x_{n},v_{n}) )$ which starts at the pre-defined initial state $(x_0, v_0)$. 
Note that, while the state of the system at iteration $k$ is defined by the tuple $(x_k,v_k) \in \Omega_L$, the progress to $(x_{k+1},v_{k+1})$ depends
on $(x_k,v_k)$, and on the value of the random number $\eta_{L,k}$. 
The random number sequence that was
used to generate a specific $\omega_L$ is denoted by $\eta_L = (\eta_{L,0}, \dots, \eta_{L,n-1})$.
The position $x_{k+1}$ is treated as a random variable, because it directly depends on a random number
(eq.~\ref{equ:iteration_scheme_1D_1}), while the  velocity $v_{k+1}$ is calculated from the new position $x_{k+1}$ and the preceding position $x_k$. 
Because the velocity $v_k$ in eq.~\ref{equ:iteration_scheme_1D_1} is determined by the positions $x_k$ and $x_{k-1}$ (eq.~\ref{equ:iteration_scheme_1D_2}), it carries a small memory effect into the time-evolution of $x$.
%

\subsection{Path probability and path probability ratio}
The probability to generate a path $\omega_L$ by the ISP scheme (eqs.~\ref{equ:iteration_scheme_1D_1} and \ref{equ:iteration_scheme_1D_2}) at the simulation potential $V(x)$ is 
\begin{eqnarray}
&&    P_L(\omega_L;\Delta t | (x_0, v_0)) \cr \cr
    &=& \left[\prod_{k=0}^{n-1} \delta\left(v_{k+1} - \frac{x_{k+1}-x_k}{\Delta t}\right) \right] \cdot \left[ \sqrt{\frac{m}{2 \pi k_B T \Delta t^2 (1 - \exp ( -2 \xi \Delta t ))}} \right]^n \nonumber\\
    &&\times \exp \left( - \sum_{k=0}^{n-1}\frac{m \left( x_{k+1} - x_k - \exp (-\xi \Delta t) v_k \Delta t + (1 - \exp(-\xi \Delta t )) \frac{\nabla V(x_k)}{\xi m} \Delta t \right)^2}{2 k_B T (1 - \exp(-2\xi \Delta t)) \Delta t^2} \right)  \, . \label{equ:multiple_step_path_prob}
\end{eqnarray}
The derivation of eq.~\ref{equ:multiple_step_path_prob} is shown in appendices \ref{appendix:path_probability}
and \ref{appendix:double_integral}.
Appendix \ref{appendix:path_probability} explains the strategy for the derivation, 
and appendix \ref{appendix:double_integral} shows how to solve the integrals that appear in the derivation.
The path probability $\widetilde{P}_L(\omega_L;\Delta t | (x_0,v_0))$ to generate a path $\omega_L$ by the ISP scheme at the target potential is obtained by inserting $\widetilde V(x)$ (eq.~\ref{equ:def_perturbed_pot})
into eq.~\ref{equ:multiple_step_path_prob}.
The path probability ratio for overdamped Langevin dynamics is
\begin{eqnarray}
    && M_{L}(\omega_L; \Delta t|(x_0,v_0)) \cr \cr
    &=& \frac{\widetilde{P}_L(\omega_L;\Delta t | (x_0, v_0))}{P_L(\omega_L;\Delta t | (x_0, v_0))} \nonumber \\
    &=& \exp \left( - \frac{\sum\limits_{k=0}^{n-1}(x_{k+1}-x_k)\left( \nabla \widetilde{V}(x_k) - \nabla V(x_k) \right)}{k_B T \xi (1+ \exp(-\xi \Delta t)) \Delta t} \right) 
    \cdot \exp \left( \frac{ \, \sum\limits_{k=0}^{n-1} \, v_k \left( \nabla \widetilde{V}(x_k) - \nabla V(x_k) \right)}{k_B T \xi (1+ \exp(\xi \Delta t))} \right) \nonumber \\
    && \cdot \exp \left( - \frac{\exp(\xi \Delta t)-1}{\exp(\xi \Delta t)+1} \cdot \frac{ \sum\limits_{k=0}^{n-1} \left( \nabla \widetilde{V}^2(x_k) - \nabla V^2(x_k) \right)}{2 k_B T \xi^2 m} \right) \, .  \label{equ:M_Langevin_from_path}
\end{eqnarray}
Analogous to eq.~\ref{equ:M_odamped_Langevin_from_path}, eq.~\ref{equ:M_Langevin_from_path} is a function of the path $\omega_L$ and and does not depend on the random number sequence $\eta_L$. 


\subsection{Random numbers and random number probability ratio}
Given $\omega_L$, the sequence of random numbers $\eta_{L}$, that was used to generate $\omega_L$ at the simulation potential $V(x)$, can be back-calculated by rearranging eq.~\ref{equ:iteration_scheme_1D_1} for $\eta_{L,k}$
\begin{eqnarray}
    \eta_{L,k} 
    &=&\sqrt{\frac{m}{k_B T (1-\exp(-2\xi \Delta t)) \Delta t^2}}  \cr
    && \times \left( x_{k+1} - x_k - \exp(-\xi \Delta t)v_k\Delta t + (1-\exp(-\xi \Delta t)) \frac{\nabla V(x_k)}{\xi m} \Delta t \right) \, . \label{equ:rand_numb_Langevin_simulated_sys}
\end{eqnarray}
The random number sequence $\widetilde \eta_L$ needed to generate $\omega_L$ at a target potential $\widetilde V(x)$ is calculated by inserting eq.~\ref{equ:def_perturbed_pot} into eq.~\ref{equ:rand_numb_Langevin_simulated_sys}
\begin{eqnarray}
    \widetilde{\eta}_{L,k} 
    &=& \sqrt{\frac{m}{k_B T (1-\exp(-2\xi \Delta t)) \Delta t^2}} \cr
    && \times\left( x_{k+1} - x_k - \exp(-\xi \Delta t)(x_k - x_{k-1}) + (1-\exp(-\xi \Delta t)) \frac{\nabla V(x_k)}{\xi m} \Delta t \right) \nonumber \\
    && + \, \sqrt{\frac{1}{k_B T \xi^2 m}} \cdot \frac{1-\exp(-\xi \Delta t)}{\sqrt{1-\exp(-2\xi \Delta t)}} \nabla U(x_k) \nonumber \\
    &=& \eta_{L,k} + \Delta \eta_{L,k}
    \, . \label{equ:rand_numb_Langevin_target_sys}
\end{eqnarray}
Eq.~\ref{equ:rand_numb_Langevin_simulated_sys} defines the change of variables from $\omega$ to $\eta_L$ for the ISP scheme at the simulation potential. 
Likewise eq.~\ref{equ:rand_numb_Langevin_target_sys} defines the change of variables from $\omega$ to $\widetilde \eta_L$ at the target potential.
The random number difference is
\begin{equation}
    \Delta \eta_{L,k} = \sqrt{\frac{1}{k_B T \xi^2 m}} \cdot \frac{1-\exp(-\xi \Delta t)}{\sqrt{1-\exp(-2\xi \Delta t)}} \nabla U(x_k) \, . \label{equ:delta_eta_Langevin}
\end{equation}
Again, the random number difference depends on the perturbation potential $U(x)$, but not on the simulation potential $V(x)$.
Inserting $\Delta \eta_{L,k}$ (eq.~\ref{equ:delta_eta_Langevin}) into eq.~\ref{eq:M_eta} yields the random number probability ratio
\begin{eqnarray}
    M_{L}(\omega_L, \eta_L; \Delta t|(x_0, v_0)) 
    &=& \exp \left( - \frac{1-\exp(-\xi \Delta t)}{\sqrt{1-\exp(-2 \xi \Delta t)}} \cdot \frac{\sum\limits_{k=0}^{n-1} \nabla U(x_k) \, \eta_{L,k}}{\sqrt{k_B T \xi^2 m}} \right) \cr
    &&\times \exp \left( - \frac{(1-\exp(-\xi \Delta t))^2}{1-\exp(-2\xi \Delta t)} \cdot \frac{\sum\limits_{k=0}^{n-1} \nabla U^2(x_k)}{2 k_B T \xi^2 m} \right) \, . \label{M_Langevin_from_rand_numb}
\end{eqnarray}
Analogous to the path probability ratio for overdamped Langevin dynamics, 
$M_{L}(\omega_L; \Delta t|(x_0,v_o))$ (eq.~\ref{equ:M_Langevin_from_path})
and $M_{L}(\omega_L, \eta_L; \Delta t|(x_0,v_0))$ (eq.~\ref{M_Langevin_from_rand_numb})
yield the same path probability ratio for a given path $\omega_L$  that has been generated using the random number sequence $\eta_L$,
but they use different arguments. 
Again, the path probability from random numbers $M_{L}(\omega_L, \eta_L; \Delta t|(x_0,v_0))$
requires an additional force calculation $-\nabla U(x)$ only along the coordinates that are affected by the perturbation, 
making it computationally more efficient than $M_{L}(\omega_L; \Delta t|(x_0,v_0))$ in most cases.


\section{Comparing Langevin and overdamped Langevin dynamics}
\label{sec:Comparing_Langevin_and_overdamped_Langevin_dynamics}
\subsection{Test system}
\label{sec:test_system}
\begin{figure}[h!]
    \includegraphics[width=8cm]{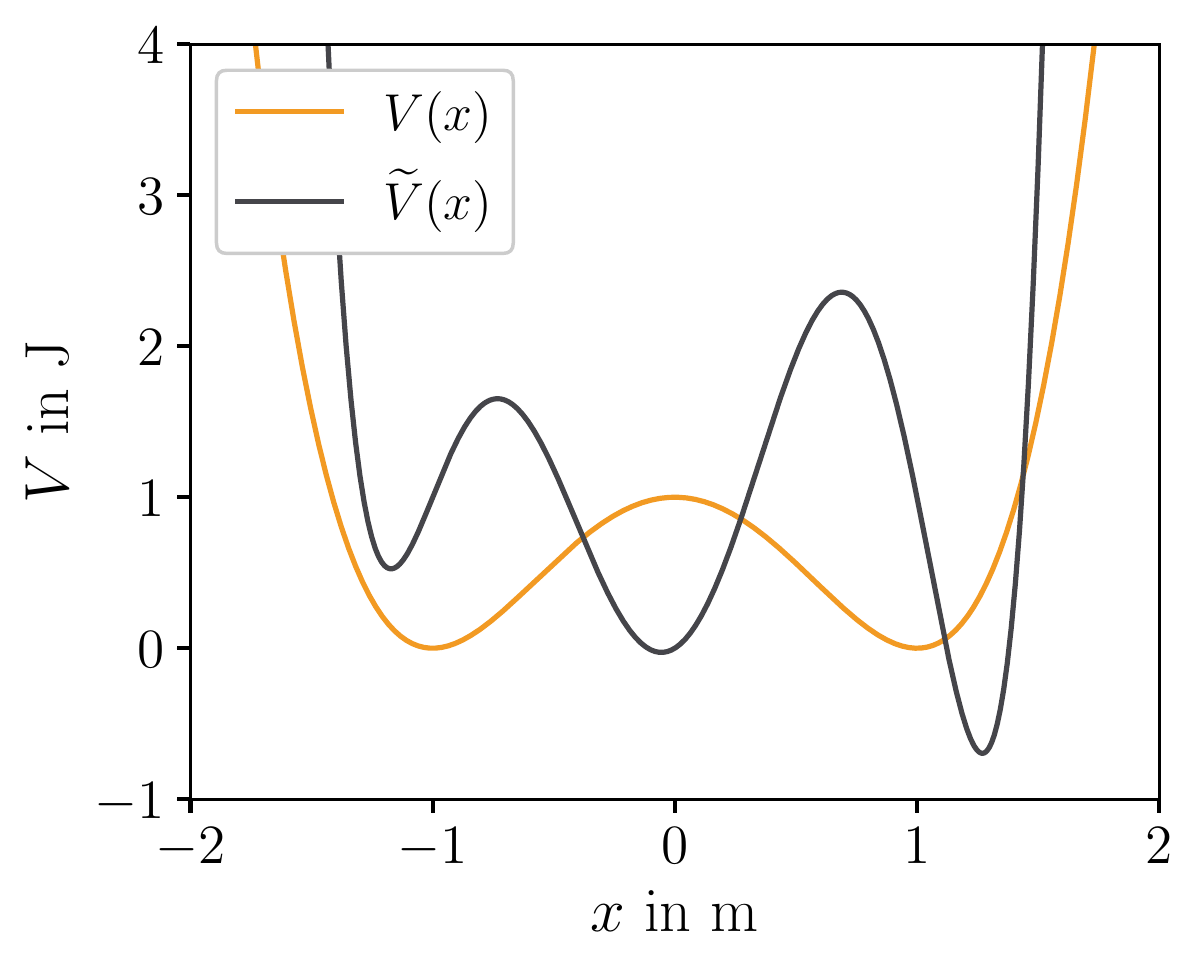}
    \caption{
    Simulation potential $V(x)$ (orange) and target potential $\widetilde V(x)$ (black).
    \label{fig:test_system} 
    }
\end{figure}
Our test system is a one-dimensional one particle system at the simulation potential $V(x)$ (fig.~\ref{fig:test_system}, orange line) and at the target potential $\widetilde{V}(x)$ (fig.~\ref{fig:test_system}, black line). 
The trajectories generated at $V(x)$ will be reweighted to the target potential $\widetilde{V}(x)$. 
The black lines in Fig.~\ref{fig:path_probabilities}.B represent the first three dominant MSM eigenfunctions \cite{Prinz2011} associated to the target potential.
The implied timescales\cite{Swope2004} are 
$t_0 = \infty$, 
$t_1 = 20.5\, \mathrm{s}$, and 
$t_2 = 6.0\, \mathrm{s}$, which are shown as black lines in Fig.~\ref{fig:path_probabilities}.C.
Computational details are reported in section \ref{sec:methods}.

\subsection{From random numbers $\eta$ to paths $\omega_o$ and $\omega_L$}
\label{sec:EulerMaruyamaVsISP}
Given a random number sequence $\eta=(\eta_0, \dots, \eta_{n-1})$ and a starting state $(x_0,v_0)$, one can use the Euler-Maruyama scheme to generate an overdamped Langevin path $\omega_o$, or else one can use the ISP scheme to generate a Langevin path $\omega_L$.
We discuss briefly how the difference between $\omega_o$ and $\omega_L$ depends on the combined parameter $\xi \Delta t$, which can be interpreted as the number of collisions per time step.
In the limit of high friction $\xi m \dot{x} \gg m \ddot{x}$, the Langevin dynamics (eq.~\ref{equ:LD_1D}) approaches the overdamped Langevin dynamics (eq.~\ref{equ:OLD_1D}).
More specifically: in eq.~\ref{equ:LD_1D} set $m \ddot{x} = 0$, rearranging yields eq.~\ref{equ:OLD_1D}. 
However, even though the equation of motion for Langevin dynamics converges to the equation of motion for overdamped Langevin dynamics, the ISP scheme (eq.~\ref{equ:iteration_scheme_1D_1} and \ref{equ:iteration_scheme_1D_2}) does not converge to the Euler-Maruyama scheme (eq.~\ref{equ:Euler-Maruyama_1D}) in the limit of high friction. 
By ``high friction'' we denote the range of collision rates $\xi$ for which $e^{-\xi\Delta t} \approx 0$ in eq.~\ref{equ:iteration_scheme_1D_1}, but $\big| \frac{\nabla V}{\xi m} \big| > 0$.
(As reference: 
$e^{-0.1} = 0.904$,
$e^{-1} = 0.368$, and
$e^{-5} = 0.007$.)
If $e^{-\xi\Delta t} \approx 0$, then also $e^{-2\xi\Delta t} \approx 0$, and eq.~\ref{equ:iteration_scheme_1D_1}
becomes
\begin{eqnarray}
x_{k+1} &\approx& x_k - \, \frac{\nabla V(x_k)}{\xi m} \Delta t  + \sqrt{\frac{k_B T}{m} \, } \, \eta_{L,k} \, \Delta t \, .
\label{equ:iteration_scheme_1D_1_highFriction}
\end{eqnarray}
The first two terms on the right-hand side are identical to the Euler-Maruyama scheme (eq.~\ref{equ:Euler-Maruyama_1D}), but the random number term differs from the Euler-Maruyama scheme. 
Thus, even in the limit of high friction the two algorithms yield different paths for a given random number sequence $\eta$.
The difference between a Langevin path $\omega_L$ and an overdamped Langevin path $\omega_o$ can be scaled by the combined parameter $\xi \Delta t$.
For some value $\xi \Delta t>1$ the difference between the two paths becomes minimal before increasing again, but for no value of $\xi \Delta t$ the two paths fully coincide.

When Langevin integration schemes are used as thermostat in MD simulations, the optimal friction coefficient should reproduce the expected temperature fluctuations, and therefore depends on the system and the simulation box \cite{Hunenberger:2005}.
Reported collision rates\cite{Izaguirre2010, Basconi:2013, Goga:2012} (while keeping the time step at $\Delta t = 0.002\, {\rm ps}$) range from 0.1 ${\rm ps}^{-1}$ to $\sim$ 100 ${\rm ps}^{-1}$, corresponding to 
$\xi \Delta t = 0.0002$ to
$\xi \Delta t = 0.2$.
But even for a large collision rate of 100 ${\rm ps}^{-1}$, $e^{-\xi \Delta t} = e^{-0.2} = 0.819 \not\approx 0$.
For these two reasons - MD simulations are not conducted in the high-friction regime, and even in the high-friction regime $\omega_o$ differs from $\omega_L$ - a simulation with the ISP scheme yields a materially different path ensemble than a simulation with the Euler-Maruyama scheme. 

\subsection{From a path $\omega$ to random numbers $\eta_o$ and $\eta_L$}
\begin{figure}[h]
    \centering
    \includegraphics[width=8cm]{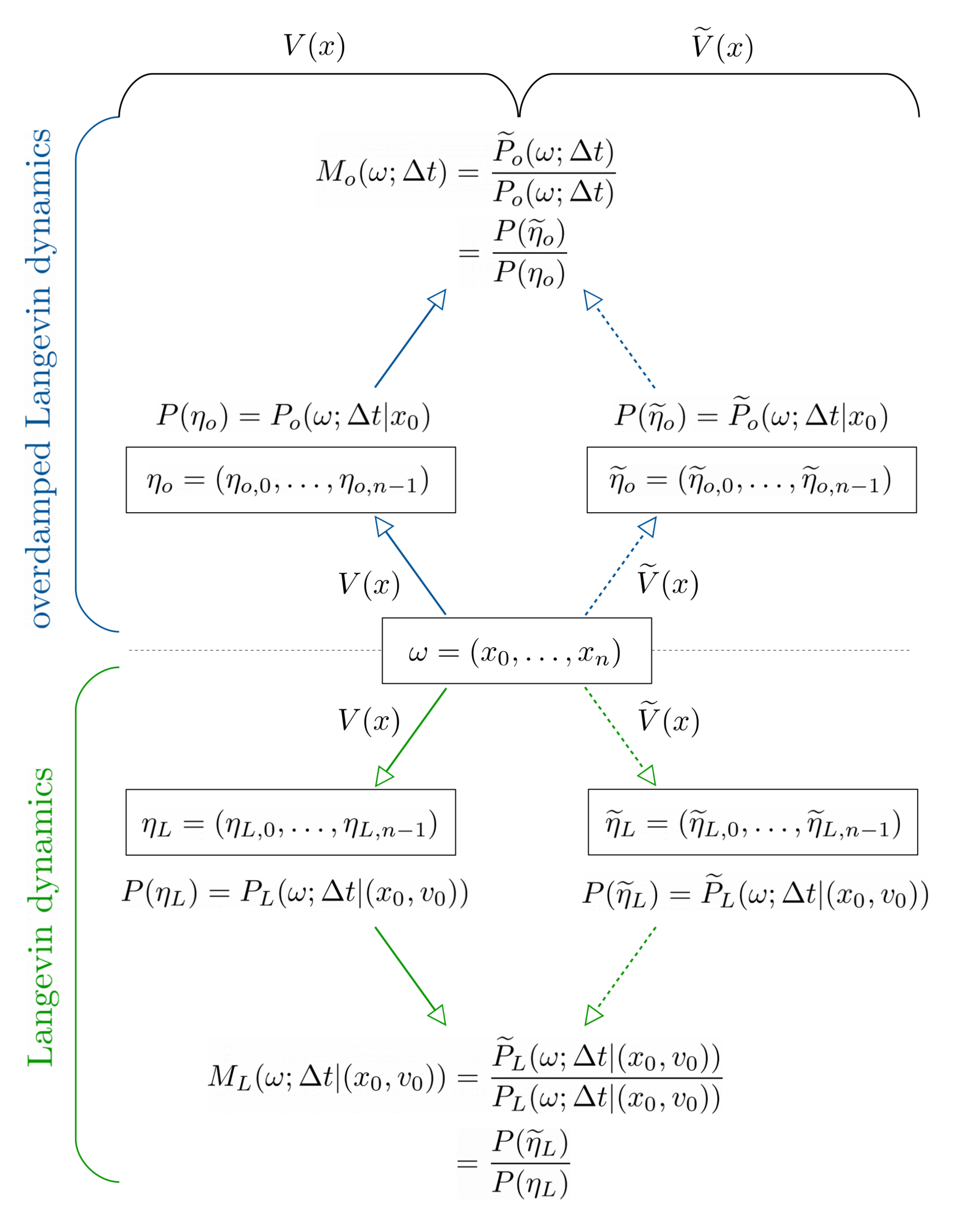} 
    \caption{Overview of path probabilities and path probability ratios for a sample path $\omega =(x_0, \dots x_n)$.}
    \label{fig:wokflow}
\end{figure} 
In the previous section, we showed that, given a random number sequence $\eta$, the path generated by the Euler-Maruyama integration scheme for overdamped Langevin dynamics differs from the path generated by the ISP integration scheme for Langevin dynamics.
More relevant for path reweighting is the reverse situation: 
Given a sample path $\omega =(x_0, \dots, x_n)$ in position space and the parameters of the dynamics ($m$, $V$, $T$, $\xi$, $k_B$, and $\Delta t$), how does the random number sequence $\eta_o$ needed to generate $\omega$ with the Euler-Maruyama scheme (eq.~\ref{equ:Euler-Maruyama_1D}) differ from the random number sequence $\eta_L$ needed to generate the same $\omega$ with the ISP scheme (eqs.~\ref{equ:iteration_scheme_1D_1} and \ref{equ:iteration_scheme_1D_2})?
An equivalent question is: How does the path probability that $\omega$ has been generated by the Euler-Maruyama scheme differ from the path probability that $\omega$ has been generated by the ISP scheme, and how does this difference affect the path probability ratios between the simulation and a target potential.
Fig.~\ref{fig:wokflow} gives an overview of the quantities we will compare.
Note that we dropped the index $o$ or $L$ from the path $\omega$, because $\omega$ is a given data set which will be analyzed using various approaches to calculate the path probabilities.

\begin{figure}[h]
    \centering
    \includegraphics[width=16cm]{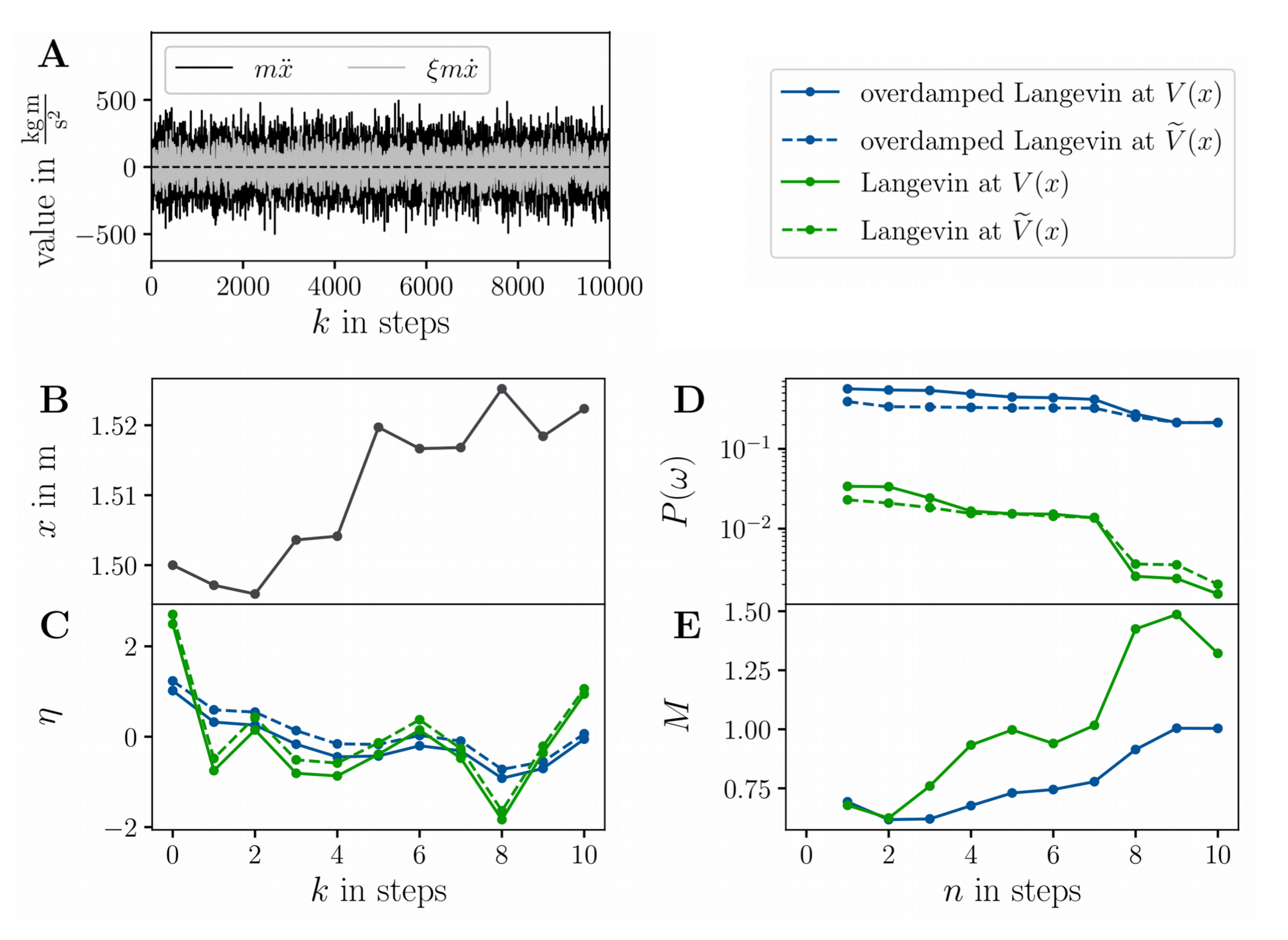}
    \caption{
    \textbf{A} The acceleration term $m \ddot{x}$ and the friction $\xi m \dot{x}$ for the test system at $V(x)$;
    \textbf{B} Example path $\omega$ of length $n=10$;
    \textbf{C} Random number sequences $\eta_L$ (solid green), $\eta_o$ (solid blue), $\widetilde{\eta}_L$ (dashed green) and $\widetilde{\eta}_o$ (dashed blue) that correspond to $\omega$;
    \textbf{D} Path probabilities $P_L(\omega; \Delta t|(x_0,v_0))$ (solid green), $P(\omega; \Delta t | x_0)$ (solid blue), $\widetilde{P}_L(\omega; \Delta t | (x_0,v_0))$ (dashed green) and $\widetilde{P}_o(\omega ; \Delta t | x_0)$ (dashed blue);
    \textbf{E} Path probability ratios: $M_L(\omega, \Delta t | (x_0,v_0))$ (green) and $M_o(\omega; \Delta t | x_0)$ (blue).
    }
\label{fig:example_quantities}    
\end{figure}

First, we need to discuss whether such a comparison between ISP scheme and Euler-Maruyama scheme is even possible. 
From an algorithmic view point this is clearly possible, 
because both integrators (eq.~\ref{equ:Euler-Maruyama_1D} and eq.~\ref{equ:iteration_scheme_1D_1}) use a single random number 
per integration time step. 
The path probabilities are then equal to the probability of the different random number sequences $\eta_L$ and $\eta_o$ needed to generate $\omega$.
From a physical view point the answer is not as clear, because overdamped Langevin dynamics evolves in position space $(x_k)$, 
whereas Langevin dynamics evolves in phase space $(x_k,v_k)$. 
The velocity $v_k$ enters the integration scheme (eq.~\ref{equ:iteration_scheme_1D_1}) as well as the path probability (eq.~\ref{equ:multiple_step_path_prob}).
However, $v_k$ is fully determined by the current position $x_k$ and the previous position $x_{k-1}$ (eq.~\ref{equ:iteration_scheme_1D_2}). 
Thus, if the initial velocity $v_0$ is known, the position trajectory is enough to evaluate the path probability (eq.~\ref{equ:multiple_step_path_prob}), 
and the comparison to overdamped Langevin dynamics is possible.
We consider the test system described in section \ref{sec:test_system} at the simulation potential $V(x)$ (double-well potential) simulated by the ISP scheme for Langevin dynamics.
With  $\xi = 50 \, {\rm s}^{-1}$ and $\Delta t = 0.01 \, {\rm s}$, we have $e^{-\xi \Delta t} = e^{-0.5} = 0.607 \not \approx 0$, meaning the system is not in the high-friction limit.
Fig.~\ref{fig:example_quantities}.A additionally shows that with these parameters $\mathcal{O}(\xi m \dot{x}) \approx \mathcal{O}(m \ddot{x})$, 
and also according to the criterion for the stochastic differential equation the system is not in the high-friction limit.
Fig.~\ref{fig:example_quantities}.B shows a sample path $\omega = (x_0, x_1, \cdots, x_{10})$. 
Fig.~\ref{fig:example_quantities}.C shows the random numbers $\eta_o$
needed to generate $\omega$ with the Euler-Maruyama scheme (solid blue line, calculated using eq.~\ref{equ:odamped_rand_numb_simulated_sys}) and the random numbers $\eta_L$ needed to generate $\omega$ with the ISP scheme (solid green line, calculated using eq.~\ref{equ:rand_numb_Langevin_simulated_sys}). 
As expected for the low-friction regime, these two random number sequences differ markedly. 
Consequently, the path probabilities differ. 
Fig.~\ref{fig:example_quantities}.D shows the unnormalized path probability for generating $\omega$ with the Euler-Maruyama scheme (blue solid line)
\begin{eqnarray}
P_o(\omega;\Delta t | x_0) 
&\sim& \exp \left( - \frac{\xi m}{4 k_B T \Delta t} \sum\limits_{k=0}^{n-1} \left( x_{k+1} - x_k + \frac{\Delta t}{\xi m} \nabla V(x_k) \right)^2 \right) \ ,
\end{eqnarray}
and for generating $\omega$ with the ISP scheme (green solid line)
\begin{eqnarray}
&&    P_L(\omega;\Delta t | (x_0, v_0)) \cr 
    &\sim& \exp \left( - \sum_{k=0}^{n-1}\frac{m \left( x_{k+1} - x_k - \exp (-\xi \Delta t) v_k \Delta t + (1 - \exp(-\xi \Delta t )) \frac{\nabla V(x_k)}{\xi m} \Delta t \right)^2}{2 k_B T (1 - \exp(-2\xi \Delta t)) \Delta t^2} \right)  \, ,
\end{eqnarray}
where we omitted those factors from eqs.~\ref{equ:multiple_step_path_prob_odamped_langevin} and
\ref{equ:multiple_step_path_prob} that cancel in the path probability ratio.
We checked that the path probabilities are consistent with $P(\eta_o)$ and $P(\eta_L)$. 
The two path probabilities diverge from the first simulation step on.
After ten integration time steps they differ by two orders of magnitude.
Clearly, $P_L(\omega;\Delta t | (x_0, v_0))$ cannot be used as an approximation for $P_o(\omega;\Delta t | x_0)$.
However, an interesting observation arises when we consider reweighting $\omega$ to the target potential $\widetilde V(x)$ (triple-well potential).
Fig.~\ref{fig:example_quantities}.C shows the random numbers $\widetilde\eta_o$
needed to generate $\omega$ with the Euler-Maruyama scheme at $\widetilde V(x)$ (dashed blue line, calculated using eq.~\ref{equ:P2}), and the random numbers $\widetilde\eta_L$ needed to generate $\omega$ with the ISP scheme at $\widetilde V(x)$  (dashed green line, calculated using eq.~\ref{equ:rand_numb_Langevin_target_sys}). 
The corresponding unnormalized path probabilities $\sim\widetilde P_o(\omega;\Delta t | x_0)$ and $\sim \widetilde P_L(\omega;\Delta t | (x_0, v_0))$ are shown as dashed lines in Fig.~\ref{fig:example_quantities}.D.
Strikingly, a change of the integration scheme from Euler-Maruyama to ISP has a much stronger influence on the random numbers and the path probability than the modification of the potential energy function.
Fig.~\ref{fig:example_quantities}.E shows the path probability ratios, i.e. the ratio between the dashed and the solid lines in Fig.~\ref{fig:example_quantities}.D, for the Euler-Maruyama scheme $M_o=M_{o}(\omega; \Delta t|x_0) = M_{o}(\omega, \eta_o; \Delta t|x_0)$
(blue line) and the ISP scheme
$M_L=M_{L}(\omega; \Delta t|(x_0, v_0)) = M_{L}(\omega, \eta_L; \Delta t|(x_0, v_0))$
(green line).
Because, within an integration scheme, the path probability does not change drastically when going from the simulation potential $V(x)$ to the target potential $\widetilde V(x)$, both path probability ratios remain at $\approx 1$ throughout the path and follow similar curves.  
That is, the path probability ratios for Langevin and overdamped Langevin dynamics are much more similar than the underlying path probabilities. 
%

\subsection{Path reweighting}
\begin{figure}[h]
    \centering
    \includegraphics[width=16cm]{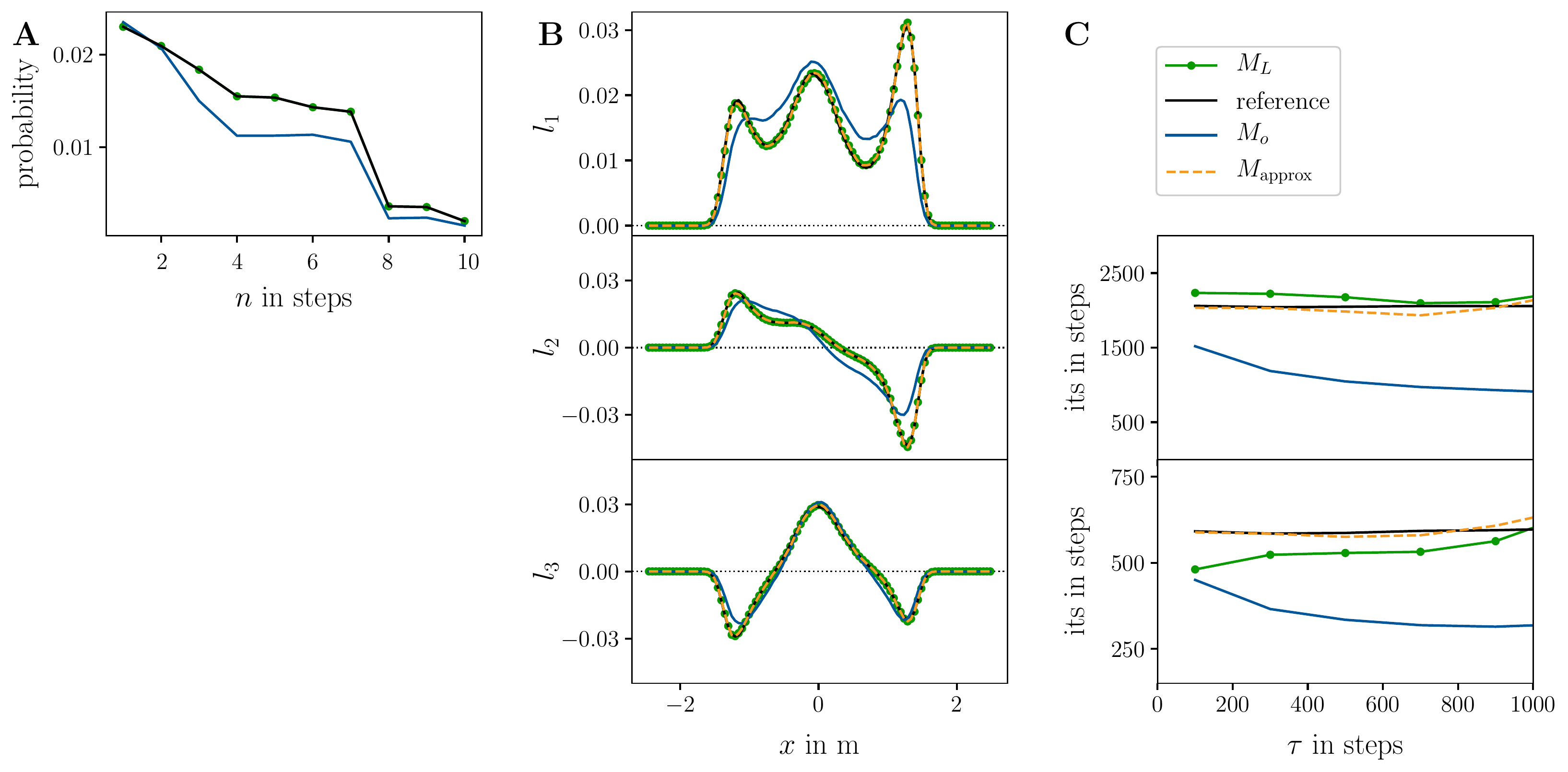}
    \caption{
    \textbf{A} Reference and reweighted path probabilities for $\omega$ for Langevin dynamics; 
    \textbf{B} Reference and reweighted first three dominant MSM left eigenfunctions $l_1,l_2$ and $l_3$ associated to $\widetilde V(x)$ for Langevin dynamics; 
    \textbf{C} Reference and reweighted implied timescales corresponding to $l_2$ and $l_3$.
    }
\label{fig:path_probabilities}    
\end{figure}
We return to the scenario described in the introduction, and ask: are the two path probability ratios similar enough that we can use $M_{o}$ as an approximation to $M_{L}$ in eq.~\ref{equ:path_reweighing}?
Fig.~\ref{fig:path_probabilities}.A compares different ways to calculate the path probability $\widetilde{P}_L(\omega; \Delta t | (x_0, v_0))$, i.e.~the probability with which example path $\omega$ would have been generated at the target potential $\widetilde V(x)$.
The black line is the reference solution calculated by inserting $\widetilde V(x)$ into eq.~\ref{equ:multiple_step_path_prob}.
It is identical to the dashed green line in Fig.~\ref{fig:example_quantities}.D. 
The green line in Fig.~\ref{fig:path_probabilities}.A shows the reweighted path probability, where we used the exact path probability ratio for the ISP scheme, 
$M_L(\omega; \Delta t | (x_0,v_0))$ (eq.~\ref{equ:M_Langevin_from_path}), in eq.~\ref{equ:path_reweighing}.
As expected, this reweighted path probability coincides with the directly calculated path probability. 
The blue line shows the reweighted path probability, where we used the path probability ratio for the Euler-Maruyama scheme, 
$M_o(\omega; \Delta t | x_0)$ (eq.~\ref{equ:M_odamped_Langevin_from_path}), as an approximation to $M_L$ in eq.~\ref{equ:path_reweighing}.
The path probability deviates from the reference solution, but overall follows a similar curve.
Fig.~\ref{fig:path_probabilities}.A merely serves to illustrate the concepts. 
With only ten steps the example path $\omega$ is far too short to judge the accuracy of the two path probability ratios for reweighting dynamic properties. 
We therefore constructed MSMs for the target potential $\widetilde V(x)$. 
The reference solution has been generated from simulations at the target potential $\widetilde V(x)$ using the ISP scheme. 
The dominant MSM eigenfunctions and associated implied timescales are shown as black lines in Fig.~\ref{fig:path_probabilities}.B and \ref{fig:path_probabilities}.C.
Next, we ran simulations at the simulation potential $V(x)$ using the ISP scheme and constructed a reweighted MSM using the exact reweighting factor $M_L(\omega; \Delta t | (x_0,v_0))$ (eq.~\ref{equ:M_Langevin_from_path}).
The dominant MSM eigenfunctions are shown as green lines in Fig.~\ref{fig:path_probabilities}.B. 
They exactly match the reference solution. 
The reweighted implied timescales are shown as green lines in Fig.~\ref{fig:path_probabilities}.C and are in good agreement with the reference solution. 
Finally, we used the simulation at $V(x)$ to construct a reweighted MSM using the reweighting factor for the Euler-Maruyama scheme $M_o(\omega; \Delta t | x_0)$ (eq.~\ref{equ:M_odamped_Langevin_from_path}).
The dominant MSM eigenfunctions are shown as blue lines in Fig.~\ref{fig:path_probabilities}.B. 
The eigenfunctions differ considerably from the reference solution.
Most notably, the stationary distribution is not reproduced correctly (blue line in the upper panel in Fig.~\ref{fig:path_probabilities}.B). 
The left peak is reduced to a shoulder of the central peak, and the relative heights of central peak and the right peak do not match those of the reference solution.
Likewise, the implied timescales (blue line in Fig.~\ref{fig:path_probabilities}.C) are severely underestimated. 
This indicates that using the path probability ratio for overdamped Langevin dynamics, $M_o(\omega; \Delta t | x_0)$, to reweight Langevin trajectories does not yield acceptable results. 
%

\section{Approximate path probability ratio}
\label{sec:Approximate_path_probability_ratio}

\begin{figure}[!h]
\centering
    \includegraphics[width=16cm]{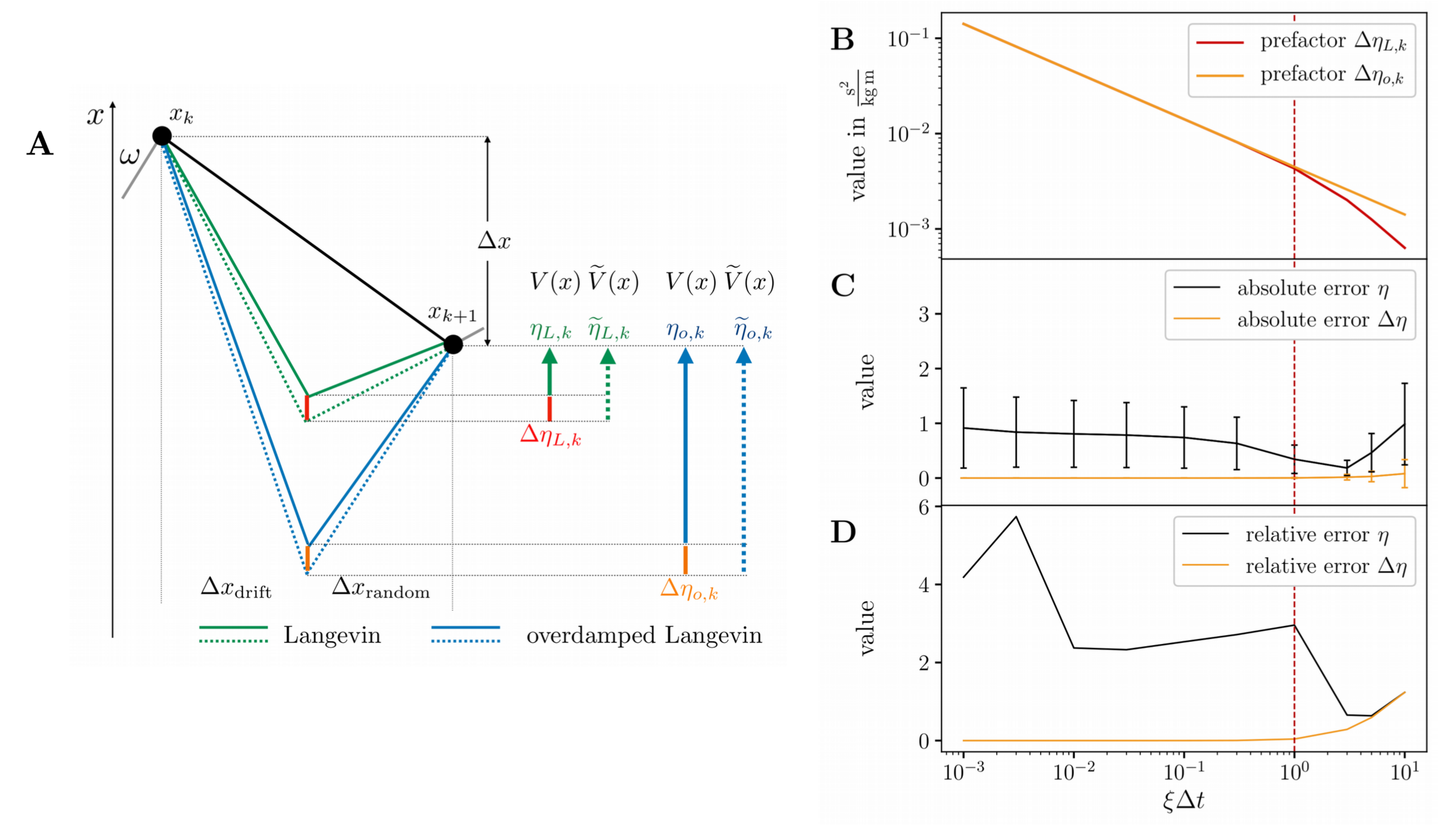}
    \caption{
    \textbf{A} Sketch of a step $x_k \to x_{k+1}$ and the quantities of influence for Langevin and overdamped Langevin dynamics.;
    \textbf{B} Prefactors of $\Delta \eta_{L,k}$ and $\Delta \eta_{o,k}$ as a function of $\xi \Delta t$;
    \textbf{C} Absolute difference (absolute error) between the random numbers $\langle |\eta_{o,k} -\eta_{L,k}|\rangle$ and the random number differences $\langle |\Delta \eta_{o,k}-\Delta \eta_{L,k}|\rangle$ as a function of $\xi \Delta t$; 
    \textbf{D} Relative difference (relative error) between the random numbers $\langle |(\eta_{o,k}-\eta_{L,k})/\eta_{L,k}|\rangle$ and the random number differences $\langle |(\Delta \eta_{o,k}-\Delta \eta_{L,k})/\Delta \eta_{L,k}|\rangle$ as a function of $\xi \Delta t$.
    \label{fig:M_approx}}
\end{figure}

\subsection{Derivation and numerical results\label{sec:M_approx}}
With the results from section \ref{sec:Langevin}, the exact random number probability ratio $M_{L}(\omega, \eta_L; \Delta t|(x_0, v_0))$ (eq.~\ref{eq:M_eta}) for the ISP scheme is straight-forward to evaluate from a simulation at $V(x)$: 
the random number sequence $\eta = \eta_L$ can be recorded during the simulation, and the random number difference $\Delta \eta = \Delta \eta_L$ is given by eq.~\ref{equ:delta_eta_Langevin}.
Inserting $\eta_L$ and $\Delta \eta_L$ into eq.~\ref{eq:M_eta} yields 
$M_{L}(\omega, \eta_L; \Delta t|(x_0, v_0))$.
However, $\Delta \eta_{L,k}$ in eq.~\ref{equ:delta_eta_Langevin} is specific to the ISP scheme. 
If one uses a different Langevin integration scheme to simulate the dynamics at $V(x)$, one needs to adapt eq.~\ref{equ:delta_eta_Langevin} via the strategy outlined in section \ref{sec:Langevin}.
Fortunately, the random number difference for overdamped Langevin dynamics $\Delta \eta_{o,k}$ (eq.~\ref{equ:delta_eta_overdamped_Langevin}) is approximately equal to $\Delta \eta_{L,k}$ for any given perturbation $U(x)$.
Fig.~\ref{fig:example_quantities}.C already suggests that. 
In appendix \ref{appendix:Delta_eta} we show that the difference between $\Delta \eta_{L, k}^2$ and  $\Delta \eta_{o, k}^2$ is in fact only of $\mathcal{O}(\xi^4 \Delta t^4)$, so that for $\xi \Delta t < 1$ we can assume with high accuracy that
\begin{eqnarray}
    \Delta \eta_{L, k} &\approx& \Delta \eta_{o, k}  \nonumber \\[2pt]
    \sqrt{\frac{1}{k_B T \xi^2 m}} \frac{1-\exp(-\xi \Delta t)}{\sqrt{1-\exp(-2 \xi \Delta t)}} \cdot \nabla U(x_k)
     &\approx& 
    \sqrt{\frac{\Delta t}{2k_B T \xi m}} \cdot \nabla U(x_k) \, .
\label{eq:Delta_eta_approx}    
\end{eqnarray}
The difference between $\Delta \eta_{L, k}$ and $\Delta \eta_{o, k}$ is determined by the prefactors in front of $\nabla U(x_k)$ in eq.~\ref{eq:Delta_eta_approx}, which are shown as a function of $\xi\Delta t$ in Fig.~\ref{fig:M_approx}.B.
For $\xi\Delta t<1$, the two curves are virtually identical.
With the approximation in eq.~\ref{eq:Delta_eta_approx}, we can derive an approximate random number probability ratio, by using the recorded $\eta_L$, but substituting $\Delta \eta_{L,k}$ (eq.~\ref{equ:delta_eta_Langevin}) by  $\Delta \eta_{o,k}$ (eq.~\ref{equ:delta_eta_overdamped_Langevin}) in eq.~\ref{eq:M_eta}:
\begin{eqnarray}
    &&M_{L}(\omega, \eta_L; \Delta t|(x_0, v_0)) \cr
    &\approx& M_{\mathrm{approx}}(\omega, \eta_L; \Delta t|x_0) \cr
    &=& \exp \left( -   \sum\limits_{k=0}^{n-1}\sqrt{\frac{\Delta t}{2k_B T \xi m}}  \nabla U(x_k) \cdot \eta_{L,k} \right) 
    \cdot \exp \left( - \frac{1}{2} \sum\limits_{k=0}^{n-1} \frac{\Delta t}{2 k_B T \xi m} \left(  \nabla U(x_k) \right)^2 \right) \, .
\label{eq:M_Langevin_approx}    
\end{eqnarray}
Eq.~\ref{eq:M_Langevin_approx} has the same functional form as the random number probability ratio for the Euler-Maruyama scheme $M_o(\omega, \eta_o; \Delta t|x_0)$ (eq.~\ref{M_odamped_Langevin_from_rand_numb}), but it uses $\eta_L$, the random numbers generated during the ISP simulation, instead of $\eta_o$. 
Eq.~\ref{eq:M_Langevin_approx} is the approximation that we used in refs.~\onlinecite{Donati2017} and \onlinecite{Donati2018}, because we had not yet derived $M_{L}(\omega, \eta_L; \Delta t|(x_0, v_0))$ (eqs.~\ref{equ:M_Langevin_from_path} and \ref{M_Langevin_from_rand_numb}).
Fig.~\ref{fig:path_probabilities} demonstrates the accuracy of the approximate random number probability ratio $M_{\mathrm{approx}}(\omega, \eta_L; \Delta t|x_0)$ (eq.~\ref{eq:M_Langevin_approx}) for our test system.
The dashed orange line in Fig.~\ref{fig:path_probabilities}.A shows the reweighted path probability for the short example path, where we used $M_{\mathrm{approx}}(\omega, \eta_L; \Delta t|x_0)$ (eq.~\ref{eq:M_Langevin_approx}), in eq.~\ref{equ:path_reweighing}.
It exactly matches the reference solution (black line). 
Next, we constructed a reweighted MSM for the target potential $\widetilde V(x)$ based on our simulations at the simulation potential $V(x)$ using $M_{\mathrm{approx}}(\omega, \eta_L; \Delta t|x_0)$ (eq.~\ref{eq:M_Langevin_approx}) to reweight the transition counts.
The dominant MSM eigenfunctions of the reweighted MSM are shown as dashed orange lines in Fig.~\ref{fig:path_probabilities}.B. 
They exactly match the reference solution. 
The reweighted implied timescales are shown as dashed orange lines in Fig.~\ref{fig:path_probabilities}.C and seem to match the reference solution even better than the ones calculated using the exact path probability ratio (green line in Fig.~\ref{fig:path_probabilities}.C).
However, the difference between the dashed orange line and the green line is likely within statistical uncertainty.
In summary, $M_{\mathrm{approx}}(\omega, \eta_L; \Delta t|x_0)$ is a highly accurate approximation to $M_{L}(\omega, \eta_L; \Delta t|x_0)$ for $\xi \Delta t <1$. 
Using $M_{\mathrm{approx}}(\omega, \eta_L; \Delta t|x_0)$ instead of  $M_{L}(\omega, \eta_L; \Delta t|x_0)$ could even have the following advantages:
($i$) the implementation is less error-prone, because the functional form of $M_{\mathrm{approx}}$ is simpler than that of  $M_{L}$; 
($ii$) $M_{\mathrm{approx}}$ might be numerically more stable because the calculation of exponential function on the left-hand-side of eq.~\ref{eq:Delta_eta_approx} is avoided.
%

%

\subsection{Intuition}
We discuss why $M_{\mathrm{approx}}(\omega, \eta_L; \Delta t|x_0)$ is a better approximation to $M_{L}(\omega, \eta_L; \Delta t|x_0)$ than 
$M_{o}(\omega; \Delta t|x_0) = M_{o}(\omega, \eta_o; \Delta t|x_0)$.
Fig.~\ref{fig:M_approx}.A shows one integration time step of a stochastic integration scheme from $x_k$ to $x_{k+1}$ (black line). 
From $k$ to $k+1$ the system has progressed by $\Delta x = x_{k+1}-x_k$.
In the ISP scheme, this progress is composed of a progress 
\begin{eqnarray}
    \Delta x_{\mathrm{drift},L} &=&  \exp \left( - \xi \, \Delta t \right) \, v_k \Delta t - \bigg[ 1 - \exp \left( - \xi \, \Delta t\right) \bigg] \, \frac{\nabla V(x_k)}{\xi m} \Delta t   
\end{eqnarray}
due to the drift force and the velocity of the system (2nd and 3rd term on the right-hand side of eq.~\ref{equ:iteration_scheme_1D_1}), 
and a progress
\begin{eqnarray}
    \Delta x_{\mathrm{random},L} &=& \sqrt{\frac{k_B T}{m} \, \bigg[ 1 - \exp \left( - 2 \xi \, \Delta t\right) \bigg]} \, \eta_{L,k} \, \Delta t 
\end{eqnarray}
due to the random force 
(4th term on the right-hand side of eq.~\ref{equ:iteration_scheme_1D_1}), 
such that $\Delta x = \Delta x_{\mathrm{drift},L} + \Delta x_{\mathrm{random},L}$.
$\Delta x_{\mathrm{drift},L}$ and $\Delta x_{\mathrm{random},L}$ are illustrated as solid green lines in Fig.~\ref{fig:M_approx}.A.  
The probability of generating the step $x_k \rightarrow x_{k+1}$ is determined by  $\Delta x_{\mathrm{random},L}$ which is proportional to the random number $\eta_{L,k}$ (solid green arrow).
With a different potential energy function $\widetilde V(x)$ at $x_k$, the displacement due to the drift force differs from the original $\Delta x_{\mathrm{drift},L}$. 
To achieve the same overall displacement $\Delta x$, $\Delta x_{\mathrm{random},L}$ needs to be adjusted (dotted green line).  
The corresponding random number $\widetilde \eta_{L,k}$ is shown as a dotted green arrow, and the difference between the two random numbers $\Delta \eta_{L,k}$ is shown as a red line. 
In path reweighting, one constructs $\widetilde \eta_{L,k}$ by adding $\Delta \eta_{L,k}$ to $\eta_{L,k}$
\begin{eqnarray}
    \widetilde \eta_{L,k} &=& \eta_{L,k} + \Delta \eta_{L,k} 
\label{eq:eta_L}
\end{eqnarray}
(analogous to eq.~\ref{eq:eta_plus_Delta_eta}), 
which then yields the general form of the random number probability ratio in eq.~\ref{eq:M_eta}.
An analogous analysis applies to the Euler-Maruyama scheme, where the progress due to the drift force is
\begin{eqnarray}
    \Delta x_{\mathrm{drift}, o} = - \frac{\nabla V(x_k)}{\xi m} \, \Delta t
\end{eqnarray}
(2nd term on the right-hand side of eq.~\ref{equ:Euler-Maruyama_1D}), and the progress due to the random force is
\begin{eqnarray}
    \Delta x_{\mathrm{random}, o} = \sqrt{\frac{2k_B T}{\xi m}} \, \sqrt{\Delta t} \, \eta_{o,k}
\end{eqnarray}
(3rd term on the right-hand side of eq.~\ref{equ:Euler-Maruyama_1D}).
In Fig.~\ref{fig:M_approx}.A $\Delta x_{\mathrm{drift}, o}$ and $\Delta x_{\mathrm{random}, o}$ are illustrated as solid blue lines, and the random number as a solid blue arrow.  
With a different potential energy function $\widetilde V(x)$ at $x_k$, the progress due to the drift force differs from the original $\Delta x_{\mathrm{drift}, o}$. 
To achieve the same overall progress $\Delta x$, $\Delta x_{\mathrm{random}, o}$ needs to be adjusted (dotted blue line).  
The corresponding random number $\widetilde \eta_{o,k}$ is shown as a dotted blue arrow, and the difference between the two random numbers $\Delta \eta_{o,k}$ is shown as an orange line. 
In section \ref{sec:M_approx} we have shown that $\Delta \eta_{L,k} \approx \Delta \eta_{o,k}$ (for $\xi\Delta t < 1$).
Thus, approximating $\Delta \eta_{L,k}$ by $\Delta \eta_{o,k}$ in eq.~\ref{eq:eta_L}, or visually: approximating the red line by the orange line in Fig.~\ref{fig:M_approx}.A, is valid.
However, the displacement due to the drift $\Delta x_{\mathrm{drift},o}$ in the Euler-Maruyama scheme can differ strongly from the $\Delta x_{\mathrm{drift},L}$ in the ISP scheme, and consequently the random numbers needed to generate the same overall progress $\Delta x$ differ
\begin{eqnarray}
    \eta_{L,k} \not\approx \eta_{o,k} 
\end{eqnarray}
(solid blue and solid green arrow in Fig.~\ref{fig:M_approx}.A).
Consequently, approximating $\eta_{L,k}$ by $\eta_{o,k}$ in eq.~\ref{eq:eta_L}, or visually: approximating the solid green arrow by the solid blue arrow in Fig.~\ref{fig:M_approx}.A, is not valid.
The exact random number probability ratio  $M_{L}(\omega, \eta_L; \Delta t|(x_0, v_0))$ (eq.~\ref{M_Langevin_from_rand_numb}) uses the exact $\eta_L$ recorded during the simulation and the exact $\Delta \eta_L$
(eq.~\ref{equ:delta_eta_Langevin}).
It therefore yields results that exactly match the reference solutions
(green lines in Fig.~\ref{fig:path_probabilities}).
$M_{\mathrm{approx}}(\omega, \eta_L; \Delta t|x_0)$ uses the exact $\eta_L$ recorded during the simulation, but approximates $\Delta \eta_{L,k}$ by $\Delta \eta_{o,k}$. 
This introduces only a small error, but still yields excellent reweighting results in our test system (dashed orange lines in Fig.~\ref{fig:path_probabilities}).
However, in $M_{o}(\omega; \Delta t|x_0) = M_{o}(\omega, \eta_o; \Delta t|x_0)$ one additionally approximates $\eta_L$ by $\eta_o$.
The difference between $\eta_L$ and $\eta_o$ is much larger than the difference between $\Delta \eta_L$ and $\Delta \eta_o$, 
and this additional approximation leads to the distorted reweighting results we observed as the blue lines in Fig.~\ref{fig:path_probabilities}.
The proportions in Fig.~\ref{fig:M_approx}.A are not exaggerated. 
The black line in Fig.~\ref{fig:M_approx}.C shows the average absolute difference between the random numbers $\langle|\eta_{o,k}-\eta_{L,k}|\rangle$ as a function of $\xi\Delta t$.
Visually this is the difference between the solid green arrow and the solid blue arrow in Fig.~\ref{fig:M_approx}.A. 
The orange line in Fig.~\ref{fig:M_approx}.C shows the average absolute difference between the random number differences $\langle|\Delta \eta_{o,k}-\Delta\eta_{L,k}|\rangle$, i.e. the difference between the orange and the red line in Fig.~\ref{fig:M_approx}.A.
The graph has been calculated by averaging over a path with $10^6$ time steps. The standard deviations are shown as vertical bars. 
$\langle|\Delta \eta_{o,k}-\Delta\eta_{L,k}|\rangle$ is close to zero for all values of $\xi\Delta t$, whereas there is a substantial difference between $\eta_L$ and $\eta_o$. 
$\langle|\eta_{o,k}-\eta_{L,k}|\rangle$ has a minimum at $\xi\Delta t\approx2$, because the difference between the Euler-Maruyama scheme and the ISP scheme is minimal for $\xi\Delta t \approx 2$
(see discussion in \ref{sec:EulerMaruyamaVsISP}).
Fig.~\ref{fig:M_approx}.D shows the corresponding average relative errors. 
For $\xi \Delta t > 1$, $\langle |(\eta_o-\eta_L)/\eta_L|\rangle$ (black line) decreases in accordance with the decrease of the absolute difference $\langle |(\eta_o-\eta_L)|\rangle$, 
and $\langle |(\Delta\eta_o-\Delta\eta_L)/ \Delta \eta_L|\rangle$ (orange line) increases, reflecting the fact that the approximation (eq.~\ref{eq:Delta_eta_approx}) does not hold for $\xi\Delta t > 1$.
However, for $\xi\Delta t < 1$, the region in which MD simulations are conducted, the relative error for the random numbers is much larger than the relative error for the random number difference. 
This reinforces that the random numbers $\eta_{L,k}$ should not be approximated in the path probability ratio, but instead should be recorded from the simulation at $V(x)$. 
By contrast, the random number difference $\Delta \eta_{L,k}$ can reliably be approximated by eq.~\ref{eq:Delta_eta_approx}.

\section{Molecular example: butane}
\label{sec:molecular_example}

\begin{figure}[!h]
\centering
    \includegraphics[width=16cm]{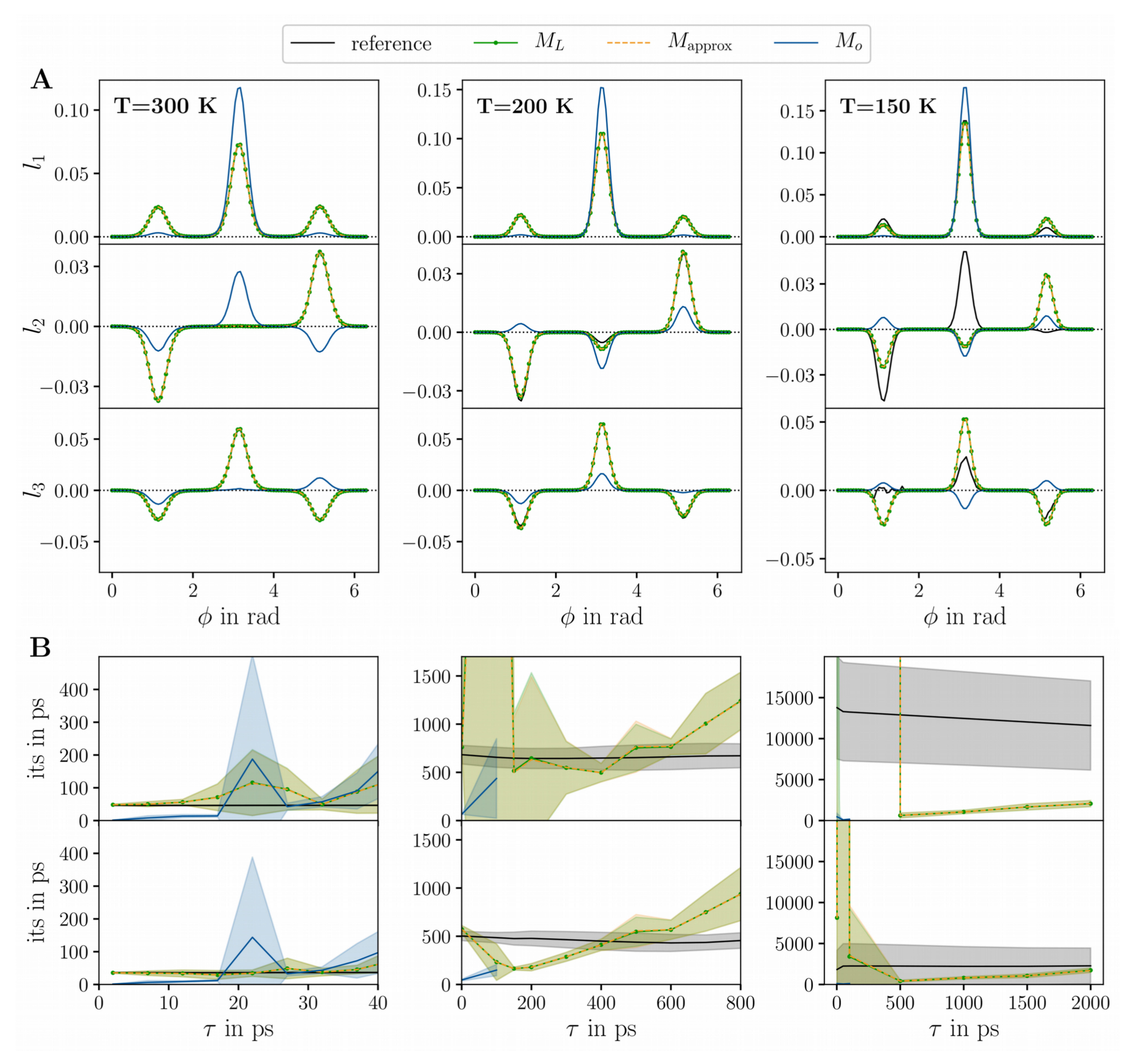}
    \caption{Dynamics of the torsion angle in butane at $T=300,\, 200$ and $150$ K.
    \textbf{A} Dominant left eigenfunctions $l_1,l_2$ and $l_3$ of the MSM along the torsion angle $\phi$, obtained by evaluating direct simulations at the target potential, 
    as well as by reweighting biased simulations. 
    \textbf{B} Implied timescales corresponding to $l_2$ and $l_3$ in panel A.  
    Solid lines: mean, shaded area: standard deviation. Standard deviations for the eigenvectors are too small to be shown.}
\label{fig:butane_MSMs}
\end{figure}

The slowest degree of freedom in butane is the torsion around the C$_2$-C$_3$ bond, which exhibits 
three metastable states: the trans-conformation at $\phi = \pi$, and the two gauche-conformations at $\phi = \pm \frac{1}{3}\pi$.
Consequently, butane has three dominant MSM eigenvectors,  where $l_1$ corresponds to the stationary density, and $l_2$ and $l_3$ represent slow transitions along $\phi$ (Fig.~\ref{fig:butane_MSMs}.A).
Because, the two gauche-conformations are equally populated, $l_2$ and $l_3$ are degenerate
(Fig.~\ref{fig:butane_MSMs}.B).
We simulated butane in implicit water at three different temperatures, $T=300$ K, $T=200$ K and $T=150$ K, using direct and biased simulations.
As we lower the temperature, we expect that the relative population of the trans-conformation increases, but that otherwise the overall shape and sign-structure of the eigenvectors remains unchanged.

At $T=300$ K and $T=200$ K, the reweighting results using $M_{\mathrm{approx}}(\omega, \eta_L; \Delta t|x_0)$ (eq.~\ref{eq:M_Langevin_approx}, dashed orange line) or $M_{L}(\omega, \eta_L; \Delta t|(x_0, v_0))$ (eq.~\ref{equ:M_Langevin_from_path}, solid green line) match the MSM obtained by direct simulation. In particular, the eigenvectors are reproduced with very high precision. 
By contrast, the reweighted results using $M_o(\omega; \Delta t | x_0)$ (eq.~\ref{equ:M_odamped_Langevin_from_path}, blue line) deviate considerable from the reference MSMs obtained by direct simulations.
The stationary distribution $l_1$ is not reproduced correctly, which then leads to further errors in the dominant eigenvectors $l_2$ and $l_3$. The associated implied timescales are underestimated. 
Moreover, for $T=200$ K and $T=150$ K the use of $M_o(\omega; \Delta t | x_0)$ yielded numerically instable transition matrices for lag times of $\tau > 100$ ps.
This demonstrates that path reweighting with an appropriate path probability ratio, such as $M_{\mathrm{approx}}(\omega, \eta_L; \Delta t|x_0)$ or $M_{L}(\omega, \eta_L; \Delta t|(x_0, v_0))$, yields accurate results.
However, $M_o(\omega; \Delta t | x_0)$ should not be used as an approximation for the exact path probability ratio $M_{L}(\omega, \eta_L; \Delta t|(x_0, v_0))$.

Note that reweighting results using the approximate probability ratio $M_{\mathrm{approx}}(\omega, \eta_L; \Delta t|x_0)$ are virtually indistinguishable from the results using the exact probability ratio $M_{L}(\omega, \eta_L; \Delta t|(x_0, v_0))$ for all three temperatures. 
This confirms our analysis that $M_{\mathrm{approx}}(\omega, \eta_L; \Delta t|x_0)$ can be used as highly accurate approximation to $M_{L}(\omega, \eta_L; \Delta t|(x_0, v_0))$.

The variation of the temperature from 300 K, to 200 K and 150 K illustrates under which circumstances path reweighting is an efficient method. 
At $T=300$ K, many transitions across the torsion angle barriers are observed in the direct simulation. 
Path reweighting and direct simulation yield identical results. 
However, path reweighting has a larger statistical uncertainty. 
At $T=200$ K, fewer transitions are observed in the direct simulations, which results in an increased statistical uncertainty in the direct MSM. 
Finally, at $T=150$ K the transitions in the direct simulation are insufficient to correctly sample the stationary density.
The MSM of the direct simulation predicts a higher population for the gauche-conformation at $\phi = +\frac{1}{3}\pi$ than for the gauche-conformation at $\phi = -\frac{1}{3}\pi$, which is clearly a sampling error. 
This error in the stationary density then leads to vastly incorrect estimates for $l_2$ and $l_3$. 
Additionally, the direct MSM predicts that the degeneracy is lifted.
By contrast, the results of the reweighted MSM are in line with what we expect: the gauche-conformations are equally populated, the overall shapes of the dominant eigenvectors corresponds to those of the eigenvectors at higher temperatures, and $l_2$ and $l_3$ are degenerate. 
In conclusion, path reweighting in combination with enhanced sampling techniques is a promising tool in situations, where the stationary density cannot be sampled accurately by direct simulation.

\section{Methods\label{sec:methods}}

\subsection{Simulations of the test system}
\label{sec:methods_test_system}
The test system is a one-dimensional one particle system with mass $m=1\,{\rm kg}$ and 
$k_BT = 2.494\, \mathrm{J}$
(corresponding to $k_B=0.008314\, \mathrm{J/K}$ and $T=300\, \mathrm{K}$).
The simulation potential (orange line in Fig.~\ref{fig:test_system}) is
\begin{eqnarray}
    V(x) &=& (x^2 - 1)^2 
\label{eq:simulation_potential}    
\end{eqnarray}
and the target potential (black line in Fig.~\ref{fig:test_system}) is
\begin{eqnarray}
    \widetilde{V}(x) = 4(x^3 - \frac{3}{2} x)^2 - x^3 + x \, .
\label{eq:target_potential}    
\end{eqnarray}
For the results in Figs.~\ref{fig:example_quantities} - \ref{fig:M_approx}, we simulated the system using the ISP scheme (eqs.~\ref{equ:iteration_scheme_1D_1} and \ref{equ:iteration_scheme_1D_2}) with a time step of $\Delta t = 0.01 \,\mathrm{s}$. 
The initial conditions were $x_0 = 1.50\, \mathrm{m}, v_0=0\, \mathrm{m/s}$. 
The number of time steps $N_t$, the collision rate $\xi$, and the potential energy function used are summarized in Table \ref{tab:simulation_details}.
\begin{table}[h!]
    \centering
    \begin{tabular}{cccc}
        Fig.    &$N_t$ &$\xi$  &potential  \\
        \hline
        \ref{fig:example_quantities}.A& $10^5$  &$50\,\mathrm{s}^{-1}$  &$V(x)$  \\
        \hline
        \ref{fig:path_probabilities}.B-C& $10^7$  &$50\,\mathrm{s}^{-1}$  &$V(x)$  \\
        \ref{fig:path_probabilities}.B-C& $10^7$  &$50\,\mathrm{s}^{-1}$  &$\widetilde V(x)$  \\
        \hline
        \ref{fig:M_approx}.C-D  &$10^7$ & $0.1\,\mathrm{s}^{-1}$ - $1000\,\mathrm{s}^{-1}$ &$V(x)$  
    \end{tabular}
    \caption{Simulation parameters}
    \label{tab:simulation_details}
\end{table}

In Fig.~\ref{fig:example_quantities}.A, we computed the acceleration $\ddot{x}=a$ as $a_{k+1} = \frac{v_{k+1}-v_k}{\Delta t}$. 
Fig.~\ref{fig:example_quantities}.B displays the first ten steps of the simulation as example path $\omega$, and all quantities displayed in Fig.~\ref{fig:example_quantities}.C-E are calculated from this short path.
The absolute and relative difference of the random numbers in Fig.~\ref{fig:M_approx} were calculated as 
\begin{eqnarray}
    \langle |\eta_{o,k}-\eta_{L,k}|\rangle &=& \frac{1}{N_t-1} \sum_{k=0}^{N_t-1} |\eta_{o,k} - \eta_{L,k} | \, ,
\end{eqnarray}
and
\begin{eqnarray}
    \left\langle \left|\frac{\eta_{o,k} - \eta_{L,k}}{\eta_{L,k}}\right|\right\rangle &=& \frac{1}{N_t-1} \sum_{k=0}^{N_t-1} \left|\frac{\eta_{o,k} - \eta_{L,k}}{\eta_{L,k}} \right| \, .
\end{eqnarray}
Analogous equations were used for $\langle |\Delta\eta_{o,k}-\Delta\eta_{L,k}|\rangle$ and $\langle |(\Delta\eta_{o,k}-\Delta\eta_{L,k})/\Delta\eta_{L,k}|\rangle$.
$\eta_{L,k}$ was recorded during the simulation. 
We used eq.~\ref{equ:delta_eta_Langevin} to calculate $\Delta \eta_{L,k}$, 
eq.~\ref{equ:odamped_rand_numb_simulated_sys} to calculate $\eta_{o,k}$, and
eq.~\ref{equ:delta_eta_overdamped_Langevin} to calculate $\Delta \eta_{o,k}$.

The reference MSM in fig.~\ref{fig:path_probabilities}.B-C has been constructed from the simulation at the target potential $\widetilde{V}(x)$.
The state space has been discretized using a regular grid of 100 microstates ($S_1,\dots,S_{100}$) in the range
$-1.7 \le x \le 1.6$. 
Transition counts between microstates were calculated as 
\begin{eqnarray}
    c_{ij}(\tau) = \frac{1}{N_t-\tau} \sum\limits_{k=0}^{N_t-\tau} \chi_i(x_k) \chi_j(x_{k+\tau})
\label{eq:transitionCountTarget}    
\end{eqnarray}
with
\begin{eqnarray}
    \chi_i(x) = 
    \begin{cases}
    1 \quad \text{if } x \in S_i \\
    0 \quad \text{else} \, ,
    \end{cases}
\end{eqnarray}
where $x_k$ is the trajectory, and lag time $\tau=200$ steps.
The resulting count matrix $\mathbf{C}(\tau)$ was symmetrized as $\mathbf{C}(\tau) + \mathbf{C}^{\top}(\tau)$ to enforce detailed balance, 
and row-normalized to obtain the MSM transition matrix $\mathbf{T}(\tau)$. 
The dominant MSM eigenvectors $l_i$ and associated eigenvalues $\lambda_i(\tau)$ were calculated from $\mathbf{T}(\tau)$ using a standard eigenvalue solver, and the implied timescales were calculated as
$t_i =-\tau / \ln(\lambda_i(\tau))$.
The reweighted MSMs in fig.~\ref{fig:path_probabilities}.B-C have been constructed from the simulation at the simulation potential $V(x)$ using the same grid and lag time as for the reference MSM. 
Transition counts between microstates were counted and reweighted as
\cite{Donati2017, Donati2018}
\begin{eqnarray}
    \widetilde{c}_{ij}(\tau) = \frac{1}{N_t-\tau} \sum\limits_{k=0}^{N_t-\tau} W((x_k,x_{k+1},\dots,x_{k+\tau});\Delta t|(x_k,v_k)) \chi_i(x_k) \chi_j(x_{k+\tau}) \, .
\label{eq:transitionCountReweighted}    
\end{eqnarray}
The weight $W$ is defined as
\begin{eqnarray}
    W((x_k,x_{k+1},\dots,x_{k+\tau});\Delta t|(x_k,v_k)) = g(x_k) \cdot M((x_k,x_{k+1},\dots,x_{k+\tau});\Delta t|(x_k,v_k)) \, 
\end{eqnarray}
with $M$ being the path probability ratio (eq.~\ref{eq:M_omega}) and $g$ being
\begin{eqnarray}
    g(x_k) = \exp \left(-\frac{U(x_k)}{k_B T} \right)\, ,
\label{equ:Boltzmann_probability_ratio}    
\end{eqnarray}
where the perturbation $U$ is defined in eq.~\ref{equ:def_perturbed_pot}.
The remaining procedure was analogous to the reference MSM. 

\subsection{Butane - direct simulations}
\label{sec:methods_butane_direct}

We performed all-atom MD simulations of $n$-butane in implicit water using the OpenMM 7.4.1 \cite{Eastman2017} simulation package.
The GAFF (Generalized Amber Force Field) forcefield \cite{Wang2004} was used to model butane, and the GBSA (Generalized Born Surface Area) model \cite{Onufriev2004} to model implicit water. 
Interactions beyond 1 nm were truncated.
The trajectory was propagated according to the ISP integration scheme for a $3N$-dimensional system
\begin{eqnarray}
x^i_{k+1} &=& x^i_k + \exp \left( - \xi \, \Delta t \right) \, v^i_k \Delta t - \bigg[ 1 - \exp \left( - \xi \, \Delta t\right) \bigg] \, \frac{\nabla_i V(\mathbf{x}_k)}{\xi m_i} \Delta t \cr
&& + \sqrt{\frac{k_B T}{m_i} \, \bigg[ 1 - \exp \left( - 2 \xi \, \Delta t\right) \bigg]} \, \eta^i_{L,k} \, \Delta t \label{equ:iteration_scheme_3ND_1} \\
v^i_{k+1} &=& \frac{x^i_{k+1} - x^i_k}{\Delta t} \, ,
\label{equ:iteration_scheme_3ND_2}
\end{eqnarray}
with $i=1, 2, \dots, 3N$ and $N$ being the number of atoms.
$x^i_k$, $v^i_k$ and $\eta^i_k$ are the position, velocity and random number along dimension $i$ at iteration step $k$, $m_i$ is the mass of dimension $i$ and $\nabla_i V(\mathbf{x}_k)$ denotes the gradient of $V(\mathbf{x}_k)$ along dimension $i$ measured at the position $\mathbf{x}_k$ with $\mathbf{x} \in \mathbb{R}^{3N}$.
We implemented the ISP integration scheme using the simtk.openmm.openmm.CustomIntegrator \cite{OpenMM_CustomIntegrator} class of OpenMM .
The collision rate was $\xi = 10$ ps$^{-1}$. 
The simulation time step was $\Delta t=0.002$ ps.
Positions were written to disc every \texttt{txout}=50 steps = 0.1 ps. 
We generated three trajectories with 500 ns each at $T=300$ K, $T=200$ K and $T=150$ K.
These direct simulations correspond to simulations at the target potential $\widetilde V(\mathbf{x})$.

For the analysis, we cut each trajectory into 5 pieces of length 100 ns.
For each 100-ns-trajectory, we constructed a MSM following the procedure outlined in section \ref{sec:methods_test_system}.
As state space we chose the C$_2$-C$_3$ dihedral angle $\phi$, which we discretized using a regular grid of 100 microstates in the range $0 \leq \phi \leq 2\pi$. 
This resulted in five MSMs for each temperature. 
Fig.~\ref{fig:butane_MSMs} shows the mean and the standard deviation of the first three left MSM eigenvectors (evaluated at lag time $\tau = 1$ ps), and the mean and the standard deviations of the associated implied timescale.

\subsection{Butane - path reweighting}
We biased the simulations along the C$_2$-C$_3$ dihedral angle $\phi$.
To generate the bias potential $U(\phi)$, we constructed a histogram of the free-energy function $\widetilde{F}(\phi)$
\begin{equation}
    \widetilde{F}(\phi) = - k_B T \ln \left( \widetilde{p}(\phi) \right) \, ,
\end{equation}
where $\widetilde{p}(\phi)$ is the stationary density along $\phi$ as measured from the 500 ns direct simulations at $T=300$ K.
Fitting the histogram with a third order Fourier series yielded
\begin{eqnarray}
    \widetilde{F}_{300 \,\mathrm{K}}(\phi)
    &=& 8.985 + 3.122 \cos(\omega \phi) + 0.959 \cos(2 \omega \phi) + 7.742 \cos(3\omega \phi)\cr
    && + 0.095 \sin(\omega \phi) + 0.047 \sin(2\omega \phi) + 0.002 \sin(3 \omega \phi) 
    \label{equ:T_300_free_energy_fit}
\end{eqnarray}
with $\omega=0.989$. 
The same procedure for the simulation at $T=200$ K yielded
\begin{eqnarray}
    \widetilde{F}_{200 \,\mathrm{K}}(\phi) 
    &=& 8.311 + 2.847 \cos(\omega \phi) + 0.841 \cos(2 \omega \phi) +  7.697 \cos(3 \omega \phi) \cr
    &&  + 0.046 \sin(\omega \phi) + 0.026 \sin(2 \omega \phi) + 0.004 \sin(3 \omega \phi)
\end{eqnarray}
with $\omega=0.989$.
$\widetilde{F}_{300 \,\mathrm{K}}(\phi)$ and $\widetilde{F}_{200 \,\mathrm{K}}(\phi)$ are almost identical.
The simulation at $T=150$ K did not yield a converged stationary density, and thus no free-energy function was constructed for this temperature, and instead $\widetilde{F}_{300 \,\mathrm{K}}(\phi)$ was used.

The biased simulations were carried out with the potential
\begin{align}
    V_{\alpha}(\mathbf{x}) &= \widetilde{V}(\mathbf{x}) - \alpha \cdot \widetilde{F}(\phi(\mathbf{x})) \, .
\end{align}
where $\widetilde{V}(\mathbf{x})$  is the target potential, and $\alpha \in [0,1]$ specifies the bias strength.
$V_{\alpha}(\mathbf{x})$ corresponds to the ``simulation potential'' within the terminology of this paper, 
thus
\begin{equation}
    U(\phi(\mathbf{x})) = \alpha \cdot \widetilde{F}(\phi(\mathbf{x})) \, .
\end{equation}
$\alpha$ was set to 0.1 in all biased simulations, corresponding to ``10\% of the full metadynamics potential''.
We carried out biased simulations at three temperatures $T=300$ K, $T=200$ K and $T=150$ K with bias potentials
$U_{300 \,\mathrm{K}}(\phi) = 0.1 \cdot \widetilde{F}_{300 \,\mathrm{K}}(\phi)$,
$U_{200 \,\mathrm{K}}(\phi) = 0.1 \cdot \widetilde{F}_{200 \,\mathrm{K}}(\phi)$ and
$U_{150 \,\mathrm{K}}(\phi) = 0.1 \cdot \widetilde{F}_{300 \,\mathrm{K}}(\phi)$.
All other simulation parameters were as described in section \ref{sec:methods_butane_direct}.

The path probability ratios for the biased simulations were calculated on-the-fly \cite{Donati2017, Donati2018}, and were written to disc at the same frequency \texttt{txout} as the positions.
For the approximate path probability ratio $M_\mathrm{approx}$ we calculated
\begin{equation}
    \mathbb{M}_\mathrm{approx}(b) = \sum\limits_{i=1}^{3N} \sum\limits_{k=(b-1)\cdot \mathrm{txout}}^{b \cdot \mathrm{txout}-1} \left( -\sqrt{\frac{\Delta t}{2k_B T \xi m_i}}  \nabla_i U(\mathbf{x}_k) \, \eta^i_{L,k} - \frac{\Delta t}{4 k_B T \xi m_i} \left(  \nabla_i U(\mathbf{x}_k) \right)^2 \right)\, ,
\end{equation}
and constructed complete path probability ratio as
\begin{equation}
    M_\mathrm{approx}(\boldsymbol{\omega}, \boldsymbol{\eta}_L; \Delta t| \mathbf{x}_0) = \exp \left( \sum\limits_{b=1}^{A} \mathbb{M}_\mathrm{approx}(b) \right) \, 
\end{equation}
during the construction of the MSM,
where $A \in \mathbb{N}$ such that $\tau= A \cdot \texttt{txout} \cdot \Delta t$.
For the Langevin path probability ratio $M_L$ we calculated the terms
\begin{align}
    \mathbb{M}_{L,1} (b) &= \sum\limits_{i=1}^{3N} \sum\limits_{k=(b-1)\cdot \mathrm{txout}}^{b \cdot \mathrm{txout}-1} (x^i_{k+1} - x^i_k)\left(\nabla_i \widetilde{V}(\mathbf{x}_k) - \nabla_i V(\mathbf{x}_k) \right) \\
    \mathbb{M}_{L,2} (b) &= \sum\limits_{i=1}^{3N} \sum\limits_{k=(b-1)\cdot \mathrm{txout}}^{b \cdot \mathrm{txout}-1} v^i_k \left(\nabla_i \widetilde{V}(\mathbf{x}_k) - \nabla_i V(\mathbf{x}_k) \right) \\
    \mathbb{M}_{L,3} (b) &= \sum\limits_{i=1}^{3N} \sum\limits_{k=(b-1)\cdot \mathrm{txout}}^{b \cdot \mathrm{txout}-1} \frac{\left( \big(\nabla_i \widetilde{V}(\mathbf{x}_k ) \big)^2 - \big(\nabla_i V(\mathbf{x}_k) \big)^2 \right)}{m_i} \, ,
\end{align}
and constructed complete path probability ratio as
\begin{eqnarray}
    M_L(\boldsymbol{\omega}, \Delta t| \mathbf{x}_0) 
    &=& \exp \bigg[ \sum\limits_{b=1}^{A} 
        \bigg( -\frac{\mathbb{M}_{L,1}(b)}{k_BT\xi (1 + \exp(-\xi \Delta t)) \Delta t} 
        + \frac{\mathbb{M}_{L,2}(b)}{k_BT\xi (1+\exp(\xi \Delta t))} \nonumber\\[5pt]
    &&  - \frac{\exp(\xi \Delta t)-1}{\exp(\xi \Delta t)+1} \, \frac{\mathbb{M}_{L,3}(b)}{2k_BT\xi^2} \bigg) \bigg]
\end{eqnarray}
during the construction of the MSM,
where $A \in \mathbb{N}$ such that $\tau= A \cdot \texttt{txout} \cdot \Delta t$.
For the  overdamped Langevin path probability ratio $M_o$ we calculated the terms
\begin{align}
    \mathbb{M}_{o,1} (b) &= \sum\limits_{i=1}^{3N} \sum\limits_{k=(b-1)\cdot \mathrm{txout}}^{b \cdot \mathrm{txout}-1} (x^i_{k+1} - x^i_k)\left(\nabla_i \widetilde{V}(\mathbf{x}_k) - \nabla_i V(\mathbf{x}_k) \right) \\
    \mathbb{M}_{o,2} (b) &= \sum\limits_{i=1}^{3N} \sum\limits_{k=(b-1)\cdot \mathrm{txout}}^{b \cdot \mathrm{txout}-1} \frac{\left( \big(\nabla_i \widetilde{V}(\mathbf{x}_k ) \big)^2 - \big(\nabla_i V(\mathbf{x}_k) \big)^2 \right)}{m_i}
\end{align}
and constructed complete path probability ratio as
\begin{equation}
    M_o(\boldsymbol{\omega}, \Delta t| \mathbf{x}_0) = \exp \left[ \sum\limits_{b=1}^{A} \left( -\frac{\mathbb{M}_{o,1}(b)}{2k_BT} 
    - \frac{\mathbb{M}_{o,2}(b) \, \Delta t}{4k_BT\xi} \right) \right] \, 
\end{equation}
during the construction of the MSM,
where $A \in \mathbb{N}$ such that $\tau= A \cdot \texttt{txout} \cdot \Delta t$.

For the analysis, we cut each trajectory into 5 pieces of length 100 ns.
For each 100-ns-trajectory, we constructed a MSM following the procedure outlined in section \ref{sec:methods_test_system}.
As state space we chose the C$_2$-C$_3$ dihedral angle $\phi$, which we discretized using a regular grid of 100 microstates in the range $0 \leq \phi \leq 2\pi$. 
Transition counts between microstates were counted and reweighted as described in eq.~\ref{eq:transitionCountReweighted} with $x_k=\phi_k$ and
\begin{equation}
    g(\phi_k) = \exp \left( -\frac{U(\phi_k)}{k_BT} \right) = \exp \left( -\frac{0.1 \cdot \widetilde{F}(\phi_k)}{k_BT} \right) \, ,
\end{equation}
where $\phi_k$ is the first entry in the path of length $\tau$.
This resulted in five reweighted MSMs for each temperature. 
Fig.~\ref{fig:butane_MSMs} shows the mean and the standard deviation of the first three left MSM eigenvectors (evaluated at lag time $\tau = 1$ ps), and the mean and the standard deviations of the associated implied timescale. 

Example scripts for simulation and analysis are included as supplementary material.

\section{Conclusion and outlook}
We have presented two strategies to derive the path probability ratio $M_L$ for the ISP scheme. 
In the first strategy, the correctly normalized path probability is derived by integrating out the random number $\eta_k$ from the one-step transition probability. 
In the second strategy, the equations for the ISP scheme are solved for $\eta_k$, and the resulting transformation is used as a change of variables on the Gaussian probability density of the random numbers. 
This yields an unnormalized path probability.
The path probability ratio $M_L$ is then calculated as the ratio between the path probability at the target potential $\widetilde P_L(\omega_L;\Delta t | (x_0, v_0))$ and the path probability at the simulation potential
$P_L(\omega_L;\Delta t | (x_0, v_0))$.
With $M_L$ we are now able to perform exact path reweighting for trajectories generated by the ISP integration scheme. 
Moreover, the two strategies serve as a blueprint for deriving path probability ratios for other Langevin integration schemes which use Gaussian white noise \cite{vanGunsteren:1981, Brunger1984, Stoltz:2007, Bussi2007, Izaguirre2010, Goga:2012, Leimkuhler:2013, Sivak:2014, Fass2018}.
Thus, path reweighting can now readily be applied to MD simulation conducted at the NVT ensemble thermostatted by a stochastic thermostat.

We compared the approximate path probability ratio $M_{\mathrm{approx}}$ that we used in earlier publications \cite{Donati2017, Donati2018} to the exact path probability ratio $M_L$, 
both analytically and numerically.
We showed that the two expressions only differ by $\mathcal{O}(\xi^4\Delta t^4)$. 
Thus, $M_{\mathrm{approx}}$ is an excellent approximation to $M_L$ for Langevin MD simulations.
To understand why the approximation is so good, we showed that the random number $\eta_k$ needed to generate a given step $x_k \rightarrow x_{k+1}$ is highly dependent on the integration scheme.
However $\Delta \eta_{k}$, the difference between the random number $\widetilde \eta_k$ at $\widetilde V(x)$ and the random number $\eta_k$ at $V(x)$ has about the same value in the ISP scheme and in the Euler-Maruyama scheme. 

In $M_\mathrm{approx}$, one uses the random numbers directly recorded during the simulation at $V(x)$, which does not introduce any error, and approximates $\Delta \eta_{k}$ by the expression from the Euler-Maruyama scheme $\Delta \eta_{o,k}$ to construct $\widetilde \eta_k$. 

We have chosen the ISP algorithm for the present analysis in order to be consistent with our previous work \cite{Donati2017, Donati2018}.
However, the same strategy can be used to derive the path probability ratio for other Langevin integrators \cite{vanGunsteren:1981, Brunger1984, Stoltz:2007, Bussi2007, Izaguirre2010, Goga:2012, Leimkuhler:2013, Sivak:2014, Fass2018}.
Specifically: solve the integrator equations for the random number $\eta_k$; from there derive an expression for $\Delta \eta_k$; record $\eta_k$ during the simulation at $V(x)$ and calculate $\Delta \eta_k$ on the fly; insert $\eta_k$ and $\Delta \eta_k$ into eq.~\ref{eq:M_eta}.
For a large application of path reweighting, using a modern Langevin integrator is likely worthwhile, 
such as the the BAOAB method \cite{Leimkuhler:2013} (or alternatively: the VRORV method \cite{Fass2018}).
This method is exceptionally efficient at sampling the configurational stationary distribution, which allows for increasing the time step \cite{Leimkuhler:2013, Fass2018}. 

It is tempting to speculate that $\Delta \eta_{k}$ for other Langevin integration schemes could also  have about the same value as $\Delta \eta_{o,k}$ for the Euler-Maruyama scheme. 
This would open up a route to a general approximate path probability ratio $M_{\approx}$ and would eliminate the problem that the path probability needs to be adapted for each integration scheme.
On the other hand, the structure of the ISP scheme is closer to the Euler-Maruyama scheme than most other Langevin integrators.
Whether the approximate path probability can indeed be generalized to these integrators is therefore not yet obvious, and needs to be checked carefully. 

Our one-dimensional test system as well as our molecular system showed that the accuracy of the reweighting sensitively depends on an accurate representation of $\eta_k$ in the path probability ratio. 
For example, reweighting a Langevin path by the path probability ratio for the Euler-Maruyama scheme yielded very distorted results. 
Neither the MSM eigenvectors nor the implied timescales were reproduced correctly.
It is however possible that the distortion is less severe in the limit of infinite sampling of the combined space of molecular states and random numbers (probably less relevant to actual applications), or if the dynamics is projected onto a reaction coordinate before the reweighted dynamical properties are evaluated (probably very relevant to actual applications).
We used path reweighting to reweight MSMs. 
The dynamical property which is reweighted to estimate a transition probability is a correlation function. 
It is important to point out that correlation functions are a combination of path ensemble averages, where the path is conditioned on a particular initial state $(x_0, v_0)$ and a phase-space ensemble average for the initial states. 
Thus, the total reweighting factor for MSMs is combined of the path probability ratio $M$ for the path ensemble average, and the Boltzmann probability ratio for the phase-space ensemble average $g(x)$ (eq.~\ref{equ:Boltzmann_probability_ratio}) \cite{Xing:2006, Prinz:2011c, Schuette:2015, Donati2017}.
Even though the reweighting of the path ensemble average can be made exact, by averaging over the initial states within a microstate one assumes local equilibrium within this microstate \cite{Kieninger2020}.
Beyond local equilibrium, the formalism has been extended to reweighting transition probabilities from non-equilibrium steady-state simulations \cite{Bause:2019}.

When is the combination of enhanced sampling and path reweighting more efficient than a direct simulation? 
This depends on the uncertainty of the transition counts estimated from a direct simulation (eq.~\ref{eq:transitionCountTarget}) compared to the uncertainty of the reweighted transition counts (eq.~\ref{eq:transitionCountReweighted}).
The molecular example demonstrated that path reweighting is particulary useful if the stationary density cannot be sampled accurately by direct simulation with the available computer resources.
Furthermore, the efficiency of path reweightinging increases if the number of transitions at the enhanced sampling simulation is large compared to the direct simulation, and if the weights $W=g \cdot M$ are not too small.
The path probability ratio $M$ decreases with the path length $\tau$ and with the dimensionality of the bias potential $U$. 
The path length is kept short by combining path reweighting with MSMs, and can be further limited by using advanced MSM discretization techniques \cite{PerezHernandez:2013, Nueske:2014, Lemke:2016}.
The bottleneck for the dimensionality $U$ already occurs at the stage of sampling, because most enhanced sampling techniques \cite{Tuckerman2010} are limited to very low-dimensional biases in practice.
Note that increasing the dimensionality of the overall system does not lower the efficiency of the path reweighting. 
The question of how strong the bias should be is more difficult to answer. 
Strong biases increase the transitions in the biased simulation, but reduce both $g$ and $M$.
In ref.~\onlinecite{Donati2018}, we empirically found that a bias of ca.~10\% of the full metadynamics biasing potential yielded optimal results, but this will likely depend on the system. 
Here, we have restricted ourselves to systems with low barriers in the order of $k_BT$, so that we could generate reference solutions by direct simulation. 
But we believe that path reweighting is most useful for systems with large barriers that cannot be sampled by direct simulation.
An example is the $\beta$-hairpin folding equilibrium in ref.~\onlinecite{Donati2018}.
Path reweighting is closely related to path sampling techniques, in particular path sampling techniques that aim at optimizing the path action \cite{Chong:2017, Grazioli:2018, Dixit:2018, Peter:2020}.
The combination of enhanced sampling, path sampling, and path reweighting might change the way we explore the molecular state space and investigate rare events.


\section{Supplementary Material}
See supplementary material for an example OpenMM script and the corresponding Python3 scripts to construct a reweighted MSM.

\section{Dedication}
This paper is dedicated to Dr.~Irina V.~Gopich, a master of stochastic processes. Her work has influenced the way scientists in the field think about the dynamics of molecules - in simulation and in experiment.

\section{Acknowledgments}
The authors would like to thank Luca Donati and Marcus Weber for helpful comments on the manuscript.
This research has been funded by Deutsche Forschungsgemeinschaft (DFG, German Research Foundation) 
under Germany´s Excellence Strategy – EXC 2008 – 390540038 – UniSysCat, and
through grant CRC 1114 "Scaling Cascades in Complex Systems", Project Number 235221301, Project B05 "Origin of scaling cascades in protein dynamics".

\section{Data availability}
The data that support the findings of this study are available from the corresponding author upon reasonable request.

\appendix


\section{Langevin Leapfrog and the ISP scheme}
\label{appendix:derivation_ISP_scheme}
J. A.~Izaguirre, C. R.~Sweet, and V.S.~Pande developed the following Langevin Leapfrog algorithm
\begin{eqnarray}
    v_{k+\frac{1}{2}} 
    &=&     \exp\left(-\xi \frac{\Delta t}{2}\right) v_k 
        -   \left[1 -  \exp\left(-\xi \frac{\Delta t}{2}\right)\right]   \frac{\nabla V(x_k)}{\xi m}  \nonumber \\
    &&  +   \sqrt{\frac{k_B T}{m} \bigg[ 1-\exp(-\xi \Delta t)  \bigg]} \, \eta_k \label{equ:Langevin_Leapfrog_1}\\
    x_{k+1} &=& x_k + v_{k+\frac{1}{2}} \Delta t \label{equ:Langevin_Leapfrog_2}\\
    v_{k+1} 
    &=& \exp\left(-\xi \frac{\Delta t}{2}\right) v_{k+\frac{1}{2}} 
        -   \left[1 -  \exp\left(-\xi \frac{\Delta t}{2}\right)\right]   \frac{\nabla V(x_{k+1})}{\xi m}  \nonumber \\
    &&  +   \sqrt{\frac{k_B T}{m} \bigg[ 1-\exp(-\xi \Delta t)  \bigg]} \, \eta_{k+1} \label{equ:Langevin_Leapfrog_3}
\end{eqnarray}
(eqs.~14-16 in ref.~\onlinecite{Izaguirre2010}).
First, the velocity $v_{k + \frac{1}{2}}$ is updated by a half step using $v_k$, $x_k$ and a random number $\eta_k$ 
(eq.~\ref{equ:Langevin_Leapfrog_1}). 
Then, the position update to $x_{k+1}$ is computed from $x_k$ assuming constant velocity $v_{k+\frac{1}{2}}$ in the interval $[k, k+1]$ (eq.~\ref{equ:Langevin_Leapfrog_2}). 
Finally, the remaining half step of the velocities to $v_{k+1}$ is computed using $x_{k+1}$, $v_{k+\frac{1}{2}}$ and a new random number $\eta_{k+1}$  (eq.~\ref{equ:Langevin_Leapfrog_3}). 
This Langevin Leapfrog algorithm has been converted to the following full-step scheme in the C$^{++}$ CpuLangevinDynamics class of OpenMM \cite{OpenMM_CpuLangevinDynamics}
\begin{eqnarray}
    v_{k+1} &=& \exp(-\xi \Delta t) v_k - \bigg[ 1-\exp(-\xi \Delta t) \bigg] \frac{\nabla V(x_k)}{\xi m} + \sqrt{\frac{k_B T}{m} \bigg[ 1-\exp(-2\xi \Delta t)  \bigg]} \, \eta_k  \label{equ:Appendix_Leapfrog_step1}\\
    x_{k+1} &=& x_k + v_{k+1} \Delta t \, , \label{equ:Appendix_Leapfrog_step2} 
\end{eqnarray}
where the velocities are propagated by a full step (i.e.~$\Delta t / 2$ in eq.~\ref{equ:Langevin_Leapfrog_1} is replaced by $\Delta t$ and $\Delta t$ in eq.~\ref{equ:Langevin_Leapfrog_1} is replaced by $2 \Delta t$), and the position update is based on $v_k$ rather than on $v_{k+\frac{1}{2}}$. 
The second half-step for the velocities (eq.~\ref{equ:Langevin_Leapfrog_3}) is omitted.
This integration scheme only uses a single random number per iteration.
Eqs.~\ref{equ:Appendix_Leapfrog_step1} and \ref{equ:Appendix_Leapfrog_step2} is the integration scheme we used in refs.~\onlinecite{Donati2017} and \onlinecite{Donati2018}.
To distinguish it from the original Langevin Leapfrog scheme (eqs.~\ref{equ:Langevin_Leapfrog_1}-\ref{equ:Langevin_Leapfrog_3})
we will refer to eqs.~\ref{equ:Appendix_Leapfrog_step1} and \ref{equ:Appendix_Leapfrog_step2} as the ``ISP scheme''.

To be able to analyze the path probability as a function of the positions, we rearrange eqs.~\ref{equ:Appendix_Leapfrog_step1} and \ref{equ:Appendix_Leapfrog_step2} such that we first update the positions using a stochastic step (replace $v_{k+1}$ in eq.~\ref{equ:Appendix_Leapfrog_step2} by eq.~\ref{equ:Appendix_Leapfrog_step1}) and then update the velocity as finite difference (rearrange eq.~\ref{equ:Appendix_Leapfrog_step2} with respect to $v_{k+1}$).
This yields eqs.~\ref{equ:iteration_scheme_1D_1} and \ref{equ:iteration_scheme_1D_2}.


\section{Path probability for Langevin dynamics}
\label{appendix:path_probability}
We derive the closed-form expression for $P_L(\omega_L;\Delta t | (x_0, v_0))$  in eq.~\ref{equ:multiple_step_path_prob} from the integration scheme (eqs.~\ref{equ:iteration_scheme_1D_1} and ~\ref{equ:iteration_scheme_1D_2}) by following the approach in ref.~\onlinecite{Bressloff2014}.
As a first step,  
we derive a closed-form expression for the one-step probability
$P_L(x_{k+1}, v_{k+1};\Delta t | (x_k,v_k))$ of observing
a step $(x_k,v_k) \to (x_{k+1},v_{k+1})$.
According to eqs.~\ref{equ:iteration_scheme_1D_1} and \ref{equ:iteration_scheme_1D_2}, the tuple
$(x_{k+1},v_{k+1})$ at iteration step $k+1$ is entirely determined by the tuple $(x_k,v_k)$ at iteration step $k$ if additionally the random number $\eta_k$ is known.
Thus, $P_L(x_{k+1}, v_{k+1};\Delta t | (x_k,v_k, \eta_k))$, 
i.e.~the one-step probability with fixed random number $\eta_k$,
is a Dirac delta function centered at $(x_{k+1},v_{k+1})$.
Our strategy is to derive a closed-form expression for 
this Dirac delta function
using eqs.~\ref{equ:iteration_scheme_1D_1} and \ref{equ:iteration_scheme_1D_2}, 
and to integrate out the dependency on $\eta_k$.
In this appendix we omit the index $L$ in $\eta_{L,k}$ to simplify the notation.
We reformulate the two-dimensional probability $P_L(x_{k+1}, v_{k+1};\Delta t | (x_k,v_k, \eta_k))$ as a product of two one-dimensional probabilities 
\begin{eqnarray}
    P_L(x_{k+1}, v_{k+1};\Delta t | (x_k,v_k, \eta_k)) 
    &=&P_L(v_{k+1} ;\Delta t| (x_{k+1}, x_k,v_k, \eta_k)) \cdot P_L(x_{k+1} ;\Delta t| (x_k,v_k, \eta_k)) 
\label{equ:conditional_probability}      
\end{eqnarray}
using the rule $P(A, B | C) = P(A | B, C) \cdot P(B | C)$
with $A = v_{k+1}$, $B = x_{k+1}$, and $C = (x_k, v_k, \eta_k)$. 
This rule is the extension of the conditional probability
${P}(A , B) = P(A | B) \cdot P(B)$
to an additional condition $C$.
The first factor is a Dirac delta function constrained to  eq.~\ref{equ:iteration_scheme_1D_2}
\begin{eqnarray}
    P_L(v_{k+1} ;\Delta t| (x_{k+1}, x_k,v_k, \eta_k)) 
    &=&   P_L(v_{k+1} ;\Delta t| (x_{k+1}, x_k)) 
    = \delta\left(v_{k+1} - \frac{x_{k+1}-x_k}{\Delta t}\right) \, , 
\label{equ:conditional_probability_v}
\end{eqnarray}
where the first equality emphasizes that $v_{k+1}$ does not depend on $\eta_k$ or $v_k$ in eq.~\ref{equ:iteration_scheme_1D_2}.
Note, that the probability of the velocity $v_{k+1}$ (eq.~\ref{equ:conditional_probability_v}) does not depend on a random number,  which mirrors our previous observation that $v_{k+1}$ is not treated as a random variable in eq.~\ref{equ:iteration_scheme_1D_2}.
The second factor in eq.~\ref{equ:conditional_probability} is a Dirac delta function constrained to  eq.~\ref{equ:iteration_scheme_1D_1}
\begin{eqnarray}
P_L(x_{k+1}; \Delta t | (x_k, v_k, \eta_k)) 
&=&\delta \bigg( x_{k+1} -  x_k - \exp \left( - \xi \, \Delta t \right) \, v_k \Delta t + 
\bigg[ 1 - \exp \left( - \xi \, \Delta t\right) \bigg] \, \frac{\nabla V(x_k)}{\xi m} \Delta t \,  \cr
&& \quad - \sqrt{\frac{k_B T}{m} \, \bigg[ 1 - \exp \left( - 2 \xi \, \Delta t\right) \bigg]} \, \eta_k \Delta t \bigg) \, . 
\label{equ:conditional_probability_rand_numb_dependency}
\end{eqnarray}
Reinserting the two factors into eq.~\ref{equ:conditional_probability} yields the desired closed-form expression for $P_L(x_{k+1}, v_{k+1};\Delta t | (x_k,v_k; \eta_k))$.
Since we know that the random numbers $\eta_k$ are drawn from a Gaussian distribution $P(\eta_k)$ with zero mean and unit variance
\begin{equation}
P(\eta_k) = N^{-1} \exp \left( - \frac{\eta_k^2}{2} \right) \, , \quad\quad N= \sqrt{2 \pi} \, , \label{equ:Gaussian_single}
\end{equation}
we can average out the random number dependency in eq.~\ref{equ:conditional_probability} 
to obtain the one-step probability
\begin{eqnarray}
    &&P_L(x_{k+1}, v_{k+1};\Delta t | (x_k,v_k)) \cr
    &=& \int_{-\infty}^{\infty} \mathrm{d}\eta_k P(\eta_k) \, P_L(x_{k+1}, v_{k+1};\Delta t | (x_k,v_k, \eta_k)) \nonumber \\
    &=& \delta\left(v_{k+1} - \frac{x_{k+1}-x_k}{\Delta t}\right)  
    \cdot \int_{-\infty}^{\infty} \mathrm{d}\eta_k P_{\eta}(\eta_k) \,P_L(x_{k+1}; \Delta t | (x_k, v_k, \eta_k)) \, .
\label{equ:one_step_probability_01}        
\end{eqnarray}
The challenge lies in solving the integral in this equation. 
The solution, which is detailed in appendix \ref{appendix:double_integral}, yields the closed-form expression for the one-step probability
\begin{eqnarray}
&&P_L(x_{k+1}, v_{k+1}; \Delta t | (x_k, v_k)) \nonumber \\[2pt]
&=& \delta\left(v_{k+1} - \frac{x_{k+1}-x_k}{\Delta t}\right) \cdot \sqrt{\frac{m}{2 \pi k_B T \Delta t^2 (1 - \exp ( -2 \xi \Delta t ))}}  \nonumber\\
&& \times \exp \left( - \frac{m \left( x_{k+1} - x_k - \exp (-\xi \Delta t) v_k \Delta t + (1 - \exp(-\xi \Delta t )) \frac{\nabla V(x_k)}{\xi m} \Delta t \right)^2}{2 k_B T (1 - \exp(-2\xi \Delta t)) \Delta t^2} \right) \, . \label{equ:one_step_probability_final}
\end{eqnarray}
Applying the Chapman-Kolmogorov equation \cite{Gardiner1983} recursively to the one-step probability yields the closed-form expression for the path probability $P_L(\omega_L;\Delta t | (x_0, v_0))$, shown in eq.~\ref{equ:multiple_step_path_prob}.


\section{Solving the double integral}
\label{appendix:double_integral}
We compute the integral 
\begin{align}
    P_L(x_{k+1}; \Delta t | (x_k, v_k)) 
    = \int\limits_{-\infty}^{\infty} \mathrm{d} \eta_k \, P(\eta_k) \, P_L(x_{k+1}; \Delta t | (x_k, v_k, \eta_k)) \label{equ:4}
\end{align}
from eq.~\ref{equ:one_step_probability_01}.
First, we replace $P(\eta_k)$ according to eq.~\ref{equ:Gaussian_single}.
Second, we substitute $P_L(x_{k+1}; \Delta t | (x_k, v_k, \eta_k))$, which is a $\delta$-function (eq.~\ref{equ:conditional_probability_rand_numb_dependency}), with its Fourier transform
\begin{equation}
\delta (z-z') =  \int\limits_{-\infty}^{+\infty} \frac{\mathrm{d} w}{2 \pi} \, \exp \left( iw(z-z') \right) \, , \label{equ:def_delta}
\end{equation}
where $z=x_{k+1}$ and $z'$ is equal to the right-hand side of eq.~\ref{equ:iteration_scheme_1D_1}.
This yields a double integral, whose outer integral is with respect to $w$, while the inner integral is with respect to $\eta_k$
\begin{eqnarray}
    &&P_L(x_{k+1}; \Delta t | (x_k, v_k)) \nonumber \\ 
    &=& \int\limits_{-\infty}^{+\infty} \frac{\mathrm{d} w}{2 \pi}  
        \int\limits_{-\infty}^{+\infty} \frac{\mathrm{d} \eta_k}{N} \,  
        \exp \left( - \frac{\eta_k^2}{2} \right) \nonumber \\
    && \times \exp \bigg( iw \bigg[  x_{k+1} -  x_k - \exp \left( - \xi \Delta t \right) \, v_k \Delta t + 
    \bigg[ 1 - \exp \left( - \xi \Delta t \right) \bigg] \, \frac{\nabla V(x_k)}{\xi m} \Delta t \, \cr
    && \qquad\qquad -\sqrt{\frac{k_B T}{m} \, \bigg[ 1 - \exp \left( - 2 \xi \Delta t \right) \bigg]} \, \eta_k \Delta t \bigg] \bigg) \cr
&&= \int\limits_{-\infty}^{+\infty} \frac{\mathrm{d} w}{2 \pi} \exp \left( iw B \right) \int\limits_{-\infty}^{+\infty} \frac{\mathrm{d} \eta_k}{N} \exp \left( - \frac{\eta_k^2}{2} - iw R \, \eta_k \right)\, ,
\label{equ:integrals}
\end{eqnarray}
where we moved all terms that do not depend on $\eta_k$ out of the inner integral and defined 
the abbreviations
\begin{eqnarray}
    B &=& \bigg[  x_{k+1} -  x_k - \exp \left( - \xi \Delta t \right) \, v_k \Delta t
        + \left[ 1 - \exp \left( - \xi \Delta t \right) \right] \, \frac{\nabla V(x_k)}{\xi m} \Delta t \bigg] \cr
    R &=& \Delta t \sqrt{\frac{k_B T}{m} \, \bigg[ 1 - \exp \left( - 2 \xi \Delta t \right) \bigg]}\, .
\label{equ:A_def_const_R_B}    
\end{eqnarray}
Both integrals in eq.~\ref{equ:integrals} can be solved with the completing-the-square technique for Gaussian integrals. The goal of this technique is, to expand and rearrange the inner integral such that we can use the analytic solution 
\begin{equation}
\int\limits_{-\infty}^\infty \mathrm{d} x \, \exp \left( -a (x \pm b)^2 \right) = \sqrt{\frac{\pi}{a}} \quad\quad \text{ for  } a,b \in \mathbb{R}
\label{equ:A_gaussian_solution} \, .
\end{equation}
This can be achieved by a systematic step-to-step procedure, that can be applied to all Gaussian integrals of this type:
\begin{eqnarray}
\int\limits_{-\infty}^{+\infty} \frac{\mathrm{d} \eta_k}{N} \exp \left( - \frac{\eta_k^2}{2} - iw R \, \eta_k \right) 
&=& \int\limits_{-\infty}^{+\infty} \frac{\mathrm{d} \eta_k}{N} \exp \left( - \frac{1}{2} \bigg[ \eta_k^2 + 2iw R \, \eta_k \overbrace{+ i^2w^2R^2 - i^2w^2R^2}^{=0} \bigg] \right) \nonumber \\
&=& \exp \left( - \frac{w^2R^2}{2} \right) \,  \int\limits_{-\infty}^{+\infty} \frac{\mathrm{d} \eta_k}{N} \exp \left( - \frac{1}{2} \bigg( \eta_k + iwR \bigg)^2 \right) \nonumber \\
&=& \exp \left( - \frac{w^2R^2}{2} \right) \, \frac{1}{N} \, \sqrt{2 \pi} \nonumber \\
&=&  \exp \left( - \frac{w^2R^2}{2} \right)\, .
\label{equ:inner_integral}
\end{eqnarray}
In the first line, we isolate $\eta_k^2$ by factoring out $- \frac{1}{2}$, and complete the first binomial formula
by adding a zero.
Then we separate the exponent into the binomial formula and the term $\exp \left( - \frac{w^2R^2}{2} \right)$, 
which can be moved in front of the integral because it does not depend on $\eta_k$.
In the third line, we solve the remaining integral using eq.~\ref{equ:A_gaussian_solution}, 
which can be further simplified by inserting the normalization constant of the Gaussian distribution: $N= \sqrt{2 \pi}$. 
Inserting eq.~\ref{equ:inner_integral} into eq.~\ref{equ:integrals} yields the outer integral
\begin{eqnarray}
\int\limits_{-\infty}^{+\infty} \frac{\mathrm{d} w}{2 \pi} \exp \left( iw B \right) \, \exp \left( - \frac{w^2R^2}{2} \right) \nonumber =  \int\limits_{-\infty}^{+\infty} \frac{\mathrm{d} w}{2 \pi} \exp \left(  - \frac{w^2R^2}{2} + iwB \right) 
\end{eqnarray}
which is solved using the same procedure:
\begin{eqnarray}
\int\limits_{-\infty}^{+\infty} \frac{\mathrm{d} w}{2 \pi} \exp \left(  - \frac{w^2R^2}{2} + iwB \right)
&=&  \int\limits_{-\infty}^{+\infty} \frac{\mathrm{d} w}{2 \pi} \exp \left(  -\frac{R^2}{2} \bigg[ w^2 + \frac{2iwB}{R^2} \overbrace{+ \frac{i^2B^2}{R^4} - \frac{i^2B^2}{R^4}}^{=0} \bigg] \right) \nonumber \\
&=&  \exp \left(- \frac{B^2}{2 R^2} \right)  \, \int\limits_{-\infty}^{\infty} \frac{\mathrm{d} w}{2 \pi} \exp \left(  -\frac{R^2}{2} \bigg( w + \frac{iB}{R^2} \bigg)^2 \right) \nonumber \\
&=&  \exp \left(- \frac{B^2}{2 R^2} \right)  \, \frac{1}{2\pi} \, \sqrt{\frac{2 \pi}{R^2}} \nonumber \\
 &=&   \sqrt{\frac{1}{2 \pi R^2}}  \, \exp \left(- \frac{B^2}{2 R^2} \right) \, .
\label{equ:A_outer_integral}
\end{eqnarray}
Inserting the expressions for the constants $R$ and $B$ (eq.~\ref{equ:A_def_const_R_B}) yields
\begin{eqnarray}
&&P_L(x_{k+1}; \Delta t | (x_k, v_k)) \nonumber \\[2pt]
&=& \sqrt{\frac{m}{2 \pi k_B T \Delta t^2 (1 - \exp ( -2 \xi \Delta t ))}} \nonumber\\
&& \times \exp \left( - \frac{m \left( x_{k+1} - x_k - \exp (-\xi \Delta t) v_k \Delta t + (1 - \exp(-\xi \Delta t )) \frac{\nabla V(x_k)}{\xi m} \Delta t \right)^2}{2 k_B T (1 - \exp(-2\xi \Delta t)) \Delta t^2} \right) \, .
\label{equ:positional_one_step_probability}
\end{eqnarray}
This is inserted into eq.~\ref{equ:one_step_probability_01} to yield eq.~\ref{equ:one_step_probability_final}.

\section{Proof of eq.~\ref{eq:Delta_eta_approx}}
\label{appendix:Delta_eta}

\begin{eqnarray}
    \frac{\left(1-e^{-x}\right)^2}{x\cdot \left(1-e^{-2 x}\right)}
    &=& \frac{1}{2} - \frac{x^2}{24} + \frac{x^4}{240} \pm \mathcal{O}(x^5) \cr
    \left(1-e^{-x}\right)^2 
    &=& \frac{x}{2}\cdot \left(1-e^{-2 x}\right) - \frac{x^2}{24} \cdot x\cdot \left(1-e^{-2 x}\right)  \pm \mathcal{O}(x^4) \cdot x\cdot \left(1-e^{-2 x}\right) \nonumber\\[4pt]
    \left(1-e^{-x}\right)^2 
    &=& \frac{x}{2}\cdot \left(1-e^{-2 x}\right) - \mathcal{O}(x^4) \, .
\end{eqnarray}
The first line shows the Taylor expansion of the expression on the right-hand side. 
To obtain the second line, we multiplied by $x\cdot \left(1-e^{-2 x}\right)$.
In the third line we used the fact that the leading term of the Taylor expansion of $x\cdot \left(1-e^{-2 x}\right)$ is $2x^2$, thus yielding an error of $\mathcal{O}(x^4)$.
Substituting $x = \xi \Delta t$ yields
\begin{eqnarray}
    \left(1-e^{-\xi \Delta t}\right)^2 
    &=& \frac{\xi \Delta t}{2}\cdot \left(1-e^{-2 \xi \Delta t}\right) - \mathcal{O}(\xi^4 \Delta t^4) \cr
    \left(1-e^{-\xi \Delta t}\right)^2 
    &\approx& \frac{\xi \Delta t}{2}\cdot \left(1-e^{-2 \xi \Delta t}\right) \, ,
\end{eqnarray}
and multiplying by $\frac{1}{k_BT\,\xi^2\, m \, \left(1-e^{-2\xi\Delta t} \right)} \left(\nabla U(x_k) \right)^2$
yields
\begin{eqnarray}
    \frac{1}{k_B T \xi^2 m} \frac{\left(1-e^{-\xi \Delta t}\right)^2}{1-e^{-2 \xi \Delta t}} \left( \nabla U(x_k) \right)^2 &\approx& \frac{\Delta t}{2k_B T \xi m} \left( \nabla U(x_k) \right)^2  \nonumber\\[8pt]
    \Delta \eta_{L,k}^2 &\approx& \Delta \eta_{o,k}^2 \, .
\label{equ:A_delta_eta_squared}    
\end{eqnarray}
Thus, the difference between  $\Delta \eta^2_{L,k}$ (eq.~\ref{equ:delta_eta_Langevin}) and $\Delta \eta^2_{o, k}$ (eq.~\ref{equ:delta_eta_overdamped_Langevin}) is of order $\mathcal{O}(\xi^4 \Delta t^4)$. 
Eq.~\ref{equ:A_delta_eta_squared} is eq.~\ref{eq:Delta_eta_approx} squared.
$\square$

\nocite{*}
\bibliography{literature.bib}

\begin{thebibliography}{72}%
\makeatletter
\providecommand \@ifxundefined [1]{%
 \@ifx{#1\undefined}
}%
\providecommand \@ifnum [1]{%
 \ifnum #1\expandafter \@firstoftwo
 \else \expandafter \@secondoftwo
 \fi
}%
\providecommand \@ifx [1]{%
 \ifx #1\expandafter \@firstoftwo
 \else \expandafter \@secondoftwo
 \fi
}%
\providecommand \natexlab [1]{#1}%
\providecommand \enquote  [1]{``#1''}%
\providecommand \bibnamefont  [1]{#1}%
\providecommand \bibfnamefont [1]{#1}%
\providecommand \citenamefont [1]{#1}%
\providecommand \href@noop [0]{\@secondoftwo}%
\providecommand \href [0]{\begingroup \@sanitize@url \@href}%
\providecommand \@href[1]{\@@startlink{#1}\@@href}%
\providecommand \@@href[1]{\endgroup#1\@@endlink}%
\providecommand \@sanitize@url [0]{\catcode `\\12\catcode `\$12\catcode
  `\&12\catcode `\#12\catcode `\^12\catcode `\_12\catcode `\%12\relax}%
\providecommand \@@startlink[1]{}%
\providecommand \@@endlink[0]{}%
\providecommand \url  [0]{\begingroup\@sanitize@url \@url }%
\providecommand \@url [1]{\endgroup\@href {#1}{\urlprefix }}%
\providecommand \urlprefix  [0]{URL }%
\providecommand \Eprint [0]{\href }%
\providecommand \doibase [0]{http://dx.doi.org/}%
\providecommand \selectlanguage [0]{\@gobble}%
\providecommand \bibinfo  [0]{\@secondoftwo}%
\providecommand \bibfield  [0]{\@secondoftwo}%
\providecommand \translation [1]{[#1]}%
\providecommand \BibitemOpen [0]{}%
\providecommand \bibitemStop [0]{}%
\providecommand \bibitemNoStop [0]{.\EOS\space}%
\providecommand \EOS [0]{\spacefactor3000\relax}%
\providecommand \BibitemShut  [1]{\csname bibitem#1\endcsname}%
\let\auto@bib@innerbib\@empty
\bibitem [{\citenamefont {Gopich}(2020)}]{Gopich:2020}%
  \BibitemOpen
  \bibfield  {author} {\bibinfo {author} {\bibfnamefont {I.~V.}\ \bibnamefont
  {Gopich}},\ }\bibfield  {title} {\enquote {\bibinfo {title} {{Multisite
  reversible association in membranes and solutions: From non-Markovian to
  Markovian kinetics.}}}\ }\href@noop {} {\bibfield  {journal} {\bibinfo
  {journal} {J. Chem. Phys.}\ }\textbf {\bibinfo {volume} {152}},\ \bibinfo
  {pages} {104101} (\bibinfo {year} {2020})}\BibitemShut {NoStop}%
\bibitem [{\citenamefont {Lane}\ \emph {et~al.}(2013)\citenamefont {Lane},
  \citenamefont {Shukla}, \citenamefont {Beauchamp},\ and\ \citenamefont
  {Pande}}]{Lane:2013}%
  \BibitemOpen
  \bibfield  {author} {\bibinfo {author} {\bibfnamefont {T.~J.}\ \bibnamefont
  {Lane}}, \bibinfo {author} {\bibfnamefont {D.}~\bibnamefont {Shukla}},
  \bibinfo {author} {\bibfnamefont {K.~A.}\ \bibnamefont {Beauchamp}}, \ and\
  \bibinfo {author} {\bibfnamefont {V.~S.}\ \bibnamefont {Pande}},\ }\bibfield
  {title} {\enquote {\bibinfo {title} {{To milliseconds and beyond: challenges
  in the simulation of protein folding}},}\ }\href@noop {} {\bibfield
  {journal} {\bibinfo  {journal} {Curr. Opin. Struct. Biol.}\ }\textbf
  {\bibinfo {volume} {23}},\ \bibinfo {pages} {58} (\bibinfo {year}
  {2013})}\BibitemShut {NoStop}%
\bibitem [{\citenamefont {Dror}\ \emph {et~al.}(2012)\citenamefont {Dror},
  \citenamefont {Dirks}, \citenamefont {Grossman}, \citenamefont {Xu},\ and\
  \citenamefont {Shaw}}]{Dror:2012}%
  \BibitemOpen
  \bibfield  {author} {\bibinfo {author} {\bibfnamefont {R.~O.}\ \bibnamefont
  {Dror}}, \bibinfo {author} {\bibfnamefont {R.~M.}\ \bibnamefont {Dirks}},
  \bibinfo {author} {\bibfnamefont {J.~P.}\ \bibnamefont {Grossman}}, \bibinfo
  {author} {\bibfnamefont {H.}~\bibnamefont {Xu}}, \ and\ \bibinfo {author}
  {\bibfnamefont {D.~E.}\ \bibnamefont {Shaw}},\ }\bibfield  {title} {\enquote
  {\bibinfo {title} {{Biomolecular Simulation: A Computational Microscope for
  Molecular Biology}},}\ }\href@noop {} {\bibfield  {journal} {\bibinfo
  {journal} {Annu. Rev. Biophys.}\ }\textbf {\bibinfo {volume} {41}},\ \bibinfo
  {pages} {429} (\bibinfo {year} {2012})}\BibitemShut {NoStop}%
\bibitem [{\citenamefont {Barros}\ \emph {et~al.}(2020)\citenamefont {Barros},
  \citenamefont {Casalino}, \citenamefont {Gaieb}, \citenamefont {Dommer},
  \citenamefont {Wang}, \citenamefont {Fallon}, \citenamefont {Raguette},
  \citenamefont {Belfon}, \citenamefont {Simmerling},\ and\ \citenamefont
  {Amaro}}]{Barros2020}%
  \BibitemOpen
  \bibfield  {author} {\bibinfo {author} {\bibfnamefont {E.~P.}\ \bibnamefont
  {Barros}}, \bibinfo {author} {\bibfnamefont {L.}~\bibnamefont {Casalino}},
  \bibinfo {author} {\bibfnamefont {Z.}~\bibnamefont {Gaieb}}, \bibinfo
  {author} {\bibfnamefont {A.~C.}\ \bibnamefont {Dommer}}, \bibinfo {author}
  {\bibfnamefont {Y.}~\bibnamefont {Wang}}, \bibinfo {author} {\bibfnamefont
  {L.}~\bibnamefont {Fallon}}, \bibinfo {author} {\bibfnamefont
  {L.}~\bibnamefont {Raguette}}, \bibinfo {author} {\bibfnamefont
  {K.}~\bibnamefont {Belfon}}, \bibinfo {author} {\bibfnamefont
  {C.}~\bibnamefont {Simmerling}}, \ and\ \bibinfo {author} {\bibfnamefont
  {R.~E.}\ \bibnamefont {Amaro}},\ }\bibfield  {title} {\enquote {\bibinfo
  {title} {The flexibility of ace2 in the context of sars-cov-2 infection},}\
  }\href@noop {} {\bibfield  {journal} {\bibinfo  {journal} {Biophys. J.}\
  }\textbf {\bibinfo {volume} {120}},\ \bibinfo {pages} {1} (\bibinfo {year}
  {2020})}\BibitemShut {NoStop}%
\bibitem [{\citenamefont {Harpole}\ and\ \citenamefont
  {Delemotte}(2018)}]{Harpole2018}%
  \BibitemOpen
  \bibfield  {author} {\bibinfo {author} {\bibfnamefont {T.~J.}\ \bibnamefont
  {Harpole}}\ and\ \bibinfo {author} {\bibfnamefont {L.}~\bibnamefont
  {Delemotte}},\ }\bibfield  {title} {\enquote {\bibinfo {title}
  {Conformational landscapes of membrane proteins delineated by enhanced
  sampling molecular dynamics simulations},}\ }\href@noop {} {\bibfield
  {journal} {\bibinfo  {journal} {Biochim. Biophys. Acta, Biomembr.}\ }\textbf
  {\bibinfo {volume} {1860}},\ \bibinfo {pages} {909} (\bibinfo {year}
  {2018})}\BibitemShut {NoStop}%
\bibitem [{\citenamefont {Cournia}, \citenamefont {Allen},\ and\ \citenamefont
  {Sherman}(2017)}]{Cournia2017}%
  \BibitemOpen
  \bibfield  {author} {\bibinfo {author} {\bibfnamefont {Z.}~\bibnamefont
  {Cournia}}, \bibinfo {author} {\bibfnamefont {B.}~\bibnamefont {Allen}}, \
  and\ \bibinfo {author} {\bibfnamefont {W.}~\bibnamefont {Sherman}},\
  }\bibfield  {title} {\enquote {\bibinfo {title} {Relative binding free energy
  calculations in drug discovery: Recent advances and practical
  considerations},}\ }\href@noop {} {\bibfield  {journal} {\bibinfo  {journal}
  {Journal of Chemical Information and Modeling}\ }\textbf {\bibinfo {volume}
  {57}},\ \bibinfo {pages} {2911} (\bibinfo {year} {2017})}\BibitemShut
  {NoStop}%
\bibitem [{\citenamefont {Badaoui}\ \emph {et~al.}(2018)\citenamefont
  {Badaoui}, \citenamefont {Kells}, \citenamefont {Molteni}, \citenamefont
  {Dickson}, \citenamefont {Hornak},\ and\ \citenamefont
  {Rosta}}]{Badaoui2018}%
  \BibitemOpen
  \bibfield  {author} {\bibinfo {author} {\bibfnamefont {M.}~\bibnamefont
  {Badaoui}}, \bibinfo {author} {\bibfnamefont {A.}~\bibnamefont {Kells}},
  \bibinfo {author} {\bibfnamefont {C.}~\bibnamefont {Molteni}}, \bibinfo
  {author} {\bibfnamefont {C.~J.}\ \bibnamefont {Dickson}}, \bibinfo {author}
  {\bibfnamefont {V.}~\bibnamefont {Hornak}}, \ and\ \bibinfo {author}
  {\bibfnamefont {E.}~\bibnamefont {Rosta}},\ }\bibfield  {title} {\enquote
  {\bibinfo {title} {{Calculating Kinetic Rates and Membrane Permeability from
  Biased Simulations}},}\ }\href@noop {} {\bibfield  {journal} {\bibinfo
  {journal} {J. Phys. Chem. B}\ }\textbf {\bibinfo {volume} {122}},\ \bibinfo
  {pages} {11571} (\bibinfo {year} {2018})}\BibitemShut {NoStop}%
\bibitem [{\citenamefont {Joswig}\ \emph {et~al.}(2020)\citenamefont {Joswig},
  \citenamefont {Anders}, \citenamefont {Zhang}, \citenamefont {Rademacher},\
  and\ \citenamefont {Keller}}]{Joswig2020}%
  \BibitemOpen
  \bibfield  {author} {\bibinfo {author} {\bibfnamefont {J.-O.}\ \bibnamefont
  {Joswig}}, \bibinfo {author} {\bibfnamefont {J.}~\bibnamefont {Anders}},
  \bibinfo {author} {\bibfnamefont {H.}~\bibnamefont {Zhang}}, \bibinfo
  {author} {\bibfnamefont {C.}~\bibnamefont {Rademacher}}, \ and\ \bibinfo
  {author} {\bibfnamefont {B.~G.}\ \bibnamefont {Keller}},\ }\bibfield  {title}
  {\enquote {\bibinfo {title} {Molecular mechanism of the ph-dependent calcium
  affinity in langerin},}\ }\href@noop {} {\bibfield  {journal} {\bibinfo
  {journal} {bioRxiv}\ ,\ \bibinfo {pages} {986851}} (\bibinfo {year}
  {2020})}\BibitemShut {NoStop}%
\bibitem [{\citenamefont {Mey}\ \emph {et~al.}(2020)\citenamefont {Mey},
  \citenamefont {Allen}, \citenamefont {Macdonald}, \citenamefont {Chodera},
  \citenamefont {Kuhn}, \citenamefont {Michel}, \citenamefont {Mobley},
  \citenamefont {Naden}, \citenamefont {Prasad}, \citenamefont {Rizzi},
  \citenamefont {Scheen}, \citenamefont {Shirts}, \citenamefont {Tresadern},\
  and\ \citenamefont {Xu}}]{Mey2020}%
  \BibitemOpen
  \bibfield  {author} {\bibinfo {author} {\bibfnamefont {A.~S. J.~S.}\
  \bibnamefont {Mey}}, \bibinfo {author} {\bibfnamefont {B.}~\bibnamefont
  {Allen}}, \bibinfo {author} {\bibfnamefont {H.~E.~B.}\ \bibnamefont
  {Macdonald}}, \bibinfo {author} {\bibfnamefont {J.~D.}\ \bibnamefont
  {Chodera}}, \bibinfo {author} {\bibfnamefont {M.}~\bibnamefont {Kuhn}},
  \bibinfo {author} {\bibfnamefont {J.}~\bibnamefont {Michel}}, \bibinfo
  {author} {\bibfnamefont {D.~L.}\ \bibnamefont {Mobley}}, \bibinfo {author}
  {\bibfnamefont {L.~N.}\ \bibnamefont {Naden}}, \bibinfo {author}
  {\bibfnamefont {S.}~\bibnamefont {Prasad}}, \bibinfo {author} {\bibfnamefont
  {A.}~\bibnamefont {Rizzi}}, \bibinfo {author} {\bibfnamefont
  {J.}~\bibnamefont {Scheen}}, \bibinfo {author} {\bibfnamefont {M.~R.}\
  \bibnamefont {Shirts}}, \bibinfo {author} {\bibfnamefont {G.}~\bibnamefont
  {Tresadern}}, \ and\ \bibinfo {author} {\bibfnamefont {H.}~\bibnamefont
  {Xu}},\ }\href@noop {} {\enquote {\bibinfo {title} {Best practices for
  alchemical free energy calculations},}\ } (\bibinfo {year} {2020}),\ \Eprint
  {http://arxiv.org/abs/2008.03067} {arXiv:2008.03067 [q-bio.BM]} \BibitemShut
  {NoStop}%
\bibitem [{\citenamefont {Tuckerman}(2010)}]{Tuckerman2010}%
  \BibitemOpen
  \bibfield  {author} {\bibinfo {author} {\bibfnamefont {M.}~\bibnamefont
  {Tuckerman}},\ }\href@noop {} {\emph {\bibinfo {title} {Monte Carlo
  Statistical mechanics: Theory and molecular simulation}}}\ (\bibinfo
  {publisher} {Oxford University Press Inc.: New York},\ \bibinfo {year}
  {2010})\ pp.\ \bibinfo {pages} {300--304}\BibitemShut {NoStop}%
\bibitem [{\citenamefont {Frenkel}\ and\ \citenamefont
  {Smit}(2002)}]{Frenkel2002}%
  \BibitemOpen
  \bibfield  {author} {\bibinfo {author} {\bibfnamefont {D.}~\bibnamefont
  {Frenkel}}\ and\ \bibinfo {author} {\bibfnamefont {B.}~\bibnamefont {Smit}},\
  }\href@noop {} {\emph {\bibinfo {title} {Understanding Molecular Simulation:
  From Algorithms to Applications}}},\ \bibinfo {edition} {1st}\ ed.\ (\bibinfo
   {publisher} {Academic Press San Diego San Francisco New York Boston London
  Sydney Tokyo},\ \bibinfo {year} {2002})\BibitemShut {NoStop}%
\bibitem [{\citenamefont {de~Oliveira}, \citenamefont {Hamelberg},\ and\
  \citenamefont {McCammon}(2007)}]{deOliveira:2007}%
  \BibitemOpen
  \bibfield  {author} {\bibinfo {author} {\bibfnamefont {C.~A.~F.}\
  \bibnamefont {de~Oliveira}}, \bibinfo {author} {\bibfnamefont
  {D.}~\bibnamefont {Hamelberg}}, \ and\ \bibinfo {author} {\bibfnamefont
  {J.~A.}\ \bibnamefont {McCammon}},\ }\bibfield  {title} {\enquote {\bibinfo
  {title} {{Estimating kinetic rates from accelerated molecular dynamics
  simulations: Alanine dipeptide in explicit solvent as a case study}},}\
  }\href@noop {} {\bibfield  {journal} {\bibinfo  {journal} {J. Chem. Phys.}\
  }\textbf {\bibinfo {volume} {127}},\ \bibinfo {pages} {175105} (\bibinfo
  {year} {2007})}\BibitemShut {NoStop}%
\bibitem [{\citenamefont {Tiwary}\ and\ \citenamefont
  {Parrinello}(2013)}]{Tiwary2013}%
  \BibitemOpen
  \bibfield  {author} {\bibinfo {author} {\bibfnamefont {P.}~\bibnamefont
  {Tiwary}}\ and\ \bibinfo {author} {\bibfnamefont {M.}~\bibnamefont
  {Parrinello}},\ }\bibfield  {title} {\enquote {\bibinfo {title} {{From
  Metadynamics to Dynamics}},}\ }\href@noop {} {\bibfield  {journal} {\bibinfo
  {journal} {Phys. Rev. Lett.}\ }\textbf {\bibinfo {volume} {111}},\ \bibinfo
  {pages} {230602} (\bibinfo {year} {2013})}\BibitemShut {NoStop}%
\bibitem [{\citenamefont {Valsson}, \citenamefont {Tiwary},\ and\ \citenamefont
  {Parrinello}(2016)}]{Valsson2016}%
  \BibitemOpen
  \bibfield  {author} {\bibinfo {author} {\bibfnamefont {O.}~\bibnamefont
  {Valsson}}, \bibinfo {author} {\bibfnamefont {P.}~\bibnamefont {Tiwary}}, \
  and\ \bibinfo {author} {\bibfnamefont {M.}~\bibnamefont {Parrinello}},\
  }\bibfield  {title} {\enquote {\bibinfo {title} {{Enhancing Important
  Fluctuations: Rare Events and Metadynamics from a Conceptual Viewpoint.}}}\
  }\href@noop {} {\bibfield  {journal} {\bibinfo  {journal} {Annu. Rev. Phys.
  Chem.}\ }\textbf {\bibinfo {volume} {67}},\ \bibinfo {pages} {159} (\bibinfo
  {year} {2016})}\BibitemShut {NoStop}%
\bibitem [{\citenamefont {Casasnovas}\ \emph {et~al.}(2017)\citenamefont
  {Casasnovas}, \citenamefont {Limongelli}, \citenamefont {Tiwary},
  \citenamefont {Carloni},\ and\ \citenamefont {Parrinello}}]{Casasnovas:2017}%
  \BibitemOpen
  \bibfield  {author} {\bibinfo {author} {\bibfnamefont {R.}~\bibnamefont
  {Casasnovas}}, \bibinfo {author} {\bibfnamefont {V.}~\bibnamefont
  {Limongelli}}, \bibinfo {author} {\bibfnamefont {P.}~\bibnamefont {Tiwary}},
  \bibinfo {author} {\bibfnamefont {P.}~\bibnamefont {Carloni}}, \ and\
  \bibinfo {author} {\bibfnamefont {M.}~\bibnamefont {Parrinello}},\ }\bibfield
   {title} {\enquote {\bibinfo {title} {{Unbinding Kinetics of a p38 MAP Kinase
  Type II Inhibitor from Metadynamics Simulations}},}\ }\href@noop {}
  {\bibfield  {journal} {\bibinfo  {journal} {J. Am. Chem. Soc.}\ }\textbf
  {\bibinfo {volume} {139}},\ \bibinfo {pages} {4780} (\bibinfo {year}
  {2017})}\BibitemShut {NoStop}%
\bibitem [{\citenamefont {Wu}\ \emph {et~al.}(2014)\citenamefont {Wu},
  \citenamefont {Mey}, \citenamefont {Rosta},\ and\ \citenamefont
  {No{\'e}}}]{Wu2014}%
  \BibitemOpen
  \bibfield  {author} {\bibinfo {author} {\bibfnamefont {H.}~\bibnamefont
  {Wu}}, \bibinfo {author} {\bibfnamefont {A.~S. J.~S.}\ \bibnamefont {Mey}},
  \bibinfo {author} {\bibfnamefont {E.}~\bibnamefont {Rosta}}, \ and\ \bibinfo
  {author} {\bibfnamefont {F.}~\bibnamefont {No{\'e}}},\ }\bibfield  {title}
  {\enquote {\bibinfo {title} {{Statistically optimal analysis of
  state-discretized trajectory data from multiple thermodynamic states}},}\
  }\href@noop {} {\bibfield  {journal} {\bibinfo  {journal} {J. Chem. Phys.}\
  }\textbf {\bibinfo {volume} {141}},\ \bibinfo {pages} {214106} (\bibinfo
  {year} {2014})}\BibitemShut {NoStop}%
\bibitem [{\citenamefont {Mey}, \citenamefont {Wu},\ and\ \citenamefont
  {No{\'e}}(2014)}]{Mey:2014}%
  \BibitemOpen
  \bibfield  {author} {\bibinfo {author} {\bibfnamefont {A.~S. J.~S.}\
  \bibnamefont {Mey}}, \bibinfo {author} {\bibfnamefont {H.}~\bibnamefont
  {Wu}}, \ and\ \bibinfo {author} {\bibfnamefont {F.}~\bibnamefont {No{\'e}}},\
  }\bibfield  {title} {\enquote {\bibinfo {title} {{xTRAM: Estimating
  Equilibrium Expectations from Time-Correlated Simulation Data at Multiple
  Thermodynamic States}},}\ }\href@noop {} {\bibfield  {journal} {\bibinfo
  {journal} {Phys. Rev. X}\ }\textbf {\bibinfo {volume} {4}},\ \bibinfo {pages}
  {041018} (\bibinfo {year} {2014})}\BibitemShut {NoStop}%
\bibitem [{\citenamefont {Wu}\ \emph {et~al.}(2016)\citenamefont {Wu},
  \citenamefont {Paul}, \citenamefont {Wehmeyer},\ and\ \citenamefont
  {No{\'e}}}]{Wu2016}%
  \BibitemOpen
  \bibfield  {author} {\bibinfo {author} {\bibfnamefont {H.}~\bibnamefont
  {Wu}}, \bibinfo {author} {\bibfnamefont {F.}~\bibnamefont {Paul}}, \bibinfo
  {author} {\bibfnamefont {C.}~\bibnamefont {Wehmeyer}}, \ and\ \bibinfo
  {author} {\bibfnamefont {F.}~\bibnamefont {No{\'e}}},\ }\bibfield  {title}
  {\enquote {\bibinfo {title} {{Multiensemble Markov models of molecular
  thermodynamics and kinetics.}}}\ }\href@noop {} {\bibfield  {journal}
  {\bibinfo  {journal} {Proc. Natl. Acad. Sci. U.S.A.}\ }\textbf {\bibinfo
  {volume} {113}},\ \bibinfo {pages} {E3221} (\bibinfo {year}
  {2016})}\BibitemShut {NoStop}%
\bibitem [{\citenamefont {Stelzl}\ \emph {et~al.}(2017)\citenamefont {Stelzl},
  \citenamefont {Kells}, \citenamefont {Rosta},\ and\ \citenamefont
  {Hummer}}]{Stelzl2017}%
  \BibitemOpen
  \bibfield  {author} {\bibinfo {author} {\bibfnamefont {L.~S.}\ \bibnamefont
  {Stelzl}}, \bibinfo {author} {\bibfnamefont {A.}~\bibnamefont {Kells}},
  \bibinfo {author} {\bibfnamefont {E.}~\bibnamefont {Rosta}}, \ and\ \bibinfo
  {author} {\bibfnamefont {G.}~\bibnamefont {Hummer}},\ }\bibfield  {title}
  {\enquote {\bibinfo {title} {{Dynamic Histogram Analysis to determine free
  Energies and rates from biased simulations}},}\ }\href@noop {} {\bibfield
  {journal} {\bibinfo  {journal} {J. Chem. Theory Comput.}\ }\textbf {\bibinfo
  {volume} {13}},\ \bibinfo {pages} {6328} (\bibinfo {year}
  {2017})}\BibitemShut {NoStop}%
\bibitem [{\citenamefont {Bicout}\ and\ \citenamefont
  {Szabo}(1998)}]{Bicout:1998}%
  \BibitemOpen
  \bibfield  {author} {\bibinfo {author} {\bibfnamefont {D.~J.}\ \bibnamefont
  {Bicout}}\ and\ \bibinfo {author} {\bibfnamefont {A.}~\bibnamefont {Szabo}},\
  }\bibfield  {title} {\enquote {\bibinfo {title} {{Electron transfer reaction
  dynamics in non-Debye solvents}},}\ }\href@noop {} {\bibfield  {journal}
  {\bibinfo  {journal} {J. Chem. Phys.}\ }\textbf {\bibinfo {volume} {109}},\
  \bibinfo {pages} {2325} (\bibinfo {year} {1998})}\BibitemShut {NoStop}%
\bibitem [{\citenamefont {Rosta}\ and\ \citenamefont
  {Hummer}(2014)}]{Rosta2014}%
  \BibitemOpen
  \bibfield  {author} {\bibinfo {author} {\bibfnamefont {E.}~\bibnamefont
  {Rosta}}\ and\ \bibinfo {author} {\bibfnamefont {G.}~\bibnamefont {Hummer}},\
  }\bibfield  {title} {\enquote {\bibinfo {title} {{Free Energies from Dynamic
  Weighted Histogram Analysis Using Unbiased Markov State Model}},}\
  }\href@noop {} {\bibfield  {journal} {\bibinfo  {journal} {J. Chem. Theory
  Comput.}\ }\textbf {\bibinfo {volume} {11}},\ \bibinfo {pages} {276}
  (\bibinfo {year} {2014})}\BibitemShut {NoStop}%
\bibitem [{\citenamefont {Donati}\ \emph {et~al.}(2018)\citenamefont {Donati},
  \citenamefont {Heida}, \citenamefont {Keller},\ and\ \citenamefont
  {Weber}}]{Donati:2018b}%
  \BibitemOpen
  \bibfield  {author} {\bibinfo {author} {\bibfnamefont {L.}~\bibnamefont
  {Donati}}, \bibinfo {author} {\bibfnamefont {M.}~\bibnamefont {Heida}},
  \bibinfo {author} {\bibfnamefont {B.~G.}\ \bibnamefont {Keller}}, \ and\
  \bibinfo {author} {\bibfnamefont {M.}~\bibnamefont {Weber}},\ }\bibfield
  {title} {\enquote {\bibinfo {title} {{Estimation of the infinitesimal
  generator by square-root approximation}},}\ }\href@noop {} {\bibfield
  {journal} {\bibinfo  {journal} {J. Phys.: Condens. Matter}\ }\textbf
  {\bibinfo {volume} {30}},\ \bibinfo {pages} {425201} (\bibinfo {year}
  {2018})}\BibitemShut {NoStop}%
\bibitem [{\citenamefont {Kieninger}, \citenamefont {Donati},\ and\
  \citenamefont {Keller}(2020)}]{Kieninger2020}%
  \BibitemOpen
  \bibfield  {author} {\bibinfo {author} {\bibfnamefont {S.}~\bibnamefont
  {Kieninger}}, \bibinfo {author} {\bibfnamefont {L.}~\bibnamefont {Donati}}, \
  and\ \bibinfo {author} {\bibfnamefont {B.~G.}\ \bibnamefont {Keller}},\
  }\bibfield  {title} {\enquote {\bibinfo {title} {Dynamical reweighting
  methods for markov models},}\ }\href@noop {} {\bibfield  {journal} {\bibinfo
  {journal} {Curr. Opin. Struct. Biol.}\ }\textbf {\bibinfo {volume} {{61}}},\
  \bibinfo {pages} {124} (\bibinfo {year} {2020})}\BibitemShut {NoStop}%
\bibitem [{\citenamefont {Zuckerman}\ and\ \citenamefont
  {Woolf}(1999)}]{Zuckerman:1999}%
  \BibitemOpen
  \bibfield  {author} {\bibinfo {author} {\bibfnamefont {D.~M.}\ \bibnamefont
  {Zuckerman}}\ and\ \bibinfo {author} {\bibfnamefont {T.~B.}\ \bibnamefont
  {Woolf}},\ }\bibfield  {title} {\enquote {\bibinfo {title} {{Dynamic reaction
  paths and rates through importance-sampled stochastic dynamics}},}\
  }\href@noop {} {\bibfield  {journal} {\bibinfo  {journal} {J. Chem. Phys.}\
  }\textbf {\bibinfo {volume} {111}},\ \bibinfo {pages} {9475} (\bibinfo {year}
  {1999})}\BibitemShut {NoStop}%
\bibitem [{\citenamefont {Woolf}(1998)}]{Woolf:1998}%
  \BibitemOpen
  \bibfield  {author} {\bibinfo {author} {\bibfnamefont {T.~B.}\ \bibnamefont
  {Woolf}},\ }\bibfield  {title} {\enquote {\bibinfo {title} {{Path corrected
  functionals of stochastic trajectories: towards relative free energy and
  reaction coordinate calculations}},}\ }\href@noop {} {\bibfield  {journal}
  {\bibinfo  {journal} {Chemical Physics Letters}\ }\textbf {\bibinfo {volume}
  {289}},\ \bibinfo {pages} {433} (\bibinfo {year} {1998})}\BibitemShut
  {NoStop}%
\bibitem [{\citenamefont {Zuckerman}\ and\ \citenamefont
  {Woolf}(2000)}]{Zuckerman:2000}%
  \BibitemOpen
  \bibfield  {author} {\bibinfo {author} {\bibfnamefont {D.~M.}\ \bibnamefont
  {Zuckerman}}\ and\ \bibinfo {author} {\bibfnamefont {T.~B.}\ \bibnamefont
  {Woolf}},\ }\bibfield  {title} {\enquote {\bibinfo {title} {{Efficient
  dynamic importance sampling of rare events in one dimension}},}\ }\href@noop
  {} {\bibfield  {journal} {\bibinfo  {journal} {Phys. Rev. E}\ }\textbf
  {\bibinfo {volume} {63}},\ \bibinfo {pages} {016702} (\bibinfo {year}
  {2000})}\BibitemShut {NoStop}%
\bibitem [{\citenamefont {Xing}\ and\ \citenamefont
  {Andricioaei}(2006)}]{Xing:2006}%
  \BibitemOpen
  \bibfield  {author} {\bibinfo {author} {\bibfnamefont {C.}~\bibnamefont
  {Xing}}\ and\ \bibinfo {author} {\bibfnamefont {I.}~\bibnamefont
  {Andricioaei}},\ }\bibfield  {title} {\enquote {\bibinfo {title} {{On the
  calculation of time correlation functions by potential scaling}},}\
  }\href@noop {} {\bibfield  {journal} {\bibinfo  {journal} {J. Chem. Phys.}\
  }\textbf {\bibinfo {volume} {124}},\ \bibinfo {pages} {034110} (\bibinfo
  {year} {2006})}\BibitemShut {NoStop}%
\bibitem [{\citenamefont {Adib}(2008)}]{Adib2008}%
  \BibitemOpen
  \bibfield  {author} {\bibinfo {author} {\bibfnamefont {A.~B.}\ \bibnamefont
  {Adib}},\ }\bibfield  {title} {\enquote {\bibinfo {title} {Stochastic actions
  for diffusive dynamics: Reweighting, sampling, and minimization},}\
  }\href@noop {} {\bibfield  {journal} {\bibinfo  {journal} {J. Phys. Chem. B}\
  }\textbf {\bibinfo {volume} {{112}}},\ \bibinfo {pages} {5910} (\bibinfo
  {year} {2008})}\BibitemShut {NoStop}%
\bibitem [{\citenamefont {Girsanov}(1960)}]{Girsanov1960}%
  \BibitemOpen
  \bibfield  {author} {\bibinfo {author} {\bibfnamefont {I.~V.}\ \bibnamefont
  {Girsanov}},\ }\bibfield  {title} {\enquote {\bibinfo {title} {On
  transforming a certain class of stochastic processes by absolutely continuous
  substitution of measures},}\ }\href@noop {} {\bibfield  {journal} {\bibinfo
  {journal} {Theory Probab. Appl.}\ }\textbf {\bibinfo {volume} {5}},\ \bibinfo
  {pages} {285} (\bibinfo {year} {1960})}\BibitemShut {NoStop}%
\bibitem [{\citenamefont {{\O}ksendal}(2003)}]{Oeksendal2003}%
  \BibitemOpen
  \bibfield  {author} {\bibinfo {author} {\bibfnamefont {B.}~\bibnamefont
  {{\O}ksendal}},\ }\href@noop {} {\emph {\bibinfo {title} {Stochastic
  Differential Equations: An Introduction with Applications}}},\ \bibinfo
  {edition} {6th}\ ed.\ (\bibinfo  {publisher} {Springer Verlag, Berlin},\
  \bibinfo {year} {2003})\BibitemShut {NoStop}%
\bibitem [{\citenamefont {Onsager}\ and\ \citenamefont
  {Machlup}(1953)}]{Onsager1953}%
  \BibitemOpen
  \bibfield  {author} {\bibinfo {author} {\bibfnamefont {L.}~\bibnamefont
  {Onsager}}\ and\ \bibinfo {author} {\bibfnamefont {S.}~\bibnamefont
  {Machlup}},\ }\bibfield  {title} {\enquote {\bibinfo {title} {Fluctuations
  and irreversible processes},}\ }\href@noop {} {\bibfield  {journal} {\bibinfo
   {journal} {Phys. Rev.}\ }\textbf {\bibinfo {volume} {91}},\ \bibinfo {pages}
  {1505} (\bibinfo {year} {1953})}\BibitemShut {NoStop}%
\bibitem [{\citenamefont {Prinz}\ \emph {et~al.}(2011)\citenamefont {Prinz},
  \citenamefont {Chodera}, \citenamefont {Pande}, \citenamefont {Swope},
  \citenamefont {Smith},\ and\ \citenamefont {No{\'e}}}]{Prinz:2011c}%
  \BibitemOpen
  \bibfield  {author} {\bibinfo {author} {\bibfnamefont {J.-H.}\ \bibnamefont
  {Prinz}}, \bibinfo {author} {\bibfnamefont {J.~D.}\ \bibnamefont {Chodera}},
  \bibinfo {author} {\bibfnamefont {V.~S.}\ \bibnamefont {Pande}}, \bibinfo
  {author} {\bibfnamefont {W.~C.}\ \bibnamefont {Swope}}, \bibinfo {author}
  {\bibfnamefont {J.~C.}\ \bibnamefont {Smith}}, \ and\ \bibinfo {author}
  {\bibfnamefont {F.}~\bibnamefont {No{\'e}}},\ }\bibfield  {title} {\enquote
  {\bibinfo {title} {{Optimal use of data in parallel tempering simulations for
  the construction of discrete-state Markov models of biomolecular
  dynamics}},}\ }\href@noop {} {\bibfield  {journal} {\bibinfo  {journal} {J.
  Chem. Phys.}\ }\textbf {\bibinfo {volume} {134}},\ \bibinfo {pages} {244108}
  (\bibinfo {year} {2011})}\BibitemShut {NoStop}%
\bibitem [{\citenamefont {Sch{\"u}tte}, \citenamefont {Nielsen},\ and\
  \citenamefont {Weber}(2015)}]{Schuette:2015}%
  \BibitemOpen
  \bibfield  {author} {\bibinfo {author} {\bibfnamefont {C.}~\bibnamefont
  {Sch{\"u}tte}}, \bibinfo {author} {\bibfnamefont {A.}~\bibnamefont
  {Nielsen}}, \ and\ \bibinfo {author} {\bibfnamefont {M.}~\bibnamefont
  {Weber}},\ }\bibfield  {title} {\enquote {\bibinfo {title} {{Markov state
  models and molecular alchemy}},}\ }\href@noop {} {\bibfield  {journal}
  {\bibinfo  {journal} {Mol. Phys.}\ }\textbf {\bibinfo {volume} {113}},\
  \bibinfo {pages} {69} (\bibinfo {year} {2015})}\BibitemShut {NoStop}%
\bibitem [{\citenamefont {Donati}, \citenamefont {Hartmann},\ and\
  \citenamefont {Keller}(2017)}]{Donati2017}%
  \BibitemOpen
  \bibfield  {author} {\bibinfo {author} {\bibfnamefont {L.}~\bibnamefont
  {Donati}}, \bibinfo {author} {\bibfnamefont {C.}~\bibnamefont {Hartmann}}, \
  and\ \bibinfo {author} {\bibfnamefont {B.~G.}\ \bibnamefont {Keller}},\
  }\bibfield  {title} {\enquote {\bibinfo {title} {Girsanov reweighting for
  path ensembles and markov state models},}\ }\href@noop {} {\bibfield
  {journal} {\bibinfo  {journal} {J. Chem. Phys.}\ }\textbf {\bibinfo {volume}
  {{146}}},\ \bibinfo {pages} {244112} (\bibinfo {year} {2017})}\BibitemShut
  {NoStop}%
\bibitem [{\citenamefont {Donati}\ and\ \citenamefont
  {Keller}(2018)}]{Donati2018}%
  \BibitemOpen
  \bibfield  {author} {\bibinfo {author} {\bibfnamefont {L.}~\bibnamefont
  {Donati}}\ and\ \bibinfo {author} {\bibfnamefont {B.~G.}\ \bibnamefont
  {Keller}},\ }\bibfield  {title} {\enquote {\bibinfo {title} {Girsanov
  reweighting for metadynamics simulations},}\ }\href@noop {} {\bibfield
  {journal} {\bibinfo  {journal} {J. Chem. Phys.}\ }\textbf {\bibinfo {volume}
  {{149}}},\ \bibinfo {pages} {072335} (\bibinfo {year} {2018})}\BibitemShut
  {NoStop}%
\bibitem [{\citenamefont {Huisinga}, \citenamefont {Sch{\"u}tte},\ and\
  \citenamefont {Stuart}(2003)}]{Huisinga2003}%
  \BibitemOpen
  \bibfield  {author} {\bibinfo {author} {\bibfnamefont {W.}~\bibnamefont
  {Huisinga}}, \bibinfo {author} {\bibfnamefont {C.}~\bibnamefont
  {Sch{\"u}tte}}, \ and\ \bibinfo {author} {\bibfnamefont {A.}~\bibnamefont
  {Stuart}},\ }\bibfield  {title} {\enquote {\bibinfo {title} {Extracting
  macroscopic stochastic dynamics: model problems},}\ }\href@noop {} {\bibfield
   {journal} {\bibinfo  {journal} {Commun. Pure Appl. Math.}\ }\textbf
  {\bibinfo {volume} {{56}}},\ \bibinfo {pages} {234} (\bibinfo {year}
  {2003})}\BibitemShut {NoStop}%
\bibitem [{\citenamefont {Swope}, \citenamefont {Pitera},\ and\ \citenamefont
  {Suits}(2004)}]{Swope2004}%
  \BibitemOpen
  \bibfield  {author} {\bibinfo {author} {\bibfnamefont {W.~C.}\ \bibnamefont
  {Swope}}, \bibinfo {author} {\bibfnamefont {J.~W.}\ \bibnamefont {Pitera}}, \
  and\ \bibinfo {author} {\bibfnamefont {F.}~\bibnamefont {Suits}},\ }\bibfield
   {title} {\enquote {\bibinfo {title} {Describing protein folding kinetics by
  molecular dynamics simulations. 1. theory},}\ }\href@noop {} {\bibfield
  {journal} {\bibinfo  {journal} {J. Phys. Chem. B}\ }\textbf {\bibinfo
  {volume} {{108}}},\ \bibinfo {pages} {6571} (\bibinfo {year}
  {2004})}\BibitemShut {NoStop}%
\bibitem [{\citenamefont {Buchete}\ and\ \citenamefont
  {Hummer}(2008)}]{Buchete:2008}%
  \BibitemOpen
  \bibfield  {author} {\bibinfo {author} {\bibfnamefont {N.-V.}\ \bibnamefont
  {Buchete}}\ and\ \bibinfo {author} {\bibfnamefont {G.}~\bibnamefont
  {Hummer}},\ }\bibfield  {title} {\enquote {\bibinfo {title} {{Coarse Master
  Equations for Peptide Folding Dynamics}},}\ }\href@noop {} {\bibfield
  {journal} {\bibinfo  {journal} {J. Phys. Chem. B}\ }\textbf {\bibinfo
  {volume} {112}},\ \bibinfo {pages} {6057} (\bibinfo {year}
  {2008})}\BibitemShut {NoStop}%
\bibitem [{\citenamefont {Keller}, \citenamefont {Daura},\ and\ \citenamefont
  {van Gunsteren}(2010)}]{Keller:2010}%
  \BibitemOpen
  \bibfield  {author} {\bibinfo {author} {\bibfnamefont {B.}~\bibnamefont
  {Keller}}, \bibinfo {author} {\bibfnamefont {X.}~\bibnamefont {Daura}}, \
  and\ \bibinfo {author} {\bibfnamefont {W.~F.}\ \bibnamefont {van
  Gunsteren}},\ }\bibfield  {title} {\enquote {\bibinfo {title} {{Comparing
  geometric and kinetic cluster algorithms for molecular simulation data}},}\
  }\href@noop {} {\bibfield  {journal} {\bibinfo  {journal} {J. Chem. Phys.}\
  }\textbf {\bibinfo {volume} {132}},\ \bibinfo {pages} {074110} (\bibinfo
  {year} {2010})}\BibitemShut {NoStop}%
\bibitem [{\citenamefont {J.-H.}\ \emph {et~al.}(2011)\citenamefont {J.-H.},
  \citenamefont {Wu}, \citenamefont {Sarich}, \citenamefont {Keller},
  \citenamefont {Senne}, \citenamefont {Held}, \citenamefont {Chodera},
  \citenamefont {Sch{\"u}tte},\ and\ \citenamefont {No\'{e}}}]{Prinz2011}%
  \BibitemOpen
  \bibfield  {author} {\bibinfo {author} {\bibfnamefont {P.}~\bibnamefont
  {J.-H.}}, \bibinfo {author} {\bibfnamefont {H.}~\bibnamefont {Wu}}, \bibinfo
  {author} {\bibfnamefont {M.}~\bibnamefont {Sarich}}, \bibinfo {author}
  {\bibfnamefont {B.~G.}\ \bibnamefont {Keller}}, \bibinfo {author}
  {\bibfnamefont {M.}~\bibnamefont {Senne}}, \bibinfo {author} {\bibfnamefont
  {M.}~\bibnamefont {Held}}, \bibinfo {author} {\bibfnamefont {J.~D.}\
  \bibnamefont {Chodera}}, \bibinfo {author} {\bibfnamefont {C.}~\bibnamefont
  {Sch{\"u}tte}}, \ and\ \bibinfo {author} {\bibfnamefont {F.}~\bibnamefont
  {No\'{e}}},\ }\bibfield  {title} {\enquote {\bibinfo {title} {{Markov models
  of molecular kinetics: Generation and validation}},}\ }\href@noop {}
  {\bibfield  {journal} {\bibinfo  {journal} {J. Chem. Phys.}\ }\textbf
  {\bibinfo {volume} {134}},\ \bibinfo {pages} {174105} (\bibinfo {year}
  {2011})}\BibitemShut {NoStop}%
\bibitem [{\citenamefont {Prinz}, \citenamefont {Keller},\ and\ \citenamefont
  {No{\'e}}(2011)}]{Prinz:2011b}%
  \BibitemOpen
  \bibfield  {author} {\bibinfo {author} {\bibfnamefont {J.-H.}\ \bibnamefont
  {Prinz}}, \bibinfo {author} {\bibfnamefont {B.}~\bibnamefont {Keller}}, \
  and\ \bibinfo {author} {\bibfnamefont {F.}~\bibnamefont {No{\'e}}},\
  }\bibfield  {title} {\enquote {\bibinfo {title} {{Probing molecular kinetics
  with Markov models: metastable states, transition pathways and spectroscopic
  observables}},}\ }\href@noop {} {\bibfield  {journal} {\bibinfo  {journal}
  {Physical Chemistry Chemical Physics}\ }\textbf {\bibinfo {volume} {13}},\
  \bibinfo {pages} {16912} (\bibinfo {year} {2011})}\BibitemShut {NoStop}%
\bibitem [{\citenamefont {Husic}\ and\ \citenamefont
  {Pande}(2018)}]{Husic:2018}%
  \BibitemOpen
  \bibfield  {author} {\bibinfo {author} {\bibfnamefont {B.~E.}\ \bibnamefont
  {Husic}}\ and\ \bibinfo {author} {\bibfnamefont {V.~S.}\ \bibnamefont
  {Pande}},\ }\bibfield  {title} {\enquote {\bibinfo {title} {{Markov State
  Models: From an Art to a Science}},}\ }\href@noop {} {\bibfield  {journal}
  {\bibinfo  {journal} {J. Am. Chem. Soc.}\ }\textbf {\bibinfo {volume}
  {140}},\ \bibinfo {pages} {2386} (\bibinfo {year} {2018})}\BibitemShut
  {NoStop}%
\bibitem [{\citenamefont {Chodera}\ \emph {et~al.}(2011)\citenamefont
  {Chodera}, \citenamefont {Swope}, \citenamefont {No{\'e}}, \citenamefont
  {Prinz}, \citenamefont {Shirts},\ and\ \citenamefont {Pande}}]{Chodera:2011}%
  \BibitemOpen
  \bibfield  {author} {\bibinfo {author} {\bibfnamefont {J.~D.}\ \bibnamefont
  {Chodera}}, \bibinfo {author} {\bibfnamefont {W.~C.}\ \bibnamefont {Swope}},
  \bibinfo {author} {\bibfnamefont {F.}~\bibnamefont {No{\'e}}}, \bibinfo
  {author} {\bibfnamefont {J.-H.}\ \bibnamefont {Prinz}}, \bibinfo {author}
  {\bibfnamefont {M.~R.}\ \bibnamefont {Shirts}}, \ and\ \bibinfo {author}
  {\bibfnamefont {V.~S.}\ \bibnamefont {Pande}},\ }\bibfield  {title} {\enquote
  {\bibinfo {title} {{Dynamical reweighting: Improved estimates of dynamical
  properties from simulations at multiple temperatures}},}\ }\href@noop {}
  {\bibfield  {journal} {\bibinfo  {journal} {J. Chem. Phys.}\ }\textbf
  {\bibinfo {volume} {134}},\ \bibinfo {pages} {244107} (\bibinfo {year}
  {2011})}\BibitemShut {NoStop}%
\bibitem [{\citenamefont {van Gunsteren}\ and\ \citenamefont
  {Berendsen}(1981)}]{vanGunsteren:1981}%
  \BibitemOpen
  \bibfield  {author} {\bibinfo {author} {\bibfnamefont {W.~F.}\ \bibnamefont
  {van Gunsteren}}\ and\ \bibinfo {author} {\bibfnamefont {H.~J.~C.}\
  \bibnamefont {Berendsen}},\ }\bibfield  {title} {\enquote {\bibinfo {title}
  {{Algorithms for brownian dynamics}},}\ }\href@noop {} {\bibfield  {journal}
  {\bibinfo  {journal} {Mol. Phys.}\ }\textbf {\bibinfo {volume} {45}},\
  \bibinfo {pages} {637} (\bibinfo {year} {1981})}\BibitemShut {NoStop}%
\bibitem [{\citenamefont {Br{\"u}nger}, \citenamefont {Brooks},\ and\
  \citenamefont {Karplus}(1984)}]{Brunger1984}%
  \BibitemOpen
  \bibfield  {author} {\bibinfo {author} {\bibfnamefont {A.}~\bibnamefont
  {Br{\"u}nger}}, \bibinfo {author} {\bibfnamefont {C.~L.}\ \bibnamefont
  {Brooks}}, \ and\ \bibinfo {author} {\bibfnamefont {M.}~\bibnamefont
  {Karplus}},\ }\bibfield  {title} {\enquote {\bibinfo {title} {{Stochastic
  boundary conditions for molecular dynamics simulations of ST2 water}},}\
  }\href@noop {} {\bibfield  {journal} {\bibinfo  {journal} {Chem. Phys.
  Lett.}\ }\textbf {\bibinfo {volume} {105}},\ \bibinfo {pages} {495} (\bibinfo
  {year} {1984})}\BibitemShut {NoStop}%
\bibitem [{\citenamefont {Stoltz}(2007)}]{Stoltz:2007}%
  \BibitemOpen
  \bibfield  {author} {\bibinfo {author} {\bibfnamefont {G.}~\bibnamefont
  {Stoltz}},\ }\bibfield  {title} {\enquote {\bibinfo {title} {{Path sampling
  with stochastic dynamics: Some new algorithms}},}\ }\href@noop {} {\bibfield
  {journal} {\bibinfo  {journal} {J. Comp. Phys.}\ }\textbf {\bibinfo {volume}
  {225}},\ \bibinfo {pages} {491} (\bibinfo {year} {2007})}\BibitemShut
  {NoStop}%
\bibitem [{\citenamefont {Bussi}\ and\ \citenamefont
  {Parrinello}(2007)}]{Bussi2007}%
  \BibitemOpen
  \bibfield  {author} {\bibinfo {author} {\bibfnamefont {G.}~\bibnamefont
  {Bussi}}\ and\ \bibinfo {author} {\bibfnamefont {M.}~\bibnamefont
  {Parrinello}},\ }\bibfield  {title} {\enquote {\bibinfo {title} {Accurate
  sampling using langevin dynamics},}\ }\href@noop {} {\bibfield  {journal}
  {\bibinfo  {journal} {Phys. Rev. E}\ }\textbf {\bibinfo {volume} {75}},\
  \bibinfo {pages} {056707} (\bibinfo {year} {2007})}\BibitemShut {NoStop}%
\bibitem [{\citenamefont {Ceriotti}, \citenamefont {Bussi},\ and\ \citenamefont
  {Parrinello}(2009)}]{Ceriotti2009}%
  \BibitemOpen
  \bibfield  {author} {\bibinfo {author} {\bibfnamefont {M.}~\bibnamefont
  {Ceriotti}}, \bibinfo {author} {\bibfnamefont {G.}~\bibnamefont {Bussi}}, \
  and\ \bibinfo {author} {\bibfnamefont {M.}~\bibnamefont {Parrinello}},\
  }\bibfield  {title} {\enquote {\bibinfo {title} {Langevin equation with
  colored noise for constant-temperature molecular dynamics simulations},}\
  }\href@noop {} {\bibfield  {journal} {\bibinfo  {journal} {Phys. Rev. Lett.}\
  }\textbf {\bibinfo {volume} {102}},\ \bibinfo {pages} {020601} (\bibinfo
  {year} {2009})}\BibitemShut {NoStop}%
\bibitem [{\citenamefont {Izaguirre}, \citenamefont {Sweet},\ and\
  \citenamefont {Pande}(2010)}]{Izaguirre2010}%
  \BibitemOpen
  \bibfield  {author} {\bibinfo {author} {\bibfnamefont {J.~A.~.}\ \bibnamefont
  {Izaguirre}}, \bibinfo {author} {\bibfnamefont {C.~R.}\ \bibnamefont
  {Sweet}}, \ and\ \bibinfo {author} {\bibfnamefont {V.}~\bibnamefont
  {Pande}},\ }\bibfield  {title} {\enquote {\bibinfo {title} {Multiscale
  dynamics of macromolecules using normal mode langevin},}\ }\href@noop {}
  {\bibfield  {journal} {\bibinfo  {journal} {Pacific Symposium on
  Biocomputing}\ }\textbf {\bibinfo {volume} {{15}}},\ \bibinfo {pages} {240}
  (\bibinfo {year} {2010})}\BibitemShut {NoStop}%
\bibitem [{\citenamefont {Goga}\ \emph {et~al.}(2012)\citenamefont {Goga},
  \citenamefont {Rzepiela}, \citenamefont {de~Vries}, \citenamefont {Marrink},\
  and\ \citenamefont {Berendsen}}]{Goga:2012}%
  \BibitemOpen
  \bibfield  {author} {\bibinfo {author} {\bibfnamefont {N.}~\bibnamefont
  {Goga}}, \bibinfo {author} {\bibfnamefont {A.~J.}\ \bibnamefont {Rzepiela}},
  \bibinfo {author} {\bibfnamefont {A.~H.}\ \bibnamefont {de~Vries}}, \bibinfo
  {author} {\bibfnamefont {S.~J.}\ \bibnamefont {Marrink}}, \ and\ \bibinfo
  {author} {\bibfnamefont {H.~J.~C.}\ \bibnamefont {Berendsen}},\ }\bibfield
  {title} {\enquote {\bibinfo {title} {{Efficient Algorithms for Langevin and
  DPD Dynamics.}}}\ }\href@noop {} {\bibfield  {journal} {\bibinfo  {journal}
  {J. Chem. Theory Comput.}\ }\textbf {\bibinfo {volume} {8}},\ \bibinfo
  {pages} {3637} (\bibinfo {year} {2012})}\BibitemShut {NoStop}%
\bibitem [{\citenamefont {Leimkuhler}\ and\ \citenamefont
  {Matthews}(2013)}]{Leimkuhler:2013}%
  \BibitemOpen
  \bibfield  {author} {\bibinfo {author} {\bibfnamefont {B.}~\bibnamefont
  {Leimkuhler}}\ and\ \bibinfo {author} {\bibfnamefont {C.}~\bibnamefont
  {Matthews}},\ }\bibfield  {title} {\enquote {\bibinfo {title} {{Robust and
  efficient configurational molecular sampling via Langevin dynamics}},}\
  }\href@noop {} {\bibfield  {journal} {\bibinfo  {journal} {J. Chem. Phys.}\
  }\textbf {\bibinfo {volume} {138}},\ \bibinfo {pages} {174102} (\bibinfo
  {year} {2013})}\BibitemShut {NoStop}%
\bibitem [{\citenamefont {Sivak}, \citenamefont {Chodera},\ and\ \citenamefont
  {Crooks}(2014)}]{Sivak:2014}%
  \BibitemOpen
  \bibfield  {author} {\bibinfo {author} {\bibfnamefont {D.~A.}\ \bibnamefont
  {Sivak}}, \bibinfo {author} {\bibfnamefont {J.~D.}\ \bibnamefont {Chodera}},
  \ and\ \bibinfo {author} {\bibfnamefont {G.~E.}\ \bibnamefont {Crooks}},\
  }\bibfield  {title} {\enquote {\bibinfo {title} {{Time step rescaling
  recovers continuous-time dynamical properties for discrete-time Langevin
  integration of nonequilibrium systems.}}}\ }\href@noop {} {\bibfield
  {journal} {\bibinfo  {journal} {J. Phys. Chem. B}\ }\textbf {\bibinfo
  {volume} {118}},\ \bibinfo {pages} {6466} (\bibinfo {year}
  {2014})}\BibitemShut {NoStop}%
\bibitem [{\citenamefont {Fass}\ \emph {et~al.}(2018)\citenamefont {Fass},
  \citenamefont {Sivak}, \citenamefont {Crooks}, \citenamefont {Beauchamp},
  \citenamefont {Leimkuhler},\ and\ \citenamefont {Chodera}}]{Fass2018}%
  \BibitemOpen
  \bibfield  {author} {\bibinfo {author} {\bibfnamefont {J.}~\bibnamefont
  {Fass}}, \bibinfo {author} {\bibfnamefont {D.~A.}\ \bibnamefont {Sivak}},
  \bibinfo {author} {\bibfnamefont {G.~E.}\ \bibnamefont {Crooks}}, \bibinfo
  {author} {\bibfnamefont {K.~A.}\ \bibnamefont {Beauchamp}}, \bibinfo {author}
  {\bibfnamefont {B.}~\bibnamefont {Leimkuhler}}, \ and\ \bibinfo {author}
  {\bibfnamefont {J.~D.}\ \bibnamefont {Chodera}},\ }\bibfield  {title}
  {\enquote {\bibinfo {title} {Quantifying configuration-sampling error in
  langevin simulations of complex molecular systems},}\ }\href@noop {}
  {\bibfield  {journal} {\bibinfo  {journal} {Entropy}\ }\textbf {\bibinfo
  {volume} {20}},\ \bibinfo {pages} {318} (\bibinfo {year} {2018})}\BibitemShut
  {NoStop}%
\bibitem [{\citenamefont {Eastman}\ \emph {et~al.}(2017)\citenamefont
  {Eastman}, \citenamefont {Swails}, \citenamefont {Chodera}, \citenamefont
  {McGibbon}, \citenamefont {Zhao}, \citenamefont {Beauchamp}, \citenamefont
  {Wang}, \citenamefont {Simmonett}, \citenamefont {Harrigan}, \citenamefont
  {Stern}, \citenamefont {Wiewiora}, \citenamefont {Brooks},\ and\
  \citenamefont {Pande}}]{Eastman2017}%
  \BibitemOpen
  \bibfield  {author} {\bibinfo {author} {\bibfnamefont {P.}~\bibnamefont
  {Eastman}}, \bibinfo {author} {\bibfnamefont {J.}~\bibnamefont {Swails}},
  \bibinfo {author} {\bibfnamefont {J.~D.}\ \bibnamefont {Chodera}}, \bibinfo
  {author} {\bibfnamefont {R.~T.}\ \bibnamefont {McGibbon}}, \bibinfo {author}
  {\bibfnamefont {Y.}~\bibnamefont {Zhao}}, \bibinfo {author} {\bibfnamefont
  {K.~A.}\ \bibnamefont {Beauchamp}}, \bibinfo {author} {\bibfnamefont {L.-P.}\
  \bibnamefont {Wang}}, \bibinfo {author} {\bibfnamefont {A.~C.}\ \bibnamefont
  {Simmonett}}, \bibinfo {author} {\bibfnamefont {M.~P.}\ \bibnamefont
  {Harrigan}}, \bibinfo {author} {\bibfnamefont {C.~D.}\ \bibnamefont {Stern}},
  \bibinfo {author} {\bibfnamefont {R.~P.}\ \bibnamefont {Wiewiora}}, \bibinfo
  {author} {\bibfnamefont {B.~R.}\ \bibnamefont {Brooks}}, \ and\ \bibinfo
  {author} {\bibfnamefont {V.~S.}\ \bibnamefont {Pande}},\ }\bibfield  {title}
  {\enquote {\bibinfo {title} {Openmm 7: Rapid development of high performance
  algorithms for molecular dynamics},}\ }\href@noop {} {\bibfield  {journal}
  {\bibinfo  {journal} {PLOS Comp. Biol.}\ }\textbf {\bibinfo {volume}
  {{13}}},\ \bibinfo {pages} {1} (\bibinfo {year} {2017})}\BibitemShut
  {NoStop}%
\bibitem [{\citenamefont {Bou-Rabee}(2014)}]{Rabee2014}%
  \BibitemOpen
  \bibfield  {author} {\bibinfo {author} {\bibfnamefont {N.}~\bibnamefont
  {Bou-Rabee}},\ }\bibfield  {title} {\enquote {\bibinfo {title} {Time
  integrators for molecular dynamics},}\ }\href@noop {} {\bibfield  {journal}
  {\bibinfo  {journal} {Entropy}\ }\textbf {\bibinfo {volume} {{16}}},\
  \bibinfo {pages} {138} (\bibinfo {year} {2014})}\BibitemShut {NoStop}%
\bibitem [{\citenamefont {Chow}\ and\ \citenamefont {Buice}(2015)}]{Chow2015}%
  \BibitemOpen
  \bibfield  {author} {\bibinfo {author} {\bibfnamefont {C.~C.}\ \bibnamefont
  {Chow}}\ and\ \bibinfo {author} {\bibfnamefont {M.~A.}\ \bibnamefont
  {Buice}},\ }\bibfield  {title} {\enquote {\bibinfo {title} {Path integral
  methods for stochastic differential equations},}\ }\href@noop {} {\bibfield
  {journal} {\bibinfo  {journal} {J. Math. Neurosci.}\ }\textbf {\bibinfo
  {volume} {{5}}},\ \bibinfo {pages} {1} (\bibinfo {year} {2015})}\BibitemShut
  {NoStop}%
\bibitem [{\citenamefont {Bressloff}(2014)}]{Bressloff2014}%
  \BibitemOpen
  \bibfield  {author} {\bibinfo {author} {\bibfnamefont {P.~C.}\ \bibnamefont
  {Bressloff}},\ }\href@noop {} {\emph {\bibinfo {title} {Stochastic Processes
  in Cell Biology}}},\ \bibinfo {edition} {1st}\ ed.\ (\bibinfo  {publisher}
  {Springer, New York},\ \bibinfo {year} {2014})\BibitemShut {NoStop}%
\bibitem [{\citenamefont {H{\"u}nenberger}(2005)}]{Hunenberger:2005}%
  \BibitemOpen
  \bibfield  {author} {\bibinfo {author} {\bibfnamefont {P.~H.}\ \bibnamefont
  {H{\"u}nenberger}},\ }\bibfield  {title} {\enquote {\bibinfo {title}
  {{Thermostat Algorithms for Molecular Dynamics Simulations}},}\ }in\
  \href@noop {} {\emph {\bibinfo {booktitle} {Advanced Computer Simulation}}}\
  (\bibinfo  {publisher} {Springer, Berlin, Heidelberg},\ \bibinfo {address}
  {Berlin, Heidelberg},\ \bibinfo {year} {2005})\BibitemShut {NoStop}%
\bibitem [{\citenamefont {Basconi}\ and\ \citenamefont
  {Shirts}(2013)}]{Basconi:2013}%
  \BibitemOpen
  \bibfield  {author} {\bibinfo {author} {\bibfnamefont {J.~E.}\ \bibnamefont
  {Basconi}}\ and\ \bibinfo {author} {\bibfnamefont {M.~R.}\ \bibnamefont
  {Shirts}},\ }\bibfield  {title} {\enquote {\bibinfo {title} {{Effects of
  Temperature Control Algorithms on Transport Properties and Kinetics in
  Molecular Dynamics Simulations.}}}\ }\href@noop {} {\bibfield  {journal}
  {\bibinfo  {journal} {J. Chem. Theory Comput.}\ }\textbf {\bibinfo {volume}
  {9}},\ \bibinfo {pages} {2887} (\bibinfo {year} {2013})}\BibitemShut
  {NoStop}%
\bibitem [{\citenamefont {Wang}\ \emph {et~al.}(2004)\citenamefont {Wang},
  \citenamefont {Wolf}, \citenamefont {Caldwell}, \citenamefont {Kollman},\
  and\ \citenamefont {Case}}]{Wang2004}%
  \BibitemOpen
  \bibfield  {author} {\bibinfo {author} {\bibfnamefont {J.}~\bibnamefont
  {Wang}}, \bibinfo {author} {\bibfnamefont {R.~M.}\ \bibnamefont {Wolf}},
  \bibinfo {author} {\bibfnamefont {J.~W.}\ \bibnamefont {Caldwell}}, \bibinfo
  {author} {\bibfnamefont {P.~A.}\ \bibnamefont {Kollman}}, \ and\ \bibinfo
  {author} {\bibfnamefont {D.~A.}\ \bibnamefont {Case}},\ }\bibfield  {title}
  {\enquote {\bibinfo {title} {Development and testing of a general amber force
  field},}\ }\href@noop {} {\bibfield  {journal} {\bibinfo  {journal} {J.
  Comput. Chem.}\ }\textbf {\bibinfo {volume} {25}},\ \bibinfo {pages} {1157}
  (\bibinfo {year} {2004})}\BibitemShut {NoStop}%
\bibitem [{\citenamefont {Onufriev}, \citenamefont {Bashford},\ and\
  \citenamefont {Case}(2004)}]{Onufriev2004}%
  \BibitemOpen
  \bibfield  {author} {\bibinfo {author} {\bibfnamefont {A.}~\bibnamefont
  {Onufriev}}, \bibinfo {author} {\bibfnamefont {D.}~\bibnamefont {Bashford}},
  \ and\ \bibinfo {author} {\bibfnamefont {D.~A.}\ \bibnamefont {Case}},\
  }\bibfield  {title} {\enquote {\bibinfo {title} {Exploring protein native
  states and large-scale conformational changes with a modified generalized
  born model},}\ }\href@noop {} {\bibfield  {journal} {\bibinfo  {journal} {J.
  Comput. Chem.}\ }\textbf {\bibinfo {volume} {55}},\ \bibinfo {pages} {383}
  (\bibinfo {year} {2004})}\BibitemShut {NoStop}%
\bibitem [{Ope({\natexlab{a}})}]{OpenMM_CustomIntegrator}%
  \BibitemOpen
  \href@noop {} {}\bibinfo {howpublished}
  {\url{http://docs.openmm.org/latest/api-python/generated/simtk.openmm.openmm.CustomIntegrator.html}}
  ({\natexlab{a}}),\ \bibinfo {note} {[Online; accessed
  25-January-2021]}\BibitemShut {NoStop}%
\bibitem [{\citenamefont {Bause}\ \emph {et~al.}(2019)\citenamefont {Bause},
  \citenamefont {Wittenstein}, \citenamefont {Kremer},\ and\ \citenamefont
  {Bereau}}]{Bause:2019}%
  \BibitemOpen
  \bibfield  {author} {\bibinfo {author} {\bibfnamefont {M.}~\bibnamefont
  {Bause}}, \bibinfo {author} {\bibfnamefont {T.}~\bibnamefont {Wittenstein}},
  \bibinfo {author} {\bibfnamefont {K.}~\bibnamefont {Kremer}}, \ and\ \bibinfo
  {author} {\bibfnamefont {T.}~\bibnamefont {Bereau}},\ }\bibfield  {title}
  {\enquote {\bibinfo {title} {{Microscopic reweighting for nonequilibrium
  steady-state dynamics.}}}\ }\href@noop {} {\bibfield  {journal} {\bibinfo
  {journal} {Phys. Rev. E}\ }\textbf {\bibinfo {volume} {100}},\ \bibinfo
  {pages} {060103} (\bibinfo {year} {2019})}\BibitemShut {NoStop}%
\bibitem [{\citenamefont {P{\'e}rez-Hern{\'a}ndez}\ \emph
  {et~al.}(2013)\citenamefont {P{\'e}rez-Hern{\'a}ndez}, \citenamefont {Paul},
  \citenamefont {Giorgino}, \citenamefont {De~Fabritiis},\ and\ \citenamefont
  {No{\'e}}}]{PerezHernandez:2013}%
  \BibitemOpen
  \bibfield  {author} {\bibinfo {author} {\bibfnamefont {G.}~\bibnamefont
  {P{\'e}rez-Hern{\'a}ndez}}, \bibinfo {author} {\bibfnamefont
  {F.}~\bibnamefont {Paul}}, \bibinfo {author} {\bibfnamefont {T.}~\bibnamefont
  {Giorgino}}, \bibinfo {author} {\bibfnamefont {G.}~\bibnamefont
  {De~Fabritiis}}, \ and\ \bibinfo {author} {\bibfnamefont {F.}~\bibnamefont
  {No{\'e}}},\ }\bibfield  {title} {\enquote {\bibinfo {title} {{Identification
  of slow molecular order parameters for Markov model construction}},}\
  }\href@noop {} {\bibfield  {journal} {\bibinfo  {journal} {J. Chem. Phys.}\
  }\textbf {\bibinfo {volume} {139}},\ \bibinfo {pages} {015102} (\bibinfo
  {year} {2013})}\BibitemShut {NoStop}%
\bibitem [{\citenamefont {N{\"u}ske}\ \emph {et~al.}(2014)\citenamefont
  {N{\"u}ske}, \citenamefont {Keller}, \citenamefont {P{\'e}rez-Hern{\'a}ndez},
  \citenamefont {Mey},\ and\ \citenamefont {No{\'e}}}]{Nueske:2014}%
  \BibitemOpen
  \bibfield  {author} {\bibinfo {author} {\bibfnamefont {F.}~\bibnamefont
  {N{\"u}ske}}, \bibinfo {author} {\bibfnamefont {B.~G.}\ \bibnamefont
  {Keller}}, \bibinfo {author} {\bibfnamefont {G.}~\bibnamefont
  {P{\'e}rez-Hern{\'a}ndez}}, \bibinfo {author} {\bibfnamefont {A.~S. J.~S.}\
  \bibnamefont {Mey}}, \ and\ \bibinfo {author} {\bibfnamefont
  {F.}~\bibnamefont {No{\'e}}},\ }\bibfield  {title} {\enquote {\bibinfo
  {title} {{Variational Approach to Molecular Kinetics}},}\ }\href@noop {}
  {\bibfield  {journal} {\bibinfo  {journal} {J. Chem. Theory Comput.}\
  }\textbf {\bibinfo {volume} {10}},\ \bibinfo {pages} {1739--1752} (\bibinfo
  {year} {2014})}\BibitemShut {NoStop}%
\bibitem [{\citenamefont {Lemke}\ and\ \citenamefont
  {Keller}(2016)}]{Lemke:2016}%
  \BibitemOpen
  \bibfield  {author} {\bibinfo {author} {\bibfnamefont {O.}~\bibnamefont
  {Lemke}}\ and\ \bibinfo {author} {\bibfnamefont {B.~G.}\ \bibnamefont
  {Keller}},\ }\bibfield  {title} {\enquote {\bibinfo {title} {{Density-based
  cluster algorithms for the identification of core sets}},}\ }\href@noop {}
  {\bibfield  {journal} {\bibinfo  {journal} {J. Chem. Phys.}\ }\textbf
  {\bibinfo {volume} {145}},\ \bibinfo {pages} {164104} (\bibinfo {year}
  {2016})}\BibitemShut {NoStop}%
\bibitem [{\citenamefont {Chong}, \citenamefont {Saglam},\ and\ \citenamefont
  {Zuckerman}(2017)}]{Chong:2017}%
  \BibitemOpen
  \bibfield  {author} {\bibinfo {author} {\bibfnamefont {L.~T.}\ \bibnamefont
  {Chong}}, \bibinfo {author} {\bibfnamefont {A.~S.}\ \bibnamefont {Saglam}}, \
  and\ \bibinfo {author} {\bibfnamefont {D.~M.}\ \bibnamefont {Zuckerman}},\
  }\bibfield  {title} {\enquote {\bibinfo {title} {{Path-sampling strategies
  for simulating rare events in biomolecular systems}},}\ }\href@noop {}
  {\bibfield  {journal} {\bibinfo  {journal} {Curr. Opin. Struct. Biol.}\
  }\textbf {\bibinfo {volume} {43}},\ \bibinfo {pages} {88--94} (\bibinfo
  {year} {2017})}\BibitemShut {NoStop}%
\bibitem [{\citenamefont {Grazioli}\ and\ \citenamefont
  {Andricioaei}(2018)}]{Grazioli:2018}%
  \BibitemOpen
  \bibfield  {author} {\bibinfo {author} {\bibfnamefont {G.}~\bibnamefont
  {Grazioli}}\ and\ \bibinfo {author} {\bibfnamefont {I.}~\bibnamefont
  {Andricioaei}},\ }\bibfield  {title} {\enquote {\bibinfo {title} {{Advances
  in milestoning. I. Enhanced sampling via wind-assisted reweighted milestoning
  (WARM)}},}\ }\href@noop {} {\bibfield  {journal} {\bibinfo  {journal} {J.
  Chem. Phys.}\ }\textbf {\bibinfo {volume} {149}},\ \bibinfo {pages} {084103}
  (\bibinfo {year} {2018})}\BibitemShut {NoStop}%
\bibitem [{\citenamefont {Dixit}\ \emph {et~al.}(2018)\citenamefont {Dixit},
  \citenamefont {Wagoner}, \citenamefont {Weistuch}, \citenamefont
  {Press{\'e}}, \citenamefont {Ghosh},\ and\ \citenamefont
  {Dill}}]{Dixit:2018}%
  \BibitemOpen
  \bibfield  {author} {\bibinfo {author} {\bibfnamefont {P.~D.}\ \bibnamefont
  {Dixit}}, \bibinfo {author} {\bibfnamefont {J.}~\bibnamefont {Wagoner}},
  \bibinfo {author} {\bibfnamefont {C.}~\bibnamefont {Weistuch}}, \bibinfo
  {author} {\bibfnamefont {S.}~\bibnamefont {Press{\'e}}}, \bibinfo {author}
  {\bibfnamefont {K.}~\bibnamefont {Ghosh}}, \ and\ \bibinfo {author}
  {\bibfnamefont {K.~A.}\ \bibnamefont {Dill}},\ }\bibfield  {title} {\enquote
  {\bibinfo {title} {{Perspective: Maximum caliber is a general variational
  principle for dynamical systems}},}\ }\href@noop {} {\bibfield  {journal}
  {\bibinfo  {journal} {J. Chem. Phys.}\ }\textbf {\bibinfo {volume} {148}},\
  \bibinfo {pages} {010901} (\bibinfo {year} {2018})}\BibitemShut {NoStop}%
\bibitem [{\citenamefont {Peter}, \citenamefont {Shea},\ and\ \citenamefont
  {Schug}(2020)}]{Peter:2020}%
  \BibitemOpen
  \bibfield  {author} {\bibinfo {author} {\bibfnamefont {E.~K.}\ \bibnamefont
  {Peter}}, \bibinfo {author} {\bibfnamefont {J.-E.}\ \bibnamefont {Shea}}, \
  and\ \bibinfo {author} {\bibfnamefont {A.}~\bibnamefont {Schug}},\ }\bibfield
   {title} {\enquote {\bibinfo {title} {{CORE-MD, a path correlated molecular
  dynamics simulation method}},}\ }\href@noop {} {\bibfield  {journal}
  {\bibinfo  {journal} {J. Chem. Phys.}\ }\textbf {\bibinfo {volume} {153}},\
  \bibinfo {pages} {084114} (\bibinfo {year} {2020})}\BibitemShut {NoStop}%
\bibitem [{Ope({\natexlab{b}})}]{OpenMM_CpuLangevinDynamics}%
  \BibitemOpen
  \href@noop {} {}\bibinfo {howpublished}
  {\url{https://github.com/openmm/openmm/blob/master/platforms/cpu/src/CpuLangevinDynamics.cpp}}
  ({\natexlab{b}}),\ \bibinfo {note} {[Online; accessed
  15-November-2020]}\BibitemShut {NoStop}%
\bibitem [{\citenamefont {Gardiner}(1983)}]{Gardiner1983}%
  \BibitemOpen
  \bibfield  {author} {\bibinfo {author} {\bibfnamefont {C.~W.}\ \bibnamefont
  {Gardiner}},\ }\href@noop {} {\emph {\bibinfo {title} {Handbook of Stochastic
  Methods for Physics, Chemistry and the Natural Sciences}}},\ \bibinfo
  {edition} {2nd}\ ed.\ (\bibinfo  {publisher} {Springer Verlag, Berlin
  Heidelberg},\ \bibinfo {year} {1983})\BibitemShut {NoStop}%
\end{thebibliography}%
\end{document}